%% file: main.tex
\documentclass[conference,compsoc]{IEEEtran}
\input{./packages}

\input{./acronyms}

\input{./macros}

\ifCLASSOPTIONcompsoc
  \usepackage[nocompress]{cite}
\else
  \usepackage{cite}
\fi
\ifCLASSINFOpdf
\else
\fi
\begin{document}
%
\title{Zero-Knowledge Location Privacy \\ via Accurate Floating-Point SNARKs}

\author{\IEEEauthorblockN{Jens Ernstberger\IEEEauthorrefmark{1}\textsuperscript{\textsection},
Chengru Zhang\IEEEauthorrefmark{2}\textsuperscript{\textsection},
Luca Ciprian\IEEEauthorrefmark{1}, 
Philipp Jovanovic\IEEEauthorrefmark{3}, and
Sebastian Steinhorst\IEEEauthorrefmark{1}}
\IEEEauthorblockA{\IEEEauthorrefmark{1}Technical University of Munich, Germany}
\IEEEauthorblockA{\IEEEauthorrefmark{2}The University of Hong Kong, Hong Kong}
\IEEEauthorblockA{\IEEEauthorrefmark{3}University College London, United Kingdom}
}

\maketitle
\begingroup\renewcommand\thefootnote{\textsection}
\footnotetext{Both authors contributed equally to this research.}
\endgroup


\maketitle

\begin{abstract}



We introduce \ZKLP, enabling users to prove to third parties that they are within a specified geographical region while not disclosing their exact location. 
\ZKLP supports varying levels of granularity, allowing for customization depending on the use case.
To realize \ZKLP, we introduce the first set of \ZKP circuits that are fully compliant to the IEEE 754 standard for floating-point arithmetic.

Our results demonstrate that our floating point circuits amortize efficiently, requiring only $64$ constraints per operation for $2^{15}$ single-precision floating-point multiplications.
We utilize our floating point implementation to realize the \ZKLP paradigm.
In comparison to a baseline, we find that our optimized implementation has $\factorSingleBetterThanBaselineConstraints$ less constraints utilizing single precision floating-point values, and $\factorDoubleBetterThanBaselineConstraints$ less constraints when  utilizing double precision floating-point values.
We demonstrate the practicability of \ZKLP by building a protocol for \textit{privacy preserving peer-to-peer proximity testing}~---~Alice can test if she is close to Bob by receiving a single message, without either party revealing any other information about their location.
In such a setting, Bob can create a proof of (non-)proximity in $\ProverTimeSinglePrecisionBNGSixteenForCZKLP$, whereas Alice can verify her distance to about $\numberProofsPerSecond$ peers per second.

\end{abstract}


%
\IEEEpeerreviewmaketitle

\section{Introduction}


Location-based services have become integral in the digital age. The widespread use of geolocation-enabled devices, including smartphones and trackers like \textit{Tile} and \textit{AirTag}, has fueled advancements in services reliant on spatial data. 
However, these developments also raise significant privacy concerns, as location data reveals sensitive information about an individual's habits and preferences. 

As a motivating example, consider digital contact tracing. Applications that collect geolocation records can, unintentionally or deliberately, expose sensitive user information, leading to potential privacy breaches.
Moreover, these systems face the dual challenge of ensuring the privacy of honest users while preventing malicious actors from spoofing data to provide deliberately incorrect information.

In response to the first problem of protecting user location privacy,
various \LPPM protocols emerged~\cite{boukoros2019lack, jiang2021location}.
These solutions either apply \emph{(i)} obfuscation~\cite{andres2013geo, lee2011protecting} or \emph{(ii)} cryptographic methods~\cite{khoshgozaran2007blind, popa2009vpriv, narayanan2011location, vsedvenka2014privacy}, where obfuscation reduces the precision of location data, while cryptographic approaches utilize secure computing and encryption to protect privacy. 
However, both solutions have shortcomings: differential privacy \LPPM struggles with correlated user locations~\cite{olteanu2016quantifying}, and both cloaking~\cite{lee2011protecting} and \MPC-based cryptography~\cite{vsedvenka2014privacy} depend on third-party data anonymization.

Similarly, ensuring the authenticity of location information has gained importance in recent years, proliferating numerous solutions to mitigate spoofing~\cite{zhang2017secure}. For example, recent advancements in \GNSS address legacy satellite system vulnerabilities lacking signal authentication~\cite{yuan2023authenticating}. Additionally, efforts to inject fabricated location reports into offline finding networks like Apple's ``Find My'' are countered by manufacturers~\cite{heinrich2021can}.

Considering these challenges, the ideal mechanism would allow clients to submit their own location proof to protect personalized location information without relying on third parties, it would preserve the utility of information with custom granularity, and it would prevent clients from deliberately providing wrong location information.

In this paper, we thus explore the following research question:
\textit{
How can we obtain a short proof of location that allows for customized privacy preservation, while retaining accuracy and utility?
}

We give an affirmative answer to our research question by introducing \textit{Zero-Knowledge Location Privacy}. 
Our approach is driven by recent advancements in \ZKPs, which enable practical use in applications previously deemed too costly.
With \ZKLP, users can prove to any third party that they are within a specific geographical region while obfuscating their exact location for utility and privacy.
To do so, we rely on \DGGS~\cite{sahr2003geodesic} with hexagons as geometric representation, which hierarchically divide the earth into progressively finer resolution grids. 
Any third party can only obtain an obfuscated location, whose correctness can be verified in milliseconds.
To demonstrate the practicability of \ZKLP, we develop a protocol for privacy-preserving peer-to-peer proximity testing and evaluate its practicality. 
While our main protocols focus on providing privacy to user-provided locations, we introduce three possible solutions to ensure non-falsifiable location proofs in Appendix~\ref{appendix:authenticate}. 

\noindent \textbf{Importance of Floating-Point Arithmetic for \ZKLP. }
The necessity of floating-point arithmetic arises from the nature of data types and operations involved in geographic applications.
Geolocation data (such as coordinates) and its associated computations (such as square roots \(\sqrt{\cdot}\) and trigonometric functions \(\sin(\cdot)\), \(\cos(\cdot)\)) are both in the domain of real numbers \(\mathbb{R}\).
In computers, these numbers and operations are typically represented using either fixed-point or floating-point formats.
Generally, fixed-point arithmetic is more efficient and is thus commonly used in \ZKP circuits~\cite{kang2022scaling, kang2022zk}, while computer hardware and software often use floating-point arithmetic, which provides greater flexibility and supports a wider range of values.

For \ZKLP, fixed-point arithmetic presents two key limitations.
First, we need to simultaneously handle large values (e.g., $\sqrt{7}^{\resolutionShort}$, where the hexagonal resolution $\resolutionShort$ can be up to $15$) and small values (e.g., the product of $\sin{\theta}$ and $\cos{\theta}$) (cf. \S\ref{section:zkLocation}), resulting in either increased data size or reduced accuracy in fixed-point representation.
In fact, compared with \FPDouble that requires $64$ bits, fixed-point costs nearly twice as many bits as \FPDouble while achieving lower accuracy (cf. \S\ref{section:evaluation-zklp}).
Second, our \ZKLP circuits need to be compatible with existing geographic applications that are already implemented using IEEE 754~\cite{ieee754fp} floating-point arithmetic, such as Uber H3~\cite{UberH3}.
The use of incompatible formats would compromise the completeness of \ZKLP, since tiny turbulence in computations may yield significantly different results~\cite{srivastava2024optimistic}.
Looking ahead, in \S\ref{section:evaluation-zklp}, with randomly generated test cases, fixed-point representation fails some tests whereas floating-point passes all.
Even worse, this inconsistency can be exploited by adversaries to generate valid proofs for maliciously crafted statements, thereby breaking the soundness of \ZKLP.

\noindent \textbf{Challenges for Floating-Point. }
The major challenge in addressing our research question is achieving efficient floating-point arithmetic that is fully compliant with IEEE 754 in a \SNARK circuit, which is nontrivial.
A \SNARK commonly operates over a finite field $\mathbb{F}_{p}$, where $p$ is usually a large prime depending on the underlying elliptic curve (e.g., $|p| = 254$ for BN254). 
By contrast, floating-point numbers require integer operations in $\mathbb{Z}_{2^k}$ that are circuit-unfriendly, such as comparison and shifts.

Previous works have investigated the use of floating-point values in \MPC~\cite{aliasgari2012secure,kamm2015secure,archer2021cost} and \ZKP~\cite{weng2021mystique,garg2022succinct,garg2023experimenting}. However, they are either \emph{(i)} not optimized for circuit size, \emph{(ii)} inefficient in terms of communication cost and verification time, \emph{(iii)} incapable of handling complex operations, or \emph{(iv)} not fully compliant with the IEEE 754 standard~\cite{ieee754fp} (see \S\ref{section:related-work} for a detailed comparison).

For instance, \cite{archer2021cost,weng2021mystique} naively convert software floating-point implementations to circuits, leading to a huge number of gates.
Garg et. al~\cite{garg2022succinct} address this problem by proving the upper bound of the relative error, instead of transforming rounding operations on floating-point values to in-circuit bitwise operations. In spite of this, their work only supports addition and multiplication while lacking a concrete practical implementation.
To realize \ZKLP, we require complex operations such as division, taking square roots, and trigonometric functions.
In addition, we also need to guarantee that these operations produce correctly rounded results.
Hence, emulating floating-point numbers precisely, whilst attaining efficiency in the operations involved, is challenging and requires significant optimization.

\noindent \textbf{Challenges for \ZKLP. }
Transforming a location described as latitude and longitude to an index in a hexagonal grid system requires extensive use of trigonometric operations.
Common approaches to approximating trigonometric functions by following the standard three-step recipe of \textit{range reduction}, \textit{polynomial approximation}, and \textit{output compensation}, rely on high bitwidths to obtain precise results~\cite{rathee2022secfloat}.
Further, naive polynomial approximation through, e.g., Taylor Series, is prohibitively expensive in-circuit, as precise results demand for many iterations, which is costly to represent in an arithmetic circuit that demands for linearization.




\input{./Figures/grid.tex}

\subsection{Contributions \& Results}

We provide a full implementation of IEEE 754 compliant floating-point operations in \SNARKs and apply them in our implementation \ZKLP.
Our implementation of floating-point arithemtic is agnostic to the underlying SNARK arithemtization and applicable in orthogonal domains.

Our main technical contributions are \emph{(i)} novel optimizations for computing floating-point \SNARKs and  \emph{(ii)} optimizations to eliminate trigonometric operations that make the \ZKLP paradigm practical.
Throughout, we leverage lookup arguments and nondeterministic programming, enabling cost-effective representation of computations that are typically resource-intensive when executed in-circuit.
To efficiently operate on floating-point values, we convert their integer components to an equivalent but circuit-efficient form, build optimized sub-circuits for integer operations, and minimize the number of costly range checks in key steps, e.g., rounding.
For compliance with IEEE 754, we  take additional care of edge cases, such as NaN, $\pm \infty$ and subnormal numbers.
To efficiently instantiate \ZKLP over primitive floating-point operations, we introduce shortcuts, eliminating expensive math operations through trigonometric identities and nondeterministic programming.

Our experiments show that our circuits for primitive floating-point operations are precise and performant. 
We test our implementation with the Berkeley TestFloat library~\cite{TestFloat} to ensure full compliance. 
The extensive use of lookups leads to amortization~---~ $2^1$ \FPSingle multiplications require $209$ constraints, whereas $2^{15}$ \FPSingle multiplications require $64$ constraints per operation.
When applied to the \ZKLP paradigm, our resulting circuits are highly efficient. 
In comparison to an unoptimized fixed-point baseline, our implementation has $\factorSingleBetterThanBaselineConstraints$ less constraints for \FPSingle values, and $\factorDoubleBetterThanBaselineConstraints$ less constraints for \FPDouble values.
We apply \ZKLP and show that it can realize \textit{privacy-preserving peer-to-peer proximity testing}, through which a user can evaluate its proximity to $\numberProofsPerSecond$ peers per second.

\noindent 
In summary, our contributions are as follows:



\begin{itemize}
    \item \textbf{\ZKLP. } We introduce \ZKLP, a novel application of \ZKPs, along with a full implementation and evaluation. \ZKLP unlocks a novel class of \ZKP-empowered applications, which enable personalized location privacy through geo-in\-dis\-ting\-uish\-ability. In particular, it allows individuals to prove that they have visited a certain location whilst ensuring privacy for the exact location. 
     \item \textbf{IEEE 754 Compliant Floating-Point \SNARKs. } We are the first to introduce optimized \SNARK circuits for floating-point operations that are fully compliant to IEEE 754~\cite{ieee754fp} (\S\ref{section:fp-primitive-ops}). Our optimization for primitive floating-point operations are universally applicable, independent of specific arithmetizations, and fulfill the precision requirements of IEEE 754. We thus deem this contribution of independent interest.
    \item \textbf{Optimizations for \ZKLP. } Transformations on geographic coordinates extensively demand for trigonometric functions. We introduce \ZKP circuits that entirely eliminate trigonometric functions for \ZKLP (\S\ref{section:zkLocation}).
    \item \textbf{Evaluation \& Application. } We provide a full implementation and evaluation of all algorithms and optimizations, show the compliance of our floating-point circuits with IEEE 754, and showcase the \ZKLP paradigm for peer-to-peer proximity testing (\S\ref{section:evaluation}).
\end{itemize}



\noindent \textbf{Limitations of \ZKLP. }
Despite our optimizations making both floating-point and \ZKLP circuits practical, there are notable limitations to our current evaluation. Specifically, our assessment does not factor in the additional overhead required to bridge the cyber-physical gap for obtaining provably authentic location information~---~a malicious prover could potentially falsify their location. To address this, we give three possible solutions to prevent the forgery of proofs attesting to incorrect locations in Appendix~\ref{appendix:authenticate}, and argue about their estimated additional cost.



\section{Preliminaries}


\subsection{Notation}



We denote the bitwidth of $x \in \mathbb{Z}$ as $|x|$, and the absolute value as $\abs(x)$.
$x \concat y$ is the concatenation of $x, y \in \mathbb{Z}$.
$x \shl n$ and $x \shr n$ shift $x$ to the left and right by $n$ bits, respectively, where $x \shl n = x \cdot 2^n$, $x \shr n = \lfloor \frac{x}{2^n} \rfloor$.
For $x, y \in \mathbb{Z}$ with $y$'s bitwidth known to be $n$, $\overline{x.y}$ is a shorthand for $\frac{x \concat y}{2^n} \in \mathbb{R}$.

We define the position of a point on the Earth's surface in spacial coordinates by radial distance (\(r\)), latitude (\(\latitudeShort\)), and longitude (\(\longitudeShort\)). 
Here, \(r\) approximates the Earth's radius, \(\latitudeShort\) represents the latitude, ranging from \(-90^\circ\) at the South Pole to \(+90^\circ\) at the North Pole with \(0^\circ\) at the Equator, and \(\longitudeShort\) represents the longitude, ranging from \(-180^\circ\) to \(+180^\circ\) with \(0^\circ\) at the Prime Meridian.
We represent a vertex of a hexagon in two-dimensional Cartesian coordinates as $v_i = (x_i, y_i)$. 
We denote a hexagon as a vector (in lower-case bold symbols) of $6$ coordinates, i.e., $\mathbf{h} = [v_1, v_2, ..., v_6]$.
Let \( \mathcal{H} = \{\mathbf{h_1}, \mathbf{h_2}, ..., \mathbf{h_n}\} \) be a set of regular hexagons in the Euclidean plane, each hexagon \(\mathbf{h_i} \) having a center point \( c_i \) and equal side length \( a \). For each hexagon \( \mathbf{h_i}\), there is a region \( R_i \) in the plane, defined by the vertices of \( \mathbf{h_i} \), with the property that every point within \( R_i \) is closer to \( c_i \) than to the center point of any other hexagon in \( \mathcal{H} \). The region \( R_i \) is referred to as a \textit{hexagonal cell}.

\subsection{Background On Floating-Point Values}

Fixed-point and floating-point are common methods for representing real numbers in computers.
In fixed-point representation, we fix a scaling factor $N$ and represent a number $x$ as an integer $v$ of bitwidth $M > N$, whose lowest $N$ bits are treated as the fractional part with an implicit decimal point in front, i.e., $x = \sum_{i=0}^{M-1} v_i \cdot 2^{i - N} = \frac{v}{2^N}$.
Arithmetic operations on fixed-point numbers are equivalent to directly operating on their underlying integers, and only a small overhead is required to keep the scaling factor unchanged. The main advantage of fixed-point is that it can be more efficient in computation and storage, especially on low-cost hardware that lacks native support for floating-point arithmetic. However, it can only handle numbers with similar orders of magnitude, and lacks the ability to encode very large or very small values simultaneously and precisely.

Floating-point representation does not have a fixed scaling factor. It uses only a portion of the available bits (\textit{mantissa}) to store the number, and the remaining bits (\textit{exponent}) are reserved for dynamically tracking the scaling factor. When operating on floating-point numbers, the exponent and mantissa need to be correctly updated. While increasing cost, floating-point representation is more general than fixed-point representation, as it can represent very small and very large numbers with greater precision.

IEEE 754~\cite{ieee754fp} is the de facto standard for floating-point numbers and is widely adopted by modern hardware and software. It defines the encoding formats of floating-point numbers and a set of arithmetic operations on these numbers. 
A floating-point number in IEEE 754 consists of three components, the sign $s$, the exponent $e$, and the mantissa (or significand) $m$. The \textit{binary} encoding format encodes these components as a binary string $s \concat e \concat m$, where $s, e, m$ have $1, E, M$ bit(s) respectively.
In this paper, we are interested in \FPSingle, the single precision binary encoding format with $E = 8, M = 23$, and \FPDouble, the double precision one with $E = 11, M = 52$.
Here, we briefly review the encoding itself and discuss arithmetic operations in \S\ref{section:fp-primitive-ops}.

The binary encoding format is capable of representing 5 types of values: \textit{signed zeros} ($\pm 0$), \textit{subnormal numbers}, \textit{normal numbers}, \textit{infinities} ($\pm \infty$), and \textit{``not a number" values} (NaNs). We use \textit{abnormal} to denote a number that is $\pm \infty$ or NaN. Below we list how the IEEE 754 specification maps the encoded value $s \concat e \concat m$ to the real number $\alpha$.

\begin{enumerate}[label=(\roman*)]
\item $e = 0, m = 0$ ($\pm 0$)~---~$\alpha = 0$ if $s = 0$ and $\alpha = -0$ if $s = 1$.

\item $e = 0, m \neq 0$ (subnormal)~---~$\alpha = (-1)^s \cdot 2^{-2^{E - 1} + 2} \cdot \overline{0.m}$.

\item $e \in [1, 2^E - 2]$ (normal)~---~$\alpha = (-1)^s \cdot 2^{e - 2^{E - 1} + 1} \cdot \overline{1.m}$. 

In this case, $e$ is the \textit{biased form} of the actual exponent with $2^{E - 1} - 1$ as the bias, and $m$, together with an implicit leading 1, describes the actual mantissa.

\item $e = 2^E - 1, m = 0$ ($\pm\infty$)~---~$\alpha = \infty$ if $s = 0$ and $\alpha = -\infty$ if $s = 1$.

\item $e = 2^E - 1, m \neq 0$ (NaN)~---~$\alpha$ isn't a numeric value.
\end{enumerate}


As such, the encoding allows storing a normal number in the range $[2^{-2^{E - 1} + 2}, 2^{2^{E - 1}})$ with $(M + 1)$-bit precision and a subnormal number that is even smaller, i.e., in the range $[2^{-2^{E - 1} + 2 - M}, 2^{-2^{E - 1} + 2})$, but with lower precision.

\subsection{Discrete Global Grid Systems}

A \DGGS divides the Earth into a hierarchy of progressively finer resolution grids. Alternatively, a solution for dividing the Earth's surface would be to apply a simple latitude-longitude grid, where the Earth is divided into a grid based on lines of latitude and longitude, creating a series of rectangular cells that cover the entire globe. 
However, such an encoding leads to lines of longitude that are in closer proximity at the poles than they are at the equator.
While a latitude-longitude grid is a simple way to partition the Earth's surface, it lacks the uniformity, efficiency, and scalability of a \DGGS~\cite{sahr2003geodesic}, particularly for complex spatial analyses and global-scale applications. 

\noindent \textbf{Hexagonal Hierarchical Geospatial Indexing.}
A long line of research suggests that defining a DGGS primarily based on hexagonal tiles exhibits superior properties than DGGS based on other geometric shapes for algorithmic efficiency~\cite{sahr2003geodesic, sahr2019central}. 
This benefit is evident in practical tools~---~Uber, for example, introduced the Hexagonal Hierarchical Spatial Index (H3),  a geospatial index that partitions the globe into hexagons for more accurate analysis of movement patterns~\cite{UberH3}.
The global grid system relies on a gnomonic projection centered on icosahedron faces.
In the Uber H3 indexing system, hexagons are utilized to create a grid on each icosahedron face. 
The H3 indexing system supports differing granularity, with $16$ resolutions. At the highest resolution, $122$ hexagons span the sphere of the earth, with $10$ hexagons per icosahedron face. 
As hexagons cannot tile a icosahedron face, $12$ pentagons are introduced at each of the icosahedron vertices to tile the full spherical projection. For enhanced intuition, Figure~\ref{figure:sphere} depicts an Icosahedral Goldberg Polyhedron of hexagons and pentagons with $92$ faces.


\subsection{SNARKs}
A zero-knowledge Succinct Non-Interactive Argument of Knowledge (zk-SNARK)~\cite{bitansky2012extractable} is a cryptographic protocol, where a prover $\mathcal{P}$ convinces a verifier $\mathcal{V}$ that a certain NP-statement is true, without disclosing any information besides the veracity of the statement.
Common \SNARKs target the problem of \textit{circuit-satisfiability}, i.e., providing \SNARKs for arbitrary NP-statements, represented as arithmetic circuits (R1CS~\cite{gennaro2013quadratic}, Plonkish~\cite{gabizon2019plonk}, AIR~\cite{ben2018scalable}).
Informally, a zk-SNARK for circuit satisfiability satisfies the following:

\noindent \textbf{Succinct:} The verification cost and the size of the proof are sublinear in the size of the circuit.

\noindent \textbf{Non-Interactive:} The prover $\mathcal{P}$ can provide a proof that can be independently verified without further communication. 


Beyond the above properties, the security properties of a zk-SNARK can be informally described as follows:

\noindent \textbf{Perfect Completeness:} An honest $\mathcal{P}$ can always convince $\mathcal{V}$ of the correctness of a true statement.

\noindent \textbf{Knowledge Soundness:} A dishonest $\mathcal{P}$ cannot convince $\mathcal{V}$ of an invalid statement, except with negligible probability. Furthermore, an extractor can successfully extract the witness to a valid statement except with negligible probability.

\noindent \textbf{Zero-Knowledge:} The proof reveals nothing to $\mathcal{V}$ besides that $\mathcal{P}$ knows an assignment satisfying the circuit predicate.


We say an \textit{argument of knowledge} retains knowledge soundness against a computationally bounded prover.

\subsection{SNARK Optimizations}
\label{subsection:SNARKOptimizations}

We introduce generic \SNARK optimizations, applied in our protocols in Section~\ref{section:fp-primitive-ops} and~\ref{section:zkLocation}, in the following.

\noindent \textbf{Lookup Arguments. }
Most \SNARKs can efficiently represent computations that can be expressed as an arithmetic circuit.
However, there are some non-arithmetic operations, such as range checks, XOR or logical AND operations, that are unfriendly to the circuit and cost more constraints. Lookup arguments aim to reduce the prover complexity for these non-arithmetic operations by simply checking that a \textit{query} is contained in a \textit{lookup table}~\cite{xiong2023verizexe}.
They were first introduced by Bootle~\emph{et. al}~\cite{bootle2018arya}, and optimized in several successive works~\cite{LookupBriefHist}. 
In this work, we consider read-only lookup tables.
To build a lookup table $\mathcal{T}$, we precompute all valid values of a function and treat them as a vector of table entries $\textbf{t} \coloneqq (t_1, t_2, ..., t_n) \in \mathbb{F}^n$.
Later, the prover is going to convince the verifier that a vector of queries $\textbf{f} \coloneqq (f_1, f_2, ..., f_m) \in \mathbb{F}^m$ is in the lookup table, i.e., $\textbf{f} \subseteq \textbf{t}$.

Based on logarithmic derivatives, LogUp~\cite{habock2022multivariate} shows that it is sufficient to check the below identity for set inclusion:
$$\sum^{m - 1}_{i = 0} \frac{1}{X - f_i} = \sum^{n - 1}_{j = 0} \frac{o_j}{X - t_j},$$
where $o_j$ is the number of $t_j$'s occurrences in the query vector $\textbf{f}$.
By Schwartz-Zippel Lemma, we can check this polynomial identity by evaluating it at a random point $X = c$.
While LogUp~\cite{habock2022multivariate} provides a dedicated protocol for this check, we instead adopt the approach in gnark~\cite{gnark-v0.9.0}, which enforces the identity in arithmetic circuits.
Also, note that the identity can be extended to support lookup tables with $w > 1$ columns, where each element in the queries and entries is now a vector of length $w$ (i.e., $f_i, t_i \in \mathbb{F}^{w}$).

\noindent \textbf{Nondeterministic Programming. }
An in-circuit computation proves that the input data satisfies a given compliance predicate.
The local input data can provide arbitrary \textit{hints} (or nondeterministic advice~\cite{naveh2016photoproof}), which are not trusted to be correct, but whose verification is more efficient in-circuit than the emulation of the plain computation.
Hints leverage the fact that certain calculations are hard to compute, but easy to verify in an arithmetic circuit.

In consonance with related work~\cite{angel2022efficient}, we formalize a \textit{hint} as the computation $\Computation(X) \xrightarrow{} Y$ done by the prover outside the arithmetic circuit. An in-circuit nondeterministic predicate $\Predicate: X \times Y \xrightarrow{} \{0,1\}$ for $\Computation$ ensures that $\forall x \in X, y \in Y$ the relations $\Computation(x) = y \iff \Predicate(x,y) = 1$ and $\Computation(x) \neq y \iff \Predicate(x,y) = 0$ hold.
Note, that in practical applications, the variable returned by a hint function is  equivalent to a prover-supplied witness.
We call those variables hints instead of witnesses for separation of concerns, highlighting that the computation is done outside the circuit, and that their values are provided by the prover.


For example, consider extracting the most significant bit of the mantissa of a floating point value.
Let the computation to find the MSB of a mantissa be $\Computation_{\mathsf{MSB}}(m) = \text{MSB}_m$ and the nondeterminism predicate be \( \Predicate_{\mathsf{MSB}}: \mathbb{Z}_p \times \mathbb{Z}_p \rightarrow \{0, 1\} \).
In the circuit, it is verified that \( \text{MSB}_m \) is indeed the MSB of \( m \) by the predicate $P_{\text{MSB}}$:
$$ 
\Predicate_{\text{MSB}}(m, \text{MSB}_m) = 
\begin{cases} 
    1 & \text{if } m - \text{MSB}_m \cdot 2^{|m|-1} \in  [0,2^{m-1}], \\
    0 & \text{otherwise.}
\end{cases}
$$

If $\text{MSB}_m = 0$ but the actual $\text{MSB}_m$ is $1$, then the expression $m - \text{MSB}_m \cdot 2^{|m|-1}$ will result in at least \( |m| \) bits.
If $\text{MSB}_m = 1$  but the actual MSB is $0$, the subtraction will result in a negative value.

\section{Primitive Floating-Point Operations}\label{section:fp-primitive-ops}

In the following, we introduce optimized floating-point circuits defined over a prime field $\mathbb{F}_p$ of order $p$ for primitive operations (addition, subtraction, multiplication, division, square root, and comparison).
Here, we denote integer-typed variables as Latin letters $(a,b,\dots)$, and floating-point values as Greek letters $(\alpha, \beta,\dots)$.
To avoid verbosity, we omit the in-circuit constructions for equality constraint ($x = y$), equality check ($result \coloneqq \CircuitIsEq(x, y)$), conditional selection ($\var{condition} \mathbin{?} \var{true\_value} : \var{false\_value}$), and boolean operations ($\land$, $\lor$, $\neg$, $\oplus$), etc.
Further, we introduce circuits for integer operations in Appendix~\ref{appendix:Integer}, including $\AssertBitLength$ for range checks, $\CircuitAbs$ for extracting signs and absolute values, $\CircuitMax$ and $\CircuitMin$ for computing maximum and minimum values, and $\shl$ and $\shr$ for left and right shifting.

\input{./Protocols/gadget_InitFloat.tex}

\subsection{Initializing Floating-Point Numbers}\label{subsection:create-var}

Now we discuss how to initialize a floating-point number $\alpha$ inside the circuit, whose original representation is $\alpha = (\hat{s}, \hat{e}, \hat{m})$, where $\hat{s} \in \{0,1\}$ is the sign bit,  $\hat{e}$ is an $E$-bit exponent, and $\hat{m}$ is an $M$-bit mantissa (or significand).
Note that although it is possible to represent $\alpha$ in circuit as-is, we convert it to a compatible format $\alpha = (s, e, m, a)$ to save as much constraints as possible, where $s, e, m$ are circuit-efficient form of $\hat{s}, \hat{e}, \hat{m}$, and $a \in \{0,1\}$ is an additional bit indicating whether the number is abnormal.


\noindent \textbf{Shape Check. } 
On unchecked input $\hat{\varsign}$, $\hat{\varexponent}$, and $\hat{\varmantissa}$ (which are usually secret witnesses), we first enforce that $\hat{\varsign}$, $\hat{\varexponent}$ and $\hat{\varmantissa}$ are well-formed in-circuit by checking $\hat{\varsign} \in \{0, 1\}, \hat{\varexponent} \in [0, 2^E - 1], \hat{\varmantissa} \in [0, 2^M - 1]$. After that, we initialize $\varsign \coloneqq \hat{\varsign}, \varexponent \coloneqq \hat{\varexponent}, \varmantissa \coloneqq \hat{\varmantissa}$.

\noindent \textbf{Convert to Circuit Efficient Forms. } 
For a normal number $\alpha = (-1)^{s} \cdot 2^{e - 2^{E - 1} + 1} \cdot \overline{1.m}$, the sign $s$ is unchanged, while we redefine $e$ as the exponent's unbiased form, i.e., $e \coloneqq e - 2^{E - 1} + 1$, and $m$ as the mantissa with an explicit leading bit 1, i.e., $m \coloneqq m + 2^M$.
In this way, we no longer need to handle the bias of $e$ and the implicit leading bit of $m$ in subsequent operations, thus improving efficiency and clarity.

We convert subnormal numbers to normal numbers by normalizing their mantissas, allowing the exponents to `underflow'. 
Specifically, for a subnormal number $\alpha = (-1)^s \cdot 2^{-2^{E - 1} + 2} \cdot (0 \concat m) \cdot 2^{-M}$, we left shift the mantissa (with leading zero) $0 \concat m$, and at the same time decrement the exponent $-2^{E - 1} + 2$ by 1, until the MSB (i.e., $M$-th bit) of mantissa becomes 1. Denoting the shift by $d \in [1, M]$, we have $\alpha = (-1)^s \cdot 2^{-2^{E - 1} + 2 - d} \cdot ((0 \concat m) \shl d) \cdot 2^{-M}$ and redefine $e \coloneqq -2^{E - 1} + 2 - d, m \coloneqq (0 \concat m) \shl d = m \shl d$.
Now,
the prover computes the hint $\var{d} \coloneqq \Computation_{\mathsf{Norm}}(\var{m})$ and provides $\var{d}$ to the circuit.
For soundness, the circuit needs to check the predicate $\Predicate_{\mathsf{Norm}}(\var{m}, \var{d})$ by calling $\AssertBitLength((\varmantissa \shl \var{d}) - (\var{m\_is\_0} \mathbin{?} 0 : 2^M), M)$, which enforces $\varmantissa$ is zero or $\varmantissa \shl \var{d} \in [2^M, 2^{M + 1} - 1]$, implying that the MSB (i.e., $M$-th bit) of a non-zero $\varmantissa \shl \var{d}$ is 1.

Normalizing unifies normal and subnormal numbers to save constraints in subsequent operations.
Although subnormal numbers now take $M + 1$ bits to represent, padding zeros do not contribute to precision, so they still have lower precision than normal numbers.

\noindent \textbf{Edge Cases. } $\pm \infty$ is represented by $\varexponent = 2^{E - 1}, \varmantissa = 2^M$. $\pm 0$ is represented by $\varexponent = -2^{E - 1} + 1 - M, \varmantissa = 0$. Although NaNs have different mantissas as per the specification, we always map them to fixed variables $\varsign = 0, \varexponent = 2^{E - 1}, \varmantissa = 0$ to simplify the handling of edge cases. We set $0$ as NaN's mantissa due to the similar behaviors between NaNs and $\pm 0$ in multiplication and division.

We compute the final exponent and mantissa inside the circuit using several conditional selections. The entire circuit for initializing floating-point variables is shown in Figure \ref{figure:init_float}.

\input{./Protocols/gadget_RoundFloat.tex}

\subsection{Rounding}
\label{subsection:round}

The IEEE 754 standard requires the resulting mantissa of an operation to be rounded correctly and defines several rules specifying how and in which direction the rounding is done.
Here, we discuss 
``rounding half to even'' for binary encoded floating-point numbers. This rule requires a decimal number to be rounded to the nearest integer, except when the fractional part is 0.5 (in base-10), the rounding should produce an even value, e.g., $0.5 \to 0, 1.5 \to 2$. 

Formally, consider an intermediate mantissa $m \in [0, 2^N - 1]$ after some operation, and we are going to compute a rounded value $m' \in [0, 2^l - 1]$.
To this end, we write $m$ as $m = u \concat v$, where $u \in [0, 2^l - 1]$ and $v \in [0, 2^{N - l} - 1]$.
The rounding direction is determined by the bits in $v$, which are the \textit{round bit}, the \textit{guard bit}, and the \textit{sticky bit} in most implementations of IEEE 754.
In our circuits, rounding is done according to only the round bit and sticky bit, where the round bit is $v_{N - l - 1}$ (the MSB of $v$), and the sticky bit is $v_0 \lor \cdots \lor v_{N - l - 2}$ (the OR value of remaining bits of $v$).
The guard bit is useful for implementations that discard the right-shifted bits but is unnecessary in our case because these bits are preserved.
Consequently, we have $m' \coloneqq u$ if round bit is 0, i.e., $v \in [0, 2^{N - l - 1})$.
If round bit is 1 and sticky bit is 0, i.e., $v = 2^{N - l - 1}$, then $m' \coloneqq u + 1$ when $u$ is odd, while $m' \coloneqq u$ when $u$ is even.
Otherwise, $v \in (2^{N - l - 1}, 2^{N - l})$, and the result $m' \coloneqq u + 1$.





Note that due to the possible increment in $m' \coloneqq u + 1$, $m'$ could become $2^l$, exceeding the upper bound $2^l - 1$.
We say $m'$ overflows in this case, and we fix this by setting $e' \coloneqq e + 1, m' \coloneqq 2^{l - 1}$.

If we follow the standard exactly, $l$ will be fixed to $M + 1$. However, in our case, $l = M + 1$ only when the result is normal. For a subnormal number with an underflown exponent $e < -2^{E - 1} + 2$, we require $l = M + 1 - \Delta e$, where $\Delta e = -2^{E - 1} + 2 - e$ (here it is guaranteed that $e \ge -2^{E - 1} + 1 - M$, implying $l \ge 0$).
The reason is that, for circuit efficiency, we pretend that subnormal numbers are normal when representing and operating on them. This works in most cases except for rounding, where subnormal numbers should actually be rounded with lower precision (i.e., $M + 1 - \Delta e$) than that of normal numbers (i.e., $M + 1$). Hence, we should only keep the first $M + 1 - \Delta e$ bits of $m$ for subnormal numbers. When the rounding is done, we left shift $m'$ by $\Delta e$ to continue disguising them as normal.

Figure \ref{figure:round_float} illustrates the rounding gadget, where
we require that $\Delta\varexponent$ has a constant upper bound $K$, and that $\Delta\varexponent = 0$ for normal numbers.
First, we need to expand $\varmantissa$ into $\var{u} \concat \var{v}$ in-circuit.
Naively, we can ask the prover to provide $u, v$ as hints, and the circuit checks $\var{u} \in [0, 2^{\var{l}} - 1], \var{v} \in [0, 2^{N - \var{l}} - 1]$. However, the upper bounds of $\var{u}, \var{v}$ depend on the variable $\var{l}$. As discussed in Appendix \ref{appendix:Integer}, a range check bounded by variables costs more constraints than a constant range check.
To maximize efficiency, the prover instead computes the hint $\Computation_{\mathsf{Split}}$ by expanding $\varmantissa \shl (K - \Delta\varexponent)$ into $\var{u}' \concat \var{b}_1 \concat \var{b}_2 \concat \var{v}'$, where $\var{u}' \in [0, 2^{M - \Delta\varexponent} - 1], \var{b}_1, \var{b}_2 \in \{0, 1\}$ and $\var{v}' \in [0, 2^{N - M - 2 + K} - 1]$. Then $\var{u}', \var{b}_1, \var{b}_2, \var{v}' \coloneqq \Computation_{\mathsf{Split}}(\varmantissa \shl (K - \Delta \varexponent))$ are fed to the circuit.
Now, $\var{u} \coloneqq \var{u}' \concat \var{b}_1$, $\var{v} \coloneqq \var{b}_2 \concat \var{v}'$.
To verify the predicate $\Predicate_{\mathsf{Split}}(\varmantissa \shl (K - \Delta \varexponent), \var{u}', \var{b}_1, \var{b}_2, \var{v}')$, the circuit checks the ranges of $\var{u}'$ and $\var{v}'$ by calling \AssertBitLength, enforces $\var{b}_1$ and $\var{b_2}$ are boolean, and asserts $\varmantissa \shl K = (\var{u}' \concat \var{b}_1 \concat \var{b}_2 \concat \var{v}') \shl \Delta\varexponent$. 
Moreover, as \varmantissa is guaranteed to have length $N$, we can loosely bound $\var{u}'$ and check $\var{u}' \in [0, 2^M - 1]$ instead. 
This is safe, since if $2^{M - \Delta\varexponent} \le \var{u}' < 2^M$, the length of $\var{u}' \concat \var{b}_1 \concat \var{b}_2 \concat \var{v}'$ will be longer than $N + K - \Delta\varexponent$, and \varmantissa's length will be longer than $N$, which is a contradiction.
Now, both range checks are bounded by constants.

After that, we compute $\var{half} \coloneqq \CircuitIsEq(\var{v}, 2^{N - M - 2 + K}) \land \var{aux}$, where $\var{aux}$ is the auxiliary information that helps determine the sticky bit in division and the computation of square root. For $\var{aux} = 1$, the sticky bit solely depends on $\var{v}$.
Then we have the rounded mantissa $\varmantissa' \coloneqq (\var{u} + (\var{half} \mathbin{?} \var{b}_1 : \var{b}_2)) \shl \Delta\varexponent$. Finally, we check $\var{overflow} \coloneqq \CircuitIsEq(\varmantissa', 2^{M + 1})$, and return the updated exponent and mantissa $\varexponent' \coloneqq \varexponent + \var{overflow}, \varmantissa' \coloneqq \var{overflow} \mathbin{?} 2^M : \varmantissa'$.

\subsection{Addition And Subtraction}

\input{./Protocols/circuit_AddFloat.tex}

Adding two IEEE 754 floating-point numbers $\alpha = (s_\alpha, e_\alpha, m_\alpha, a_\alpha)$ and $\beta = (s_\beta, e_\beta, m_\beta, a_\beta)$ is done in the following 5 steps, and we depict the corresponding in-circuit logic in Figure \ref{figure:add_float}. 
At a high level, addition requires $5$ steps, described in the following: \emph{(i)} aligning exponents, \emph{(ii)} adding mantissas, \emph{(iii)} normalizing, \emph{(iv)} rounding the intermediate mantissa and \emph{(v)} handling edge cases.
Note, that subtraction is equivalent to addition by adding $\alpha$ and $-\beta$.

\noindent \textbf{Align exponents (lines 1-3).} We first compare the exponents of $\alpha$ and $\beta$. 
If $e_\alpha \neq e_\beta$, we need to align the exponents before performing the actual addition by shifting the mantissa of the number with smaller exponent to the \textit{right} by $abs \coloneqq \abs(e_\alpha - e_\beta)$ bits, such that the common exponent is $e \coloneqq \max(e_\alpha, e_\beta)$.
To avoid separately tracking the shifted bits (which will be used later in rounding), before performing the right shift, we first shift both mantissas to the \textit{left} by $L$ bits, where $L$ is the upper bound of $abs$.
That is, if $e_\alpha > e_\beta$, we compute $x \coloneqq m_\alpha \shl L, y \coloneqq (m_\beta \shl L) \shr (e_\alpha - e_\beta)$, and otherwise, $x \coloneqq m_\beta \shl L, y \coloneqq (m_\alpha \shl L) \shr (e_\beta - e_\alpha)$.
The $M + 1$ MSB's of shifted mantissas contribute to the final result, the remaining bits determine the rounding direction.

In circuit, we achieve this by computing $\var{c}, \var{abs} \coloneqq \CircuitAbs(\varexponent_\var{\beta} - \varexponent_\var{\alpha}, E + 1)$ to obtain $\varexponent \coloneqq \var{c} \mathbin{?} \varexponent_\var{\beta} : \varexponent_\var{\alpha}$. We observe that the final sum is only determined by $\var{x}$ when $\var{y}$ is completely shifted out, i.e., when $\var{abs} \ge M + 3$. Thus, $\var{abs} > M + 3$ has the same effect as $\var{abs} = M + 3$. We improve the circuit efficiency by setting $\var{abs} \coloneqq \CircuitMin(\var{abs}, M + 3)$, so that it is no longer necessary to compute a large $2^\var{abs}$. Now, $L = M + 3$, $\var{x} \coloneqq (\var{c} \mathbin{?} \varmantissa_\var{\beta} : \varmantissa_\var{\alpha}) \shl L, \var{y} \coloneqq (\var{c} \mathbin{?} \varmantissa_\var{\alpha} : \varmantissa_\var{\beta}) \shl (L - \var{abs})$.

\noindent \textbf{Add signed mantissas (lines 4-9).} Then we add the signed mantissas of adjusted $\alpha$ and $\beta$, and obtain the resulting (signed) mantissa, i.e., $s \cdot m \coloneqq s_\alpha \cdot x + s_\beta \cdot y$ if $e_\alpha > e_\beta$, and $s \cdot m \coloneqq s_\beta \cdot x + s_\alpha \cdot y$ otherwise. The sign $s$ and unsigned mantissa $m$ are extracted from the result, where $m \le x + y < 2(2^{M + 1} \cdot 2^L) = 2^{2M + 5}$, i.e., $m$ has at most $N = 2M + 5$ bits, where the leading bit is caused by the possible carry. Thus, we also adjust the exponent as $e \coloneqq e + 1$.
For efficiency, we redefine the in-circuit variables $\var{x} \coloneqq (\var{c} \mathbin{?} \varsign_\var{\beta} \varmantissa_\var{\beta} : \varsign_\var{\alpha} \varmantissa_\var{\alpha}) \shl L, \var{y} \coloneqq (\var{c} \mathbin{?} \varsign_\var{\alpha} \varmantissa_\var{\alpha} : \varsign_\var{\beta} \varmantissa_\var{\beta}) \shl (L - \var{abs})$ to avoid extra conditional selections.
Then we compute $\var{z} \coloneqq \var{x} + \var{y}$, and compute $\varsign$ and $\varmantissa$ thanks to the $\CircuitAbs$ gadget: $\neg \varsign, \varmantissa \coloneqq \CircuitAbs(\var{z}, N)$.
The result is abnormal if either input is abnormal, i.e., $a \coloneqq a_\alpha \lor a_\beta$.

\noindent \textbf{Normalize intermediate mantissa (lines 10-13).} Normalization for $m$ is the same as in Section \ref{subsection:create-var} for normalizing subnormal numbers. $m$ is shifted to the left by $d$ bits, so that its MSB (i.e., the $N - 1$-th bit) becomes 1, unless $m = 0$, and $e$ is decreased by $d$.

\noindent \textbf{Round intermediate mantissa (line 14).} The normalized mantissa $m \shl d$ of length $N$ is then rounded as in Section \ref{subsection:round}, with $\Delta e = K = 0$, obtaining $\varexponent'$ and $\varmantissa'$. Theoretically, $\Delta e$ should be $-2^{E - 1} + 2 - e$. However, we observe that for addition, a smaller $\Delta e$ doesn't affect the result. In fact, it is safe to set $\Delta e = 0$ to maximize circuit efficiency, and we explain the reasoning below.

Recall that the purpose of $\Delta e$ is to limit the precision of subnormal numbers, but, as we will show later, the number of meaningful bits in $m \shl d$ will never exceed the required precision, and the remaining bits are guaranteed to be zero. 
Consider a subnormal result with $e \le -2^{E - 1} + 1$, and assume, without loss of generality, $e_\alpha < e_\beta$. We denote $\Delta e_\alpha = -2^{E - 1} + 2 - e_\alpha, \Delta e_\beta = -2^{E - 1} + 2 - e_\beta$. Since $e = \max(e_\alpha, e_\beta) + 1 - d = e_\beta + 1 - d$, we have $e_\alpha < e_\beta \le -2^{E - 1} + d$. According to the rounding rule, we need to keep only $M + 1 - (-2^{E - 1} + 2 - e)$ bits in the mantissa, while the remaining $N - (M + 1) + (-2^{E - 1} + 2 - e) = L + d + \Delta e_\beta$ bits determine the rounding direction. Now we prove that the number of trailing zeros in $m \shl d$ is at least $L + d + \Delta e_\beta$ if $e \le -2^{E - 1} + 1$. 
Note, that $x = s_\beta m_\beta \shl L, y = s_\alpha m_\alpha \shl (L - \min(e_\beta - e_\alpha, L)) = s_\alpha m_\alpha \shl \max(L - e_\beta + e_\alpha, 0)$.

\begin{enumerate}[label=(\roman*),wide]
    \item \textit{$\alpha$ subnormal, $\beta$ subnormal}~---~In this case, $m_\alpha$ and $m_\beta$ were left-shifted by $\Delta e_\alpha$ and $\Delta e_\beta$ bits when initialized. Also, $e_\beta - e_\alpha \le -2^{E - 1} + 1 - (-2^{E - 1} + 1 - M) = M < L$, thus $\max(L - e_\beta + e_\alpha, 0) = L - e_\beta + e_\alpha$. Now, $x$ has at least $\Delta e_\beta + L$ trailing zeros, and $y$ has at least $\Delta e_\alpha + \max(L - e_\beta + e_\alpha, 0) = \Delta e_\alpha + L - e_\beta + e_\alpha = \Delta e_\beta + L$ trailing zeros. Hence, $m \shl d$ has at least $\Delta e_\beta + L + d$ trailing zeros.

    \item \textit{$\alpha$ subnormal, $\beta$ normal}~---~Here, $m_\alpha$ was left-shifted by $\Delta e_\alpha$ bits when initialized. Now, $x$ has at least $L$ trailing zeros, and $y$ has at least $\Delta e_\alpha + \max(L - e_\beta + e_\alpha, 0) = \max(L + \Delta e_\beta, \Delta e_\alpha) < L$ trailing zeros. So $m \shl d$ has at least $\max(L + \Delta e_\beta, \Delta e_\alpha) + d \ge L + \Delta e_\beta + d$ trailing zeros.
    
    \item \textit{$\alpha$ normal, $\beta$ normal}~---~In this case, $e_\beta - e_\alpha \le -2^{E - 1} + d - (-2^{E - 1} + 2) = d - 2 < L$, thus $\max(L - e_\beta + e_\alpha, 0) = L - e_\beta + e_\alpha$. Now, $x$ has at least $L$ trailing zeros, and $y$ has at least $\max(L - e_\beta + e_\alpha, 0) = L - e_\beta + e_\alpha < L$ trailing zeros. Consequently, $m \shl d$ has at least $L - e_\beta + e_\alpha + d$ trailing zeros, and $L - e_\beta + e_\alpha + d \ge L + d + \Delta e_\beta$ because $\Delta e_\beta + e_\beta - e_\alpha = \Delta e_\alpha \le 0$.
    
\end{enumerate}
\noindent \textbf{Edge Cases (lines 15-21).} 
Finally, we need to handle the following:
\begin{enumerate}[label=(\roman*)]
    \item If the mantissa $m' = 0$ but $e' \neq -2^{E - 1} + 1 - M$ (which is possible, e.g., when computing $1.0 - 1.0$), canonicalize the exponent as $e' \coloneqq -2^{E - 1} + 1 - M$.
    \item If the exponent becomes too large, i.e., $e' \ge 2^{E - 1}$, return $\pm \infty$ (depending on the sign $s$).
    \item If $\alpha = \beta = \pm 0$, return $-0$ for $-0 - 0$ and $+0$ otherwise.
    \item If either $\alpha$ or $\beta$ is NaN, return NaN.
    \item If either $\alpha$ or $\beta$ is $\pm \infty$, return NaN for $\infty - \infty$ and $-\infty + \infty$, and otherwise, $\pm \infty$ (depending on $s$).
    \item Otherwise, return $(s, e', m', 0)$ as the result.
\end{enumerate}
To minimize the number of constraints, we unify some cases above based on the return value's exponent and mantissa in-circuit. First, the result is abnormal if either inputs is abnormal (\emph{iv}, \emph{v}), or the exponent is too large (\emph{ii}). Hence, $\varabnormal' \coloneqq \varabnormal \lor \CircuitGreaterThanZero(\varexponent' - 2^{E - 1}, E + 1)$.
Second, to support \emph{(iii)}, we set $\varsign' \coloneqq \CircuitIsEq(\varsign_{\var{\alpha}}, \varsign_{\var{\beta}}) \mathbin{?} \varsign_{\var{\alpha}} : \varsign$.
Third, the result's exponent is $2^{E - 1}$ if the result is abnormal (\emph{ii}, \emph{iv}, \emph{v}), and is $-2^{E - 1} + 1 - M$ if the result is zero (\emph{i}). Thus, $\varexponent' \coloneqq \varabnormal' \mathbin{?} 2^{E - 1} : (\var{m\_is\_0} \mathbin{?} -2^{E - 1} + 1 - M : \varexponent')$.
Finally, for the result's mantissa $\varmantissa'$, if both $\alpha$ and $\beta$ are abnormal, $\varmantissa' = 2^M$ for $\infty + \infty$ and $-\infty - \infty$ (\emph{v}), and $\varmantissa' = 0$ otherwise (\emph{iv}, \emph{v}). If only one of $\alpha$ and $\beta$ is abnormal, $\varmantissa'$ equals the abnormal one's mantissa (\emph{iv}, \emph{v}). If both $\alpha$ and $\beta$ are normal, the result is $2^M$ if the exponent becomes too large (\emph{ii}).
Lines 19-21 in Figure~\ref{figure:add_float} summarize the logic above for handling $\varmantissa'$.

\subsection{Multiplication And Division}

\input{./Protocols/circuit_MulFloat.tex}

Multiplying two IEEE 754 floating-point numbers $\alpha = (s_\alpha, e_\alpha, m_\alpha, a_\alpha)$ and $\beta = (s_\beta, e_\beta, m_\beta, a_\beta)$ is done in the following $4$ steps~---~\emph{(i)} computing the product of $\alpha$ and $\beta$, \emph{(ii)} normalizing and \emph{(iii)} rounding the intermediate mantissa and \emph{(iv)} handling edge cases. We depict the corresponding in-circuit logic in Figure \ref{figure:mul_float}.
The steps of division operation are highly similar to those of multiplication, and we defer their description to Appendix \ref{section:division} due to the space limit.

\noindent \textbf{Compute product (line 1).} The product is negative only when one of $\alpha$ and $\beta$ is negative. Therefore, the sign of the product is $s \coloneqq s_\alpha \oplus s_\beta$. The exponent and mantissa of the product are respectively $e \coloneqq e_\alpha + e_\beta$ and $m \coloneqq m_\alpha \cdot m_\beta$. Since $m_\alpha, m_\beta$ are either 0 or lie in $[2^M, 2^{M + 1} - 1]$, a non-zero $m$ should be bounded by $m \in [2^{2M}, 2^{2M + 2})$. We further compute $a \coloneqq a_{\alpha} \lor a_{\beta}$.

\noindent \textbf{Normalize intermediate mantissa (lines 2-4).} By leveraging the fact that $m$ is either 0 or in $[2^{2M}, 2^{2M + 2})$, the leading 1 of a non-zero $m$ is either the $2M$-th bit or the $2M + 1$-th bit. 
Hence, we can simplify the normalization process by checking if the $2M + 1$-th bit of $m$ is 1. If this is the case, $m$ is already normal, and otherwise, we compute $m \coloneqq m \shl 1$. Also, $m_{2M + 1} = 1$ indicates that the multiplication carries, and hence we increment $e \coloneqq e + 1$ if so. The improved normalization is done in-circuit as follows: the prover feeds $\var{b} \coloneqq \Computation_{\mathsf{MSB}}(\varmantissa) = \varmantissa_{2M + 1}$, the MSB of \varmantissa, as a hint to circuit, and the circuit checks the predicate $\Predicate_{\mathsf{MSB}}(\varmantissa, \var{b})$ in 2 steps: \emph{(i)} enforce $\var{b}$ is a boolean, and \emph{(ii)} assert $\varmantissa - (\var{b} \shl (2M + 1)) \in [0, 2^{2M + 1})$. Then $\varmantissa, \varexponent$ are updated according to $\var{b}$, i.e., $\varmantissa \coloneqq \var{b} \mathbin{?} \varmantissa : \varmantissa \shl 1$, $\varexponent \coloneqq \varexponent + \var{b}$.

\noindent \textbf{Round intermediate mantissa (lines 5-6).} The normalized mantissa $m$ of length $N = 2M + 2$ is rounded by following the steps in Section \ref{subsection:round}, with $\Delta e = \max(\min(-2^{E - 1} + 2 - e, K), 0), K = M + 2$, obtaining $\varexponent'$ and $\varmantissa'$.

\noindent \textbf{Edge Cases (lines 7-11).} 
Finally, we handle the following:
\begin{enumerate}[label=(\roman*)]
    \item If the exponent becomes too large, i.e., $e' \ge 2^{E - 1}$, return $\pm \infty$ (depending on the sign $s$).
    \item If the exponent becomes too small, i.e., $e' < -2^{E - 1} + 1 - M$, or equivalently, the rounded mantissa becomes 0, return $\pm 0$ (depending on the sign $s$).
    \item If either $\alpha$ or $\beta$ is NaN, return NaN.
    \item If either $\alpha$ or $\beta$ is $\pm \infty$, return NaN for $\pm 0 \cdot \pm \infty$ and $\pm \infty \cdot \pm 0$, and otherwise, $\pm \infty$ (depending on $s$).
    \item Otherwise, return $(s, e', m', 0)$ as the result.
\end{enumerate}
To minimize the number of constraint, we unify some cases above based on the return value's exponent and mantissa in-circuit.
First, the result is abnormal if either input is abnormal (\emph{iii}, \emph{iv}), or the exponent is too large (\emph{i}). Hence, $\varabnormal' \coloneqq \varabnormal \lor \CircuitGreaterThanZero(\varexponent' - 2^{E - 1}, E + 1)$. Second, the result's exponent is $2^{E - 1}$ if the result is abnormal (\emph{i}, \emph{iii}, \emph{iv}), and is $-2^{E - 1} + 1 - M$ if the result is zero (\emph{ii}). Thus, $\varexponent' \coloneqq \varabnormal' \mathbin{?} 2^{E - 1} : (\var{m'\_is\_0} \mathbin{?} -2^{E - 1} + 1 - M : \varexponent')$. Finally, the result's mantissa is only different from the rounded mantissa if the exponent is too large (\emph{i}), or either inputs is $\pm \infty$ and neither of them is $\pm 0$  (\emph{iv}). Both conditions are equivalent to the case where the result is abnormal but not NaN, so we have $\varmantissa' \coloneqq (\varabnormal' \land \neg \var{m'\_is\_0}) \mathbin{?} 2^M : \varmantissa'$.

\subsection{Square Root Computation}
\label{section:squareRoot}

\input{./Protocols/circuit_SqrtFloat.tex}

The approximation of square roots is often based on the iterative Newton method~\cite{suli2003introduction}. To compute $\beta = \sqrt{\alpha}$, we first estimate an initial value $\beta_0$, and improve the accuracy in each round by computing $\beta_{i + 1} = \frac{1}{2}(\beta_i + \frac{\alpha}{\beta_i})$.
However, directly translating this approach to in-circuit operations introduces two challenges: \emph{(i)} the number of iterations depends on how fast $\sqrt{\alpha}$ converges, but handling loops conditioned on a variable in-circuit is hard, and \emph{(ii)} each round of iteration requires one floating-point addition and one floating-point division, which are costly. To address \emph{(i)}, we need to run the loop for fixed number of rounds, taking the worst case for convergence into account, e.g., achieving the accuracy of \FPDouble needs $6$ rounds of iteration in the worst case. We can resolve \emph{(ii)} by computing the square root of the mantissa $m_\alpha$ rather than the floating-point value $\alpha$. $\beta$'s exponent can be obtained by halving $e_\alpha$. Since $m_\alpha$ is an integer, addition and division in each round are cheap.

Nevertheless, this improved approach would easily cost hundreds of constraints due to the range checks caused by in-circuit integer division. To further reduce circuit size, we leverage the nondeterminism of the constraint system: the prover is asked to compute the square root of $m_\alpha$ outside the circuit, and the circuit, given the square root as a hint, only needs to check its validity, thereby achieving the minimum cost. More specifically, we compute the square root of an IEEE 754 floating-point number $\alpha = (s_\alpha, e_\alpha, m_\alpha, a_\alpha)$ inside the circuit in the following 4 steps, the process of which is also shown in Figure \ref{figure:sqrt_float}.

\noindent \textbf{Compute square root (lines 1-10).} First, we halve the exponent $\varexponent_\var{\alpha}$. When $\varexponent_{\var{\alpha}}$ is even, we can simply compute $\varexponent_\var{\alpha} / 2$, and otherwise, we need to calculate $(\varexponent_\var{\alpha} - 1) / 2$. Combining both cases, the prover feeds $\var{b} \coloneqq \Computation_{\mathsf{LSB}}(\varexponent_\var{\alpha})$, the exponent's LSB, as a hint to circuit. The circuit checks the predicate $\Predicate_{\mathsf{LSB}}(\varexponent_\var{\alpha}, \var{b})$ as follows: enforce $\var{b}$ is boolean, compute $\varexponent \coloneqq (\varexponent_\var{\alpha} - \var{b}) / 2$, and assert $\varexponent \in [-2^{E - 1} + 1, 2^{E - 1} - 1]$ by calling $\CircuitAbs(\varexponent, E - 1)$ (note that $\varexponent$ might be negative). This guarantees that $\var{b}$ is indeed the LSB of $\varexponent_\var{\alpha}$, as otherwise, \varexponent would be close to $(p - 1) / 2$ and its absolute value cannot fit into $E - 1$ bits. Knowing the validity of $\var{b}$, $\varexponent$ is in fact in $[-2^{E - 2} - M/2, 2^{E - 2}]$.
Next, we compute the mantissa's square root. To this end, the prover feeds the hint $\var{n} \coloneqq \Computation_{\mathsf{Sqrt}}(\varmantissa_\var{\alpha} \shl (M + 4 + \var{b})) =  \sqrt{\varmantissa_\var{\alpha} \shl (M + 4 + \var{b})}$ to circuit, and the circuit checks the predicate $\Predicate_{\mathsf{Sqrt}}(\varmantissa_\var{\alpha} \shl (M + 4 + \var{b}), \var{n})$ by enforcing $\var{n}^2 \le (\varmantissa_\var{\alpha} \shl (M + 4 + \var{b})) < (\var{n} + 1)^2$ using two range checks. This guarantees that
$\var{n}$ is (the integer part of) the shifted mantissa's square root. We shift $\varmantissa_\var{\alpha}$ to the left before computing the square root for two reasons: \emph{(i)} when $\varexponent_{\var{\alpha}}$ is odd, we decrease it by 1, and thus the mantissa should be doubled when $\var{b} = 1$, or equivalently, $\varmantissa_\var{\alpha} \shl \var{b}$, and \emph{(ii)} the shift $M + 4$ scales $\varmantissa_\var{\alpha}$ to achieve the desired precision. Otherwise, the result $\var{n}$ would only have approx. $M / 2$ bits of precision. Successively, we obtain the intermediate mantissa $\varmantissa \coloneqq \var{n}$.
Recall that the standard requires the intermediate result to have infinite precision, but $\var{m}$ is not the exact square root. Thus, we apply the technique introduced in Appendix \ref{section:division}: we further compute $\var{r} \coloneqq (\varmantissa_\var{\alpha} \shl (M + 4 + \var{b})) - \var{n}^2$, which helps compute the sticky bit in rounding without storing the precise square root.
The result is abnormal if $\alpha$ is abnormal or negative ($-0$ is not included, as $\sqrt{-0} = -0$). Hence, we set $\varabnormal \coloneqq \varabnormal_\var{\alpha} \lor (\varsign_\var{\alpha} \land \neg \CircuitIsEq(\varmantissa_\var{\alpha}, 0))$.

\noindent \textbf{Normalize intermediate mantissa.} Now, a non-zero $\varmantissa$'s upper bound is $\sqrt{(2^{M + 1} - 1) \shl (M + 5)} < 2^{M + 3}$, and its lower bound is $\sqrt{2^M \shl (M + 4)} = 2^{M + 2}$. Hence, the MSB of $\varmantissa$ is always 1 and normalization is unnecessary.

\noindent \textbf{Round intermediate mantissa (line 11).} The mantissa \varmantissa of length $N = M + 3$ is then rounded as in Section \ref{subsection:round}, with $\Delta e = K = 0$, obtaining $\varexponent'$ and $\varmantissa'$. $\Delta e$ is fixed to 0 because the intermediate exponent \varexponent of a non-zero result is always greater than $-2^{E - 2} - M/2 > -2^{E - 1} + 2$, hence we don't need to handle the subnormal case.
In addition, the equality between $r$ and $0$ is used to determine the sticky bit, thus we set the in-circuit parameter $\var{aux} \coloneqq \CircuitIsEq(r, 0)$.

We omit Edge Cases (lines 12-14) due to space limits.

\subsection{Comparison}

\input{./Protocols/circuit_LessFloat.tex}

Finally, we discuss the comparison between two IEEE 754 floating-point values $\alpha = (s_\alpha, e_\alpha, m_\alpha, a_\alpha)$ and $\beta = (s_\beta, e_\beta, m_\beta, a_\beta)$ in-circuit.
For equality, we support two types of checks: the strict comparison and the fuzzy comparison.
The former enforces that $\alpha$ and $\beta$ are strictly equal by checking if all their components are equal, while the latter asserts that the difference $\alpha - \beta$ is less than a threshold.

Now we introduce the inequality check by using the less than operation as an example.
Other comparators like $\le, >, \ge$ can be constructed analogously.

First, we check if $\alpha$ or $\beta$ is NaN, in which case we return $0$.
Second, we compare the signs $s_\alpha$ and $s_\beta$. If $s_\alpha \neq s_\beta$, then the result is $0$ if $\alpha = -0$ and $\beta = +0$ ($-0 = +0$), is $1$ if $s_\alpha$ is true but $\alpha \neq -0$ (a negative value is always smaller than a positive one), and $0$ otherwise.
Now, $s_\alpha = s_\beta$. 
For the exponent, if $e_\alpha \neq e_\beta$, then the result equals $e_\alpha < e_\beta$ for positive $\alpha, \beta$ and $e_\alpha > e_\beta$ for negative ones. Otherwise, $\alpha$ and $\beta$ have the same sign and exponent. We return $m_\alpha < m_\beta$ for positive $\alpha, \beta$ and $m_\alpha > m_\beta$ for negative ones.
Utilizing $\CircuitIsEq$ and $\CircuitGreaterThanZero$, we translate the above process to in-circuit constraints in Figure~\ref{figure:less_float}.




\section{Zero Knowledge Location Privacy}
\label{section:zkLocation}


In this section, we discuss the technical challenges when evaluating whether a location $(\latitudeShort, \longitudeShort)$ is in a hexagonal tile $\mathbf{h}$. 
We provide a simplified description of the H3 protocol~\cite{UberH3} that transforms spherical coordinates to hexagonal indices in Figure~\ref{figure:baseline}.  In the following, we first describe how the baseline algorithm in the H3 hexagonal spatial indexing system transforms $(\latitudeShort, \longitudeShort)$ into $(i, j, k)$ coordinates, which uniquely identify a hexagonal tile $\mathbf{h}$ in the hexagonal grid. Successively, we discuss how to emulate the transformation in \SNARK circuits using all the floating-point circuits in \S~\ref{section:fp-primitive-ops}, highlight the difficulties in circuit construction, and introduce optimizations that make the \ZKLP paradigm practical.

\noindent \textbf{Transforming Spherical Coordinates to Coordinates in a Discrete Hexagon Planar Grid Systems.}
To utilize hexagonal hierarchical geospatial indexing, spherical coordinates (i.e., $(\latitudeShort, \longitudeShort)$) need to be converted to relative coordinates of the hexagon in the grid system of the geospatial index (i.e., the $(i, j, k)$ coordinates). 
The H3 coordinate system deterministically maps $(i, j, k)$ coordinates to H3 indices, and the $(i, j, k)$ coordinates, in combination with the resolution $\resolutionShort$, are sufficient to determine a unique hexagon based on the latitude and longitude. 
In the following, we therefore solely describe the transformation of latitude and longitude to $(i, j, k)$ coordinates in the hexagonal planar grid system of H3.
The transformation relies on two logical steps: 


\input{./Protocols/zkLocation_Baseline.tex}

\textit{(1) Transforming spherical coordinates to Cartesian coordinates:}
The first step transforms a point from spherical coordinates to Cartesian coordinates in the 2D plane of an icosahedral face, specifically for the hexagonal grid system used in the H3 geospatial indexing system.
Given geographic coordinates $(\latitudeShort, \longitudeShort)$, the distance from a given point on the sphere to the center of the closest face of the icosahedron is computed by evaluating the squared Euclidean distance \(d^2 = (x_2 - x_1)^2 + (y_2 - y_1)^2 + (z_2 - z_1)^2\). 
To do so, $(\latitudeShort, \longitudeShort)$ are converted to 3D Cartesian coordinates: 
\begin{equation}
\label{equation:cartesian}
\begin{aligned}
    z &= \sin(\latitudeShort) \quad & x &= \cos(\longitudeShort) \cdot a \\
    a &= \cos(\latitudeShort) \quad & y &= \sin(\longitudeShort) \cdot a
\end{aligned}
\end{equation}

To determine the closest icosahedral face to $(\latitudeShort, \longitudeShort)$, the distance to each face is calculated individually.
Given the squared Euclidean distance between two points on a sphere, the algorithm now aims to find the angular distance $r$ between the two points when projected onto a unit sphere. The angular distance between two points on a sphere can be calculated using the spherical law of cosines and relates the sine of half the angular distance to the squared Euclidean distance between the points: $\sin^2\left(\frac{r}{2}\right) = \frac{d^2}{4}$, which can be transformed as $\cos(r) = 1 - 2 \sin^2\left(\frac{r}{2}\right) = 1 - 2 \cdot \frac{d^2}{4} = 1 - \frac{d^2}{2}$. This yields the angular distance as $r = \arccos(1 - \frac{d^2}{2})$.


The algorithm performs a gnomonic projection of this angle by taking its tangent $\tan(r)$. 
The tangent function is used here to convert the angular distance $r$ into a linear distance for the 2D plane.
After the gnomonic scaling, $r$ can be thought of as analogous to the radial distance in 2D polar coordinates.
The radial distance is scaled according to the scaling factor from hexagonal grid unit length at resolution $0$ to gnomonic unit length $\scalingFactorGnomonic$ given the desired H3 resolution $\resolutionShort$, such that $r = \frac{r}{\scalingFactorGnomonic} \cdot \sqrt{7}^{\resolutionShort}$.
Once the radial distance $r$ is calculated, the counterclockwise angle $\Angle$ between a reference direction on a given face of an icosahedron (for Uber H3 the i-axis of the Class II orientation~\cite{popko2021divided}) and the direction from the center of that face to a point on the globe is determined. 
The radial distance $r$ and the angle $\theta$ together describe the polar coordinates of the $(\latitudeShort, \longitudeShort)$ relative to the icosahedron face.
The angle $\Angle$ is calculated as the difference in azimuth (a type of angular measurement in a spherical coordinate system) between a reference axis on the icosahedron face and the azimuth from the center of that face to the given point. 
The azimuth values are normalized to be between $0$ and $2 \cdot \pi$, such that $\Angle = \text{norm}\left(\azimuthShort_{F_i} - \text{norm}\left(\azimuthShort(F_{\text{center}}, (\latitudeShort, \longitudeShort))\right)\right)$.
The azimuth $\azimuthShort(p1, p2)$ from point p1 to point p2, where $\latitudeShort_1$, $\longitudeShort_1$ are the latitude and longitude of p1, and $\latitudeShort_2$, $\longitudeShort_2$ are the latitude and longitude of p2, is calculated as $\azimuthShort(p1, p2) = \arctan(\frac{a}{b})$, with $a = \cos(\latitudeShort_2) \sin(\Delta)$, $b = \cos(\latitudeShort_1) \sin(\latitudeShort_2) - \sin(\latitudeShort_1) \cos(\latitudeShort_2) \cos(\Delta)$ and $\Delta = \longitudeShort_2 - \longitudeShort_1$.



In the H3 system, Class II and Class III orientations~\cite{sahr2003geodesic} alternate with each resolution.
To adjust for the alternate orientation of hexagons at different resolutions, a constant rotation angle $\arcsin(\sqrt{\frac{3}{28}})$ is subtracted from $\Angle$ if the chosen resolution is odd. 
The remaining transformation transforms the coordinates $(r, \Angle)$ to Cartesian coordinates $(x, y)$ in the two-dimensional plane by computing $x = r \cdot \cos(\Angle)$ and $y = r \cdot \sin(\Angle)$.
The resulting coordinates $(x,y)$ are the two-dimensional coordinates of the chosen hexagon relative to the face center of the closest icosahedron.

\textit{(2) Transforming $(x,y)$ coordinates to $(i,j,k)$ coordinates:}
The second step in obtaining coordinates in the discrete hexagonal planar grid system translates two-dimensional Cartesian coordinates $(x,y)$ to three-dimensional coordinates $(i,j,k)$, uniquely identifying a hexagon at a given resolution. It is natural for the grid system to have three coordinate axis, spaced $120$ degrees apart from each other, due to the structure of the underlying hexagons.
The three-axis system allows for unique addressing without ambiguities~\cite{sahr2019central}.

\input{./Protocols/circuit_ZKLP.tex}

The algorithm initially proceeds with quantization, setting $k$ to $0$ and operating on absolute values of $(x,y)$.
If the value of $(x,y)$ is not equivalent to the center of the hexagon, the continuous variables are rounded to the nearest hexagon center.
To adjust for negative Cartesian coordinates, the Cartesian coordinates are folded across the axes to map them onto the hexagonal grid accordingly in $(i,j,k)$ coordinates.
Finally, the computed $(i,j,k)$ coordinates are normalized to ensure that coordinates are as small as possible and non-negative. Normalizing the result is essential in ensuring that each hexagon in the grid has a unique address.

\noindent \textbf{Representing the Transformation as Constraints. }
\SNARKs work by encoding the computation in an arithmetic circuit over a finite field. In contrast, the transformation of geographic coordinates works over real numbers, represented in traditional programs as floating-point values.
We address this issue by utilizing the circuits for primitive floating-point operations described in Section~\ref{section:fp-primitive-ops}.

We still face the problem that our primitive operations in Section~\ref{section:fp-primitive-ops} do not aim to provide precise math functions ($\sin, \cos, \tan, ...$) as the IEEE standard does not specify the precision of math libraries.
Even more so, a naive implementation would require approximation of math functions by the standard three-step recipe, as utilized in standard libraries~---~range reduction, polynomial approximation, and output compensation. Emulating polynomial approximations, by computing in-circuit Taylor Series, or applying the Remez algorithm~\cite{remez1934determination}, would lead to expensive increase in constraints due to high degree polynomials.
Further, \textit{precise} approximation is non-trivial, and standard techniques are inefficient without significant in-circuit optimization. Efficient algorithms for emulating accurate trigonometric functions are known for Two-Party-Computation~\cite{rathee2022secfloat}. However, we are not aware of any optimizations that lead to accurate and efficient in-circuit trigonometric approximations for \SNARKs.
We describe optimizations that fully eliminate trigonometric functions in our circuits as follows.


\input{./Protocols/circuit_R.tex}

\noindent \textbf{Avoiding Trigonometric Operations. }
Recall that transforming spherical coordinates to Cartesian coordinates demands \emph{(i)} calculating the radial distance $r$ of $\userPoint$ to the closest icosahedral face (Step 4 in Figure~\ref{figure:baseline}), \emph{(ii)} calculating the angle $\Angle$ of $\userPoint$ to the closest icosahedral face (Step 5 in Figure~\ref{figure:baseline}) and \emph{(iii)} converting $(r, \Angle)$ to Cartesian coordinates $(x, y)$ in the 2D plane by computing $x = r \cdot \cos(\Angle)$ and $y = r \cdot \sin(\Angle)$ (Step 6 in Figure~\ref{figure:baseline}).
We observe that we can avoid trigonometric operations altogether in the above steps by leveraging trigonometric identities.

To avoid evaluating trigonometric operations for computing the radial distance $r$, we substitute $r = \arccos(1 - \frac{d^2}{2})$ into $\tan(r)$. 
By the Pythagorean identity, we express $\tan(r) = \tan(\arccos(1 - \frac{d^2}{2}))$.




\begin{equation}
\label{equation:AvoidTrig}
\begin{aligned}
    r = \tan(r) = \sqrt{\frac{(-(d^2 - 4) \cdot d^2)}{(2 - d^2)}} 
\end{aligned}
\end{equation}

To minimize the number of constraints, we scale to the desired H3 resolution by applying the square and multiply algorithm for bitwise exponentiation instead of naive exponentiation (cf. Figure~\ref{figure:CircuitRadial}).

Similarly, we simplify computing $(x,y)$ with the angle $\Angle$. Recall that $\Angle = \text{norm}\left(\azimuthShort_{F_i} - \text{norm}\left(\azimuthShort(F_{\text{center}}, (\latitudeShort, \longitudeShort))\right)\right)$ where $\azimuthShort(p1, p2) = \arctan(\frac{a}{b})$.
By substituting the above values in the equations for calculating $(x,y)$ and leveraging trigonometric identities, we obtain:
$$
x = r \cdot \left( \cos(\theta_{F_i})\frac{b}{\sqrt{a^2 + b^2}} + \sin(\theta_{F_i})\frac{a}{\sqrt{a^2 + b^2}} \right)
$$
$$
y = r \cdot \left(\sin(\azimuthShort_{F_i})\frac{b}{\sqrt{a^2 + b^2}} - \cos(\azimuthShort_{F_i})\frac{a}{\sqrt{a^2 + b^2}}\right)
$$

Note, that the trigonometric identities used in the simplification are mathematically sound regardless of normalization.
As the center point of each icosahedron is fixed and known, the remaining $\sin$ and $\cos$ terms can be pre-computed.
This representation is significantly less costly, as we already derived an optimized gadget for computing the square root of a floating-point variable in Section~\ref{section:squareRoot}.


\noindent \textbf{Eliminating Trigonometric Operations with Hints. }
It remains the elimination of trigonometric operations by avoiding the initial calculation of the Cartesian coordinates $(x,y,z)$ from the user-supplied spherical coordinates $(\latitudeShort, \longitudeShort)$ (cf. Equation~\ref{equation:cartesian}).
We observe that we can mindfully construct hints, such that the in-circuit computation reduces to \emph{(i)} evaluating the hint predicate and \emph{(ii)} calculating $(x,y,z)$ without using trigonometric functions.
As such, the hint $\Computation_{\mathsf{ZKLP}}(\latitudeShort) = [\alpha_{\latitudeShort}, \beta_{\latitudeShort}, \gamma_{\latitudeShort}, \delta_{\latitudeShort}]$ is computed as $\gamma_{\latitudeShort} = \sin\left(\frac{\latitudeShort}{2}\right)$, $\delta_{\latitudeShort} = \cos\left(\frac{\latitudeShort}{2}\right)$, $\beta_{\latitudeShort} = 2 \cdot \gamma_{\latitudeShort} \cdot \delta_{\latitudeShort}$ and $\alpha_{\latitudeShort} = \tan\left(\frac{\latitudeShort}{2}\right)$.

Soundness holds as $\Predicate_{\mathsf{ZKLP}}$ evaluates that \emph{(i)} $\gamma_{\latitudeShort}^2 + \delta_{\latitudeShort}^2$ equals 1, and thereby fulfills the fundamental trigonometric identity, \emph{(ii)} $\delta_{\latitudeShort} \cdot \alpha_{\latitudeShort}$ equals $\gamma_{\latitudeShort}$, which checks if the same angle is used, and \emph{(iii)} $2 \cdot \gamma_{\latitudeShort} \cdot \delta_{\latitudeShort}$ equals $\beta_{\latitudeShort}$, confirming that $\beta_{\latitudeShort}$ is correctly related to $\gamma_{\latitudeShort}$ and $\delta_{\latitudeShort}$.
Afterwards, the x,y,z coordinates can simply be derived as $z = \beta_{\latitudeShort}$, $r = \sqrt{1 - \beta_{\latitudeShort}^2}$, $x = \sqrt{1 - \beta_{\longitudeShort}^2} \cdot r$ and $y = b_{\longitudeShort} \cdot r$.
As a result, trigonometric operations are eliminated from $\CircuitZKLP$.
Note that $\sin(i) = \beta_i$ and $\cos(i) = \delta_i^2 - \gamma_i^2$ holds for $i \in \{\latitudeShort, \longitudeShort\}$.

\input{./Protocols/circuit_Theta.tex}

\noindent \textbf{Transforming $(x,y)$ to $(i,j,k)$ coordinates. } 
The transformation is conducted by first transforming $(x,y)$ to $(i,j,k)$ coordinates and successively normalizing $(i,j,k)$ coordinates to adjust for negative coordinates.
We provide a detailed description of the sub-circuits for obtaining and normalizing $(i,j,k)$ coordinates in Figure~\ref{figure:IJK} and~\ref{figure:CircuitNormalizeIJK} in the appendix.
They directly benefit from our floating-point implementation, due to many floating-point comparisons.

\section{Empirical Evaluation}
\label{section:evaluation}


Our empirical evaluation addresses three questions: \emph{(i)} What is the performance and accuracy of our floating-point implementation?, \emph{(ii)} How effective are our optimizations for the \ZKLP paradigm?, and \emph{(iii)} How tolerable is the cost of \ZKLP in real-world use?

\noindent \textbf{Implementation. }
We implement the floating-point primitive operations (cf. \S\ref{section:fp-primitive-ops}) as a reusable library in gnark~\cite{gnark-v0.9.0}.
We provide a full implementation of the optimized circuits for \ZKLP (cf. \S\ref{section:zkLocation}) for \FPSingle and \FPDouble values.
In addition, we implement the baseline protocol, without the optimizations as described in \S\ref{section:zkLocation}, over fixed-point arithmetic.
Due to the agnostic nature of gnark, our implementation supports Groth16 and Plonk as the SNARK.
We instantiate the lookup argument as LogUp \cite{habock2022multivariate}, which is used in gnark for range checks. 
Our implementation and measurement data are open-sourced~\cite{ImplementationLink} for reproducibility.

\noindent \textbf{Test Suite. }
To ensure compliance with IEEE 754~\cite{ieee754fp}, we create a set of test values with the Berkeley TestFloat library~\cite{TestFloat},  which generates test cases to ensure that an implementation conforms to the IEEE Standard for Floating-Point Arithmetic. Specifically, 46464 test cases for each binary operation (e.g., Add), and 600 test cases for the unary operation Sqrt are created.
Our implementation passes all these test cases, including inputs with abnormal values.

To evaluate the \ZKLP circuit, we generate a series of geospatial points within hexagonal cells at various resolution levels, using Uber H3 implemented in C~\cite{UberH3}.
For each resolution $\resolutionShort \in [0, 15]$, we test 16 distances from the center to the point.
The $i$-th distance is $(1 - 2^{-i})d$, where $d$ is the center-to-boundary distance.
We further randomly sample 100 points at each distance.
Thus, we have a sparser distribution of points closer to the center and denser as it approaches the boundary (cf. Figure~\ref{figure:sphere}).
In addition, because the test cases generated by Uber H3 contain floating-point values, when evaluating the fixed-point baseline, we convert the floating-point values to fixed-point by multiplying by a scalar and rounding to the nearest integer.

\noindent \textbf{Experimental Setup.} 
When evaluating the runtime and memory consumption of a circuit, we execute all tests on an \texttt{m6i.xlarge} AWS instance with 4 vCPUs and $16$ GB of RAM.
The number of constraints is independent of the execution architecture.
For running time we report all values as the mean of 20 executions with multi-threading enabled.

\input{./Tables/table_evaluation_float.tex}

\subsection{Microbenchmarks and Comparison}\label{section:evaluation-fp}

We evaluate the cost of floating-point circuits (Section \ref{section:fp-primitive-ops}) and compare our implementation with existing works.

\noindent \textbf{Number of Constraints.} Table~\ref{table:fp_constraints} presents the number of R1CS constraints for in-circuit floating-point primitive operations.
As discussed in Appendix~\ref{appendix:Integer}, we utilize two lookup tables $\mathcal{T}_\mathsf{RC}$ and $\mathcal{T}_\mathsf{Pow2}$, where the former is for range check, and the latter is for the computation of $2^d$.
In Table~\ref{table:fp_constraints}, the size of $\mathcal{T}_\mathsf{RC}$ is fixed at $2^8$, and the size of $\mathcal{T}_\mathsf{Pow2}$ is 32 for \FPSingle and 64 for \FPDouble.
The constraints for each operation consist of two parts: native constraints (supported by the constraint system) and lookup constraints (for lookup tables).
The inclusion of lookup constraints is necessary due to gnark's implementation of LogUp~\cite{habock2022multivariate}.
Recall that gnark checks LogUp's identity for set inclusion $\sum_{i=1}^{|\mathbf{f}|} \frac{1}{X - f_i} = \sum_{j=1}^{|\mathbf{t}|} \frac{o_j}{X - t_j}$ in arithmetic circuit, resulting in a single \SNARK proof for both the original relation and the validity of set inclusion.
To eliminate lookup constraints, one can instantiate the lookup argument as standalone protocol and connect it to \SNARK in a commit-and-prove fashion, at the cost of larger proofs.

Specifically, the in-circuit verification of LogUp's identity involves three steps: \emph{(i)} compute the LHS $\sum_{i=1}^{|\mathbf{f}|} \frac{1}{X - f_i}$, \emph{(ii)} compute the RHS $\sum_{j=1}^{|\mathbf{t}|} \frac{o_j}{X - t_j}$, and \emph{(iii)} check the equality between the LHS and the RHS. As highlighted in Table~\ref{table:fp_constraints}, the costs of \emph{(ii)} and \emph{(iii)} are operation-agnostic; they only depend on the sizes of lookup tables $\mathcal{T}_\mathsf{RC}$ and $\mathcal{T}_\mathsf{Pow2}$. Hence, costs remain unchanged as the number of queries increases and they can be amortized across multiple operations.

Figure \ref{figure:amortization} shows the amortized cost of primitive operations, using multiplication as an example. With a larger $\mathcal{T}_{RC}$, the cost of step \emph{(ii)} increases, while step \emph{(i)} requires fewer constraints. This is particularly advantageous when proving the execution of many operations, as the one-time cost of steps \emph{(ii)} and \emph{(iii)} becomes negligible.

\input{./Figures/graph_float.tex}

\noindent \textbf{Comparison With Other Works.} Naively converting \FPSingle operations compliant to IEEE 754 requires $2456$ and $8854$ boolean gates for addition and multiplication~\cite{weng2021mystique}. \FPDouble addition and multiplication require $15637$ and $44899$ boolean gates respectively~\cite{archer2021cost}. Garg~\emph{et. al} \cite{garg2022succinct} provide the state-of-the-art succinct \ZKP for floating-point operations, which requires 108 non-zero entries in the R1CS instance for \FPSingle addition and 25 for \FPSingle multiplication in a circuit over BN254.
Due to the inconsistent metrics and our requirement of amortization, we are unable to present a fair comparison. 


\subsection{Zero-Knowledge Location Proofs}\label{section:evaluation-zklp}


We assess the efficiency of end-to-end \ZKLP circuits (cf. Section~\ref{section:zkLocation}) and highlight their relevance in privacy-preserving peer-to-peer proximity testing as a case study. 
Our analysis includes a baseline ($\Pi_{Base}$, cf. Figure~\ref{figure:baseline}), implemented via unoptimized fixed-point circuits that support two fixed-point data types P20 and P40.
The scaling factors for P20 and P40 are $10^6 \approx 2^{20}$ and $10^{12} \approx 2^{40}$, i.e., their last $20$ and $40$ bits are fractional parts.
To represent large intermediate results, we do not limit the number of bits used for the integer parts of P20 and P40, as long as they are smaller than the order of $\mathbb{F}_p$.
Specifically, for $\sqrt{7}^{\resolutionShort}$ with $\resolutionShort = 15$, the integer parts must use at least $23$ bits, and subsequent operations even require P20 and P40 to have $42$ bits and $81$ bits in their integer parts.
In total, the maximum bit lengths for P20 and P40 are $62$ bits and $121$ bits, respectively.
In the fixed-point circuits, we approximate trigonometric functions within the circuit: $\sin(x)$ is approximated via a Taylor Series expansion, and $\arctan(x)$ is approximated using the Remez algorithm~\cite{remez1934determination}. The mathematical function approximation follows the standard three-step method: \textit{range reduction}, \textit{polynomial approximation}, and \textit{output compensation}~\cite{cody1980software}.
In contrast, we present results for our optimized floating-point circuits, implementing the $\CircuitZKLP$ circuit (cf. Figure~\ref{figure:ZKLP}).


\noindent \textbf{Quantitative Baseline Comparison.}
Table~\ref{table:ZKLP} depicts the quantitative comparison of the fixed-point baseline and $\CircuitZKLP$ implemented over our floating-point circuits. Our results show, that our optimizations applied for $\CircuitZKLP$ are indeed very effective.
With Groth16, our single precision floating-point circuit has $\factorSingleBetterThanBaselineConstraints$ less constraints than the fixed-point baseline, whereas with double precision floating-point, our circuit has $\factorDoubleBetterThanBaselineConstraints$ less constraints than the baseline.
Note that the number of constraints in $\CircuitZKLP$ remains constant, regardless of the chosen resolution.
The proof generation time for both $\CircuitZKLP$ over \FPSingle and \FPDouble is below $\SI{1}{\second}$ for Groth16.
With Plonk, the time to generate a proof is higher, which is expected given Plonk's prover complexity $\mathcal{O}(n \log n)$ given an arithmetic circuit of $n$ gates.

While we evaluate $\CircuitZKLP$ on a server, mid-range mobile devices can achieve similar performance. On an Android phone with a Snapdragon 7+ Gen 2 CPU, the prover takes $\SI{0.256}{\second}$ for \FPSingle  and $\SI{0.333}{\second}$ for \FPDouble with Groth16 as the SNARK, demonstrating the practicality of \ZKLP.

\noindent \textbf{Qualitative Baseline Comparison.} We report the success rate of tests for fixed-point and floating-point implementations for emulating the transformation of $(\latitudeShort, \longitudeShort)$ to $(i,j,k)$ coordinates in a \ZKP for resolutions $0$ to $15$ (Figure~\ref{figure:resolutionComparison}).
We find that in the baseline protocol with P20, errors start to occur frequently at a resolution of $\resolutionShort = 3$, with most tests failing for $\resolutionShort \geq 11$. In contrast, errors in the modified protocol become frequent only for $\resolutionShort \geq 7$, and significant failures are observed only when $\resolutionShort = 15$.
After adjusting the fixed-point implementation to P40, the errors are greatly reduced, though there are still some edge cases where P40 falls short.
On the other hand, all tests pass with FP64, as its data format and rounding mode align with the Uber H3 implementation.
Overall, we find that both the precision and the rounding mode have a significant impact on the accuracy of the computations, e.g., those involving the constant scaling factor of the resolution ($\sqrt{7}^{\resolutionShort}$).

\input{./Tables/table_evaluation_zklp.tex}


\noindent\textbf{P2P Proximity Testing. } 
We utilize the \ZKLP paradigm to realize P2P proximity testing.
Let $\mathcal{H} = \{\mathbf{h_1}, \mathbf{h_2}, ..., \mathbf{h_n}\}$ be a set of hexagons generated by the H3 system, where each hexagon $\mathbf{h}$ is defined by its boundary vertices. Each vertex $v_i$ of hexagon $\mathbf{h}$ is given in geographical coordinates (latitude $\latitudeShort_i$ and longitude $\longitudeShort_i$).
Given a user's position $P_u = (\latitudeShort_u, \longitudeShort_u)$, the proximity to a hexagonal cell $\mathbf{h}_v$ is evaluating the Haversine formula to calculate the great-circle distance between $P_u$ and each vertex of $\mathbf{h}_v$.
The Haversine distance $\delta$ can be calculated as 
$c = 2 \cdot \text{atan2}\left(\sqrt{a}, \sqrt{1-a}\right)$, where
$a = \sin^2\left(\frac{\latitudeShort_2 - \latitudeShort_1}{2}\right) + \cos(\latitudeShort_1) \cdot \cos(\latitudeShort_2) \cdot \sin^2\left(\frac{\longitudeShort_2 - \longitudeShort_1}{2}\right)$.
Here, $\latitudeShort_1, \longitudeShort_1$ and $\latitudeShort_2, \longitudeShort_2$ are the latitude and longitude in radians of points $P_u$ and a vertex of $\mathbf{h}_v$, respectively, and $r$ is the radial length of the earth.
The minimum distance $\Delta_{\text{min}}$ from point $P_u$ to the boundary of hexagon $\mathbf{h}_v$ is calculated as the smallest distance $\Delta_{\text{min}} = \min(\delta_0, \delta_1, ..., \delta_i)$ (cf. Figure~\ref{figure:sphere}).
Calculating the proximity of $P_u$ to the hexagonal boundary is done in plain and hence efficient when compared to proving the \ZKLP circuit.
We implement proximity testing in Go, relying on the H3 library in C to determine the boundary points of the hexagon. We find that the computation requires $\approx \TimeProximityTesting$ to execute, which is comparable to verifying a Groth16 proof ($\approx \SI{1.54}{\millisecond}$).
Hence, a verifier can evaluate its proximity to $\approx \numberProofsPerSecond$ peers per second.

\section{Related Works}\label{section:related-work}

\noindent\textbf{Floating-Point Secure Computing. }
To the best of our knowledge, there is no prior work supporting fully IEEE 754 compliant floating point computations for succinct proof systems.
There are several prior works that investigate floating-point computations for secure \MPC~\cite{aliasgari2012secure,kamm2015secure}.
Later, \cite{pullonen2015combining,archer2021cost} build IEEE 754 compliant MPC circuits by compiling from software implementations of IEEE 754 using tools such as CBMC-GC.
However, the resulting circuits have at least thousands of gates per operation, making them inefficient in practice.
In~\cite{rathee2022secfloat}, Rathee~\emph{et. al} construct standard compliant functionalities for 2PC with dedicated optimizations.
While providing better efficiency, they can only achieve partial compliance with IEEE 754, but subnormal values and NaNs are not considered.
Another line of research focuses on proving floating-point computations using \ZKPs.
Weng~\emph{et. al}~\cite{weng2021mystique} use the IEEE 754 compliant single-precision boolean circuits from EMP-toolkit~\cite{emp-toolkit} with non-succinct proofs.
Closest to our work is the work of Garg~\emph{et. al}~\cite{garg2022succinct}, which studies succinct \ZKPs for floating-point arithmetic.
However, their approach only supports addition and multiplication, whilst not providing a concrete implementation and only theoretical performance estimates.
Also, the verifier time in~\cite{garg2022succinct} is linear. While they provide a method to achieve sub-linear verification, the best complexity they could achieve is $O(\sqrt{n})$, where $n$ is the circuit size. In comparison, due to Groth16, the verifier time of our construction is constant in $n$.
Further, the relative error model in~\cite{garg2022succinct} is not suitable for many applications. For instance, it is observed in~\cite{srivastava2024optimistic} that for machine learning, even the minor rounding errors due to non-determinism in GPUs can result in very different predictions.
Consequently, an adversary may leverage this fact to generate valid proofs for arithmetic circuits that do not produce the expected results as in IEEE 754 compliant computer hardware.

\noindent\textbf{Location Privacy. }
There is a long line of work on \LPPM via geo-indistinguishability~\cite{shokri2011quantifying,shokri2012protecting,andres2013geo, vsedvenka2014privacy}.
Narayanan~\emph{et. al}~\cite{narayanan2011location} introduce location privacy via private proximity testing.
Vsedvenka~\emph{et. al}~\cite{vsedvenka2014privacy} introduce interactive protocols for proximity testing over a spherical surface. In a similar setting, their protocols require $> \SI{1}{\second}$ and $>10$ kB communication. In contrast, our protocol is non-interactive and requires $\approx \ProverTimeSinglePrecisionBNGSixteenForCZKLP$ execution time (disregarding latency) and communicating a $\approx 200$ Byte proof with Groth16.
Babel~\emph{et. al}~\cite{babel2023bringing} show how to evaluate whether a location is in a polygon. However, their approach assumes coordinates in $(x,y)$ form in Euclidean plane. In comparison, our circuit for transforming $(x,y)$ to a hexagonal index ($\CircuitIJK$) yields $\approx 11.7\times$ less constraints.
To the best of our knowledge, \ZKLP provides the first paradigm for non-interactive, publicly verifiable and privacy preserving proofs of geolocation.

\section{Discussion \& Future Work}

This paper introduces \ZKLP (\S~\ref{section:zkLocation}) via accurate floating-point \SNARKs (\S~\ref{section:fp-primitive-ops}), identifying a novel class of applications of general-purpose succinct ZK proofs.
The main challenge was the accurate emulation of floating-point values without increasing constraints or compromising soundness.

In our experiments, we show that our implementation of floating-point arithmetic is efficient and accurate. 
We show that our instantiation of lookup tables amortizes the number of constraints per primitive operation.
We show that the \ZKLP paradigm is efficient, allowing users to generate an accurate proof in $\ProverTimeSinglePrecisionBNGSixteenForCZKLP$. We instantiate \ZKLP with P2P proximity testing and show that a verifier can verify proximity to up to $\numberProofsPerSecond$ peers per second. Note, that the verification time and proof size is inherited by Groth16~\cite{groth2016size, ernstberger2023zk}.

\ZKLP can be directly applied to scenarios where the location data is already authenticated. For example, in the workflow of C2PA~\cite{C2PA}, the location where a photo is taken is signed by a C2PA-compatible camera, and thus we can seamlessly integrate \ZKLP with C2PA to provide photo authenticity while obfuscating the accurate location, thereby preserving the privacy of the photo's author.
Further, we identify three potential solutions on how to obtain authentic location, which we detail in Appendix~\ref{appendix:authenticate}.


We expect our results to adapt naturally to other settings.
For instance, in machine learning, where parameters are often floating-point numbers~\cite{yeh2022like}, our methods can enable efficient, precise training~\cite{garg2023experimenting} and inference proofs~\cite{kang2022scaling}. 
Finally, we believe that authenticated \ZKLP could be a useful building block in applications for Proof-of-Personhood to obtain verifiable location-based Sybil-resistance~\cite{borge2017proof} and leave its exploration for future work. 

\ifCLASSOPTIONcompsoc
  \section*{Acknowledgments}
\else
  \section*{Acknowledgment}
\fi
We acknowledge the financial support by the Federal Ministry of Education and Research of Germany in the programme of “Souverän. Digital. Vernetzt.”. Joint project 6G-life, project identification number: 16KISK002.
This work was done while Chengru Zhang was visiting UCL.





\bibliographystyle{IEEEtran}
\bibliography{bibliography}


\appendices

\input{./Tables/table_test_new.tex}


\input{./Protocols/gadget_AssertBitLength.tex}

\section{Circuits for Integer Operations}
\label{appendix:Integer}
Arithmetic circuits only natively support operations in the prime field $\mathbb{F}_p$, but it is non-trivial to perform in-circuit integer operations in an efficient and sound way.
Below we introduce circuits that checks the range, extracts sign and absolute value, computes maximum and minimum values, and performs left and right shifts for integers.


\noindent \textbf{Range Check. }
Bit decomposition is a standard technique for ensuring a variable $\var{v} \in \mathbb{F}_p$ is an $L$-bit string, i.e., $\var{v} \in [0, 2^L - 1]$. 
The prover first provides the decomposition as a hint to the circuit by computing $\Computation_{\mathsf{Dec}}(\var{v}) = \{\var{v}_0, \dots, \var{v}_{L - 1}\} $.
The predicate \( \Predicate_{\mathsf{Dec}} \) for verifying the bit decomposition of \( \var{v} \) into 
\( \{\var{v}_0, \dots, \var{v}_{L - 1}\} \) asserts that all variables are boolean by checking $\var{v}_i(1 - \var{v}_i) = 0$ for all $i \in [0, L - 1]$, and that $\var{v}$ is indeed composed of $\{\var{v}_0, \dots, \var{v}_{L - 1}\}$ by asserting $\sum_{i = 0}^{L - 1}2^i \var{v}_i = \var{v}$.
$\Predicate_{\mathsf{Dec}}$ returns $1$ if these checks pass and $0$ otherwise.
Note, that $L$ should satisfy $2^L < p$ to ensure $\sum 2^i \var{v}_i$ does not overflow.
For $\var{a}$ and $\var{b}$ known to be $L$-bit strings, this method can be extended to ensure $\var{v} \in [\var{a}, \var{b}]$ by decomposing $\var{v} - \var{a}$ and $\var{b} - \var{v}$ separately into $L$ bits.

However, bit decomposition requires $L + 1$ constraints, which is not optimal. We can improve the circuit efficiency by leveraging lookup argument. We build a lookup table $\mathcal{T}_\mathsf{RC}$ with entries $\{0, \dots, 2^T - 1\}$. On inputs $\var{v}, L$, the circuit $\AssertBitLength$ now requires the prover to compute the hint $\Computation_{\mathsf{Dec}'}(\var{v})$ by decomposing $\var{v}$ into $T$-bit strings $\var{v}'_0, \dots, \var{v}'_{L / T - 1}$. Then the circuit checks the predicate $\Predicate_{\mathsf{Dec}'}(\var{v}, \{\var{v}'_0, \dots, \var{v}'_{L / T - 1}\})$ by appending the set $\{\var{v}'_0, \dots, \var{v}'_{L / T - 1}\}$ to the vector of queries $\textbf{t}_\mathsf{RC}$ and enforcing $\var{v} = \sum_{i = 0}^{L/T - 1}2^{iT} \var{v}'_i$.

The $\AssertBitLength$ circuits based on both approaches for checking $\var{v} \in [0, 2^L - 1]$ are described in Figure~\ref{figure:assert_bit_length}. 


\noindent \textbf{Sign and Absolute Value. }
Obtaining the sign and the $L$-bit absolute value of a variable $\var{v}$ presents a more complex challenge.
Intuitively, a number is positive if it is greater than 0, and is negative otherwise. However, as the field $\mathbb{F}_p$ is not \textit{ordered}, we cannot \textit{compare} between its elements. To address this, we manually define elements in the set $\{ 1, 2, \dots, (p - 1) / 2\}$ as positive, and those in the set $\{(p + 1) / 2, \dots, p - 2, p - 1\}$ as negative. 
Now, as long as $2^L < (p - 1) / 2$, we can extract the sign and the absolute value of $\var{v}$ as depicted in $\CircuitAbs$ in Figure \ref{figure:combined_circuits}. 
The prover determines if $\var{v}$ is positive by checking which set it belongs to, and provides $\Computation_{\mathsf{GEZ}} = \signBitShort$ as a hint to the circuit. 
The gadget enforces that $\var{\signBitShort}$ is boolean,
and computes $\var{v}$'s absolute value $\absoluteValueShort$, which is $\var{v}$ if $\signBitShort$ is 1, and is $-\var{v}$ otherwise.
Finally, the gadget enforces that $\absoluteValueShort$ has at most $L$ bits and returns $\absoluteValueShort$ and $\signBitShort$.
Soundness holds because if an adversary feeds the incorrect $\signBitShort$ to the circuit, then $\absoluteValueShort$'s value belongs to the negative set and is hence greater than $(p - 1) / 2$, but $\AssertBitLength$ is later used to guarantee that $\absoluteValueShort < 2^L < (p - 1) / 2$.

\input{./Protocols/gadget_Abs_Max_Min.tex}



\noindent \textbf{Maximum and Minimum. }
Given $\CircuitAbs$, it is straightforward to build circuits that find the maximum and minimum values between $\var{x}$ and $\var{y}$ whose difference has $L$ bits (cf. Figure \ref{figure:combined_circuits}), which is done by calling $\CircuitAbs$ on $\var{x} - \var{y}$ and select $\var{x}$ or $\var{y}$ based on the sign of the difference.
$\CircuitMax$ and $\CircuitMin$ forward $\var{x} - \var{y}$ and $L$ to $\CircuitAbs$, where $\CircuitAbs$ returns the sign bit $\signBitShort$ only.

\input{./Protocols/gadget_Shift.tex}

\noindent \textbf{Shifting. }
Figure~\ref{figure:shl_shr} summarizes the circuits for shifting left $\shl$ and right $\shr$.
First, we compute powers of 2 in-circuit.
For a constant exponent $D$, $2^D$ is also a constant and does not involve any constraints.
For $2^\var{d}$ with a variable exponent $\var{d} \in [0, K]$ where $2^K < p$, one approach is to leverage the square-and-multiply algorithm. That is, we decompose $\var{d}$ into $K$ bits $\var{d}_0, \dots, \var{d}_{K - 1}$ and select the term $2^{2^i}$ or 1 based on the value of $\var{d}_i$. The product of these terms is the result $2^\var{d}$.
This method costs $O(K)$ constraints. For example, in R1CS, we need $K + 1$ constraints for bit decomposition and $K - 1$ constraints for multiplying $K$ terms, resulting in $2K$ constraints in total.
This is not ideal, especially for our case where shift operations are frequently used. In order to minimize circuit size, we introduce $\mathcal{T}_\mathsf{Pow2}$, a lookup table for $2^d$. 
$\mathcal{T}_\mathsf{Pow2}$ is first populated with entries $\textbf{t}_\mathsf{Pow2} = \{(i, 2^i)\}^{K}_{i=0}$.
Then, to compute $2^d$, the prover provides a hint $r$, and the circuit appends $(d, r)$ to the vector of queries $\textbf{f}_\mathsf{Pow2}$, which is later enforced to satisfy $\textbf{f}_\mathsf{Pow2} \in \textbf{t}_\mathsf{Pow2}$.
Due to the soundness of the lookup argument, $r$ is guaranteed to be $2^d$.

Assuming that $\var{v} \in [0, 2^L - 1]$ and $\var{d} \in [0, K]$ where $2^{L + K} < p$, it is straightforward to compute the left shift $\var{v} \shl \var{d}$; we only need to compute $2^\var{d}$ and then return $\var{v} \cdot 2^\var{d}$. 

Constructing an efficient right shift gadget $\var{v} \shr \var{d}$ needs non-trivial techniques. 
Here, we also assume that $\var{v} \in [0, 2^L - 1]$, $\var{d} \in [0, K]$, and $2^{L + K} < p$. 
Naively, we could treat the right shift operation as integer division, i.e., $v \shr d = v / 2^d$. 
The prover computes the quotient $q$ and the remainder $r$ such that $v = q \cdot 2^d + r$, and feeds $\var{q}, \var{r} \coloneqq \Computation_{\mathsf{Div}}(\var{v}, 2^\var{d})$ as hints to the circuit. Then to check the predicate $\Predicate_{\mathsf{Div}}(\var{v}, 2^\var{d}, \var{q}, \var{r})$, it is required to enforce that $\var{q} \in [0, 2^L - 1]$, i.e., $\var{q} \cdot 2^\var{d}$ does not overflow, that $\var{r} \in [0, 2^\var{d} - 1]$, i.e., the remainder should be positive and smaller than the divisor, and that $\var{v} = \var{q} \cdot 2^\var{d} + \var{r}$. We can see that checking the range of $\var{q}$ requires decomposing an $L$-bit integer, and checking the range of $\var{r}$ requires decomposing two $K$-bit integers $r$ and $2^\var{d} - 1 - r$ (note that the upper bound $2^\var{d} - 1$ is a variable). 
This is suboptimal, as the above checks are equivalent to decomposing an $L + 2K$-bit integer.

We now reduce the number of bits to be decomposed.
Instead of naively computing $\var{v} \shr \var{d}$, the prover first computes $\var{v}' \coloneqq \var{v} \shl (K - \var{d}) = \var{v} \cdot 2^{K - \var{d}}$. Since $\var{d} \in [0, K]$, we have $K - \var{d} \in [0, K]$, and thus $\var{v} \cdot 2^{K - \var{d}} < 2^{L + K} < p$ is safe. Then we handle $\var{v}' \shr K$ analogously: the prover computes the quotient $q$ and the remainder $r$ for $v' / 2^K$ and provides $\var{q}, \var{r} = \Computation_{\mathsf{Div}}(\var{v}', 2^K)$ as hints, and the circuit checks the predicate $\Predicate_{\mathsf{Div}}(\var{v}', 2^K, \var{q}, \var{r})$ by asserting that $\var{q} \in [0, 2^L - 1]$, $\var{r} \in [0, 2^K - 1]$, and $\var{v}' = \var{q} \cdot 2^K + \var{r}$. Another way to understand how to check $\var{v'} \shr K$ is that the circuit first decomposes $\var{v'}$ into $L + K$ bits, and then computes $\var{q} = \sum_{i = 0}^{L - 1} 2^i\var{v}'_i$. 
The optimized approach only requires decomposing $L + K$ bits, saving $O(K)$ constraints.

\section{Circuit for Floating-Point Division}
\label{section:division}

\input{./Protocols/circuit_DivFloat.tex}

Dividing an IEEE 754 floating-point number $\alpha = (s_\alpha, e_\alpha, m_\alpha, a_\alpha)$ by another $\beta = (s_\beta, e_\beta, m_\beta, a_\beta)$ is done in the following $4$ steps~---~\emph{(i)} computing the quotient of $\alpha$ and $\beta$, \emph{(ii)} normalizing and \emph{(iii)} rounding the intermediate mantissa and \emph{(iv)} handling edge cases. We depict the corresponding in-circuit logic in Figure \ref{figure:div_float}.

\noindent \textbf{Compute quotient (lines 1-8).} Analogous to multiplication, the quotient has sign $s \coloneqq s_\alpha \oplus s_\beta$ and exponent $e \coloneqq e_\alpha - e_\beta$. However, extra care is needed for computing the mantissa. Our rounding operation necessitates intermediate mantissas with infinite precision, but it is infeasible to represent the exact quotient when $m_\beta \nmid m_\alpha$. To address this, we instead divide $\alpha$'s scaled mantissa $m_\alpha \shl (M + 2)$ by $m_\beta$ and obtain the quotient $q$ and remainder $r$, such that $m_\alpha \shl (M + 2) = q \cdot m_\beta + r$, and the intermediate mantissa is $m \coloneqq q$. Since $m_\alpha, m_\beta$ are either 0 or lie in $[2^M, 2^{M + 1} - 1]$, a non-zero, finite $m$ should be bounded by $m \in (2^{M + 1}, 2^{M + 3})$.

The shift $M + 2$ is the smallest value that allows $m$ to retain the correct round bit $m_{M + 1}$, and $r$ is used to assist the rounding process and determine the sticky bit, just as if we are rounding a mantissa with infinite precision. This is achieved by checking if $r$ is zero. If this is the case, $m_\beta \mid (m_\alpha \shl (M + 2))$, and the remaining bits after the round bit in the exact result $\frac{m_\alpha}{m_\beta}$ are all 0, implying that the sticky bit is 0. Otherwise, the sticky bit is 1.

To compute $(m_\alpha \shl (M + 2)) / m_\beta$ inside the circuit, the prover needs to do the division outside the circuit and provide $\var{q}, \var{r} \coloneqq \Computation_{\mathsf{Div}}(\varmantissa_\var{\alpha} \shl (M + 2), \varmantissa_\var{\beta})$ as hints, and the circuit will check the predicate $\Predicate_{\mathsf{Div}}(\varmantissa_\var{\alpha} \shl (M + 2), \varmantissa_\var{\beta}, \var{q}, \var{r})$ by enforcing \emph{(i)} $\var{q} \in [0, 2^{2M + 3} - 1]$, \emph{(ii)} $\var{r} \in [0, \varmantissa_\var{\beta})$, and \emph{(iii)} $\varmantissa_\var{\alpha} \shl (M + 2) = \var{q} \cdot \varmantissa_\var{\beta} + \var{r}$. We eliminate the check \emph{(i)}, which is unnecessary as $\var{q}$'s range will be narrowed to $[0, 2^{M + 3} - 1]$, as we describe later. \emph{(ii)} is converted to two range checks since the upper bound $\varmantissa_{\var{\beta}}$ is a variable.

The quotient is abnormal if the dividend is abnormal or the divisor is $\pm 0$ or NaN, i.e., $a \coloneqq a_\alpha \lor \CircuitIsEq(\varmantissa_\var{\beta}, 0)$.

\noindent \textbf{Normalize intermediate mantissa (lines 9-13).} $m$ is normalized in the same way as the normalization of multiplication. Since a non-zero and finite $m$ is in $(2^{M + 1}, 2^{M + 3})$, the leading 1 of a non-zero $m$ is either the $M + 1$-th bit or the $M + 2$-th bit, and we check if $m_{M + 2} = 1$. If so, $m$ and $e$ are unchanged. Otherwise, $m \coloneqq m \shl 1, e \coloneqq e - 1$, where $e$ is decremented as $m_{M + 2} = 0$ indicates that the division borrows. The in-circuit operation is similar to normalization for multiplication. The prover feeds $\var{b} \coloneqq \Computation_{\mathsf{MSB}}(\varmantissa) = \varmantissa_{M + 2}$, the MSB of \varmantissa, as a hint to circuit, and the circuit checks the predicate $\Predicate_{\mathsf{MSB}}(\varmantissa, \var{b})$ in 2 steps: \emph{(i)} enforce $\var{b}$ is a boolean, and \emph{(ii)} assert $\varmantissa - (\var{b} \shl (M + 2)) \in [0, 2^{M + 2})$. Note that \emph{(ii)} implies that $\varmantissa \in [0, 2^{M + 3} - 1]$, which indicates the hinted $\varmantissa \coloneqq \var{q}$ is the correct quotient mantissa and thus lies in $(2^{M + 1}, 2^{M + 3})$. Finally, the circuit updates $\varmantissa, \varexponent$ according to $\var{b}$, i.e., $\varmantissa \coloneqq \var{b} \mathbin{?} \varmantissa : \varmantissa \shl 1$, $\varexponent \coloneqq \varexponent + \var{b}$.

\noindent  \textbf{Round intermediate mantissa (lines 14-15).} The normalized mantissa $m$ of length $N = M + 3$ is rounded as in Section \ref{subsection:round}, with $\Delta e = \max(\min(-2^{E - 1} + 2 - e, K), 0), K = M + 2$, obtaining $\varexponent'$ and $\varmantissa'$. In addition, the equality between $r$ and $0$ is used to determine the sticky bit, thus we set the in-circuit parameter $\var{aux} \coloneqq \CircuitIsEq(r, 0)$.

\noindent \textbf{Edge Cases (lines 16-20).} We omit the detailed explanation due to space limit.


\input{./Protocols/circuit_IJK.tex}

\input{./Protocols/circuit_NormalizeIJK.tex}

\section{Authentic Location Information}
\label{appendix:authenticate}

Currently, the \ZKLP paradigm described in Section~\ref{section:zkLocation} only introduces efficient circuits for transforming $(\latitudeShort, \longitudeShort)$ to $(i,j,k)$ in the Uber H3~\cite{UberH3} hexagonal spatial index. Whilst $(\latitudeShort, \longitudeShort)$ are private inputs, and hence not disclosed to the verifier, the prover could still choose arbitrary values as input to the circuit, as it doesn't constrain that location information is obtained correctly, i.e., it does not ensure \textit{data provenance}.
We introduce three approaches to mitigate this issue by proving that data comes from a trusted source.

\textbf{\textit{(i) Offline Finding Networks}}
Offline Finding ecosystems, such as Apple's ``Find My'' network, allow device owners to track the location of missing offline devices via Bluetooth, and report an approximated location via the internet~\cite{beck2023abuse, heinrich2021can}. 
Each device within these networks generates unique public-private key pairs and frequently rotates its public keys to mitigate tracking risks. Lost devices broadcast their public key, which nearby Apple devices utilize to encrypt their own location. The encrypted location data is then sent to and stored on Apple's servers and can only be decrypted by device owner's private key. 
Although these ecosystems is proprietary to the manufacturer, recent work shows how to utilize these offline finding protocols (e.g., Apple's ``Find My'' network) to localize arbitrary devices~\cite{heinrich2021can}. 

To obtain authentic location information, one can leverage the network of unknown devices that post ``Location Reports'' to Apple's server. A ``Location Report'' can only be decrypted by the owner of the private key using AES-128-GCM. After decryption, the location data (latitude and longitude) is available in plain. To ensure authenticity and prevent forgery, one can prove the correct key derivation and decryption of several location reports, and further prove correct triangulation before applying the optimized \ZKLP circuits.
Given recent work that shows how to optimize non-native field arithmetic in SNARKs with lookup arguments~\cite{orru2024beyond}, this solution effectively bridges the cyber-physical gap without demanding for additional hardware or distributed networks that may not yet be in place.

\textbf{\textit{(ii) Authenticated \GNSS signals. }}
Traditionally, \GNSS services were for military use, which meant that they lacked robust security features like signal authentication. Addressing this legacy issue necessitates modifications to millions of GPS receivers. Recently, the security vulnerabilities in GPS have gained attention, leading to efforts to improve its security. One such effort is Open Service Navigation Message Authentication (OSNMA), which targets the lack of signal authentication~\cite{yuan2023authenticating}. Open-source implementations of OSNMA receivers exist and are well-documented~\cite{galan2022osnmalib}.



OSNMA combines ECDSA signatures, the \TESLA key chain mechanism, and Messages Authentication Codes (MACs) for message authentication.
The receiver's cryptographic operations in OSNMA include verifying a root key of the 
\TESLA chain, authenticating new public keys, verifying TESLA chain keys, and authenticating the MACs of navigation messages.
First, the receiver validates the authenticity of a Root Key through an ECDSA signature.
Successively, the receiver uses the authenticated root key to verify the current chain key. By successfully verifying the chain key against the root key, the receiver ensures that the chain key is part of the legitimate \TESLA key chain.
Using the verified chain key, the receiver computes the MAC of the navigation data. The computed MAC is compared with the extracted MAC. If they match, it confirms that the navigation message is authentic and has not been tampered with.

We presume that the overhead of proving an authenticated \GNSS signals with OSNMA will be dominated by the in-circuit verification of ECDSA signatures. At the time of writing, it requires $4 \cdot 10^6$ constraints for in-circuit emulation over the circuit unfriendly curve secp256k1 in gnark.
Further, there remains an open problem~---~a malicious prover may collaborate with a third party, which obtains the \GNSS signal and forwards it. The verifier needs to ensure that \emph{(i)} obtaining the location and \emph{(ii)} computing the proof is conducted \textit{atomically}~---~which remains unsolved.



\textbf{\textit{(iii) TLS Oracles. }}
Alternatively, one may trust a third party entity to verify that location information is obtained from a trusted entity.
TLS Oracles~\cite{xie2023lightweight, zhang2020deco, lauinger2023janus, ernstberger2024origo} provide the possibility to verify data provenance by extending a plain TLS session with a third party, which verifies that the data obtained by a client from a server is authentic.
As such, one could extend the \ZKLP paradigm to obtain location information from an API endpoint. 
The request includes information about nearby cell towers and WiFi access points detected by a mobile client, which allows for accurate location estimation via an external API. 
We leave the details of potential optimizations for authenticated and atomic \GNSS, and a concrete system design for integration with TLS oracles, to potential future work.

\end{document}

%% file: packages.tex
\usepackage{xcolor}

\usepackage{booktabs,tabularx,caption}
\usepackage{pifont}
\usepackage{makecell}
\usepackage{mathtools}
\usepackage{siunitx}
\DeclareSIUnit{\second}{s}
\DeclareSIUnit{\millisecond}{ms}
\DeclareSIUnit{\meter}{m}
\usepackage{listings}
\usepackage{colortbl}
\usepackage{multirow}
\usepackage{amsmath}
\usepackage{amssymb}
\usepackage{algorithm}
\usepackage[noend]{algpseudocode}
\usepackage{graphicx}
\usepackage{fancyvrb}
\usepackage{tikz}
\usetikzlibrary{calc}
\usetikzlibrary{shapes.geometric}
\usepackage{todonotes}
\usepackage{enumitem}
\setlist[enumerate]{topsep=0pt, itemsep=0pt, partopsep=0pt}
\usepackage{xspace, url} 
\usepackage{amsfonts}
\usepackage{setspace}
\usepackage[nounderscore]{syntax}
\PassOptionsToPackage{hyphens}{url}
\usepackage{hyperref} 
\usepackage{pgfplots}
\usepackage{multicol}
\usepackage{environ}
\usepackage[normalem]{ulem}
\usepackage{adjustbox}
\usepackage{nicematrix}

%% file: acronyms.tex
\usepackage[nolist]{acronym}

\begin{acronym}
\acro{GNSS}[GNSS]{Global Navigation Satellite System}
\acro{TESLA}[TESLA]{Timed Efficient Stream Loss-tolerant Authentication}
\acro{DGGS}[DGGS]{Discrete Global Grid Systems}
\acro{ZKP}[ZKP]{Zero-Knowledge Proof}
\acro{ZKPs}[ZKPs]{Zero-Knowledge Proofs}
\acro{SNARK}[SNARK]{Succinct Non-Interactive Argument of Knowledge}
\acrodefplural{SNARK}[SNARKs]{Succinct Non-Interactive Arguments of Knowledge}
\acro{ZKLP}[ZKLP]{Zero-Knowledge Location Privacy}
\acro{LPPM}[LPPM]{Location Privacy Preserving Mechanisms}
\acro{MPC}[MPC]{Multi-Party Computation}
\acro{FP32}[FP32]{single precision floating point}
\acro{FP64}[FP64]{double precision floating point}

\end{acronym}

%% file: macros.tex
\definecolor{coolblack}{rgb}{0.0, 0.18, 0.39}
\definecolor{darkbrown}{rgb}{0.4, 0.26, 0.13}
\definecolor{darkspringgreen}{rgb}{0.09, 0.45, 0.27}
\definecolor{dodgerblue}{rgb}{0.12, 0.56, 1.0}
\definecolor{navyblue}{rgb}{0.0, 0.0, 0.5}
\definecolor{burgundy}{rgb}{0.5, 0.0, 0.13}

\newtoggle{doPrintCommentsAll}
\NewEnviron{maybePrintAll}{%
  \iftoggle{doPrintCommentsAll}{\BODY}{}%
}

\toggletrue{doPrintCommentsAll}

\definecolor{deepcarrotorange}{rgb}{0.91, 0.41, 0.17}
\definecolor{darkpowderblue}{rgb}{0.0, 0.2, 0.6}

\newcounter{mycounter}

\newcounter{mycounterfunctionality}

\newcounter{mycounterAlt}

\newcommand{\GNSS}{\ac{GNSS}\xspace}
\newcommand{\TESLA}{\ac{TESLA}\xspace}
\newcommand{\DGGS}{\ac{DGGS}\xspace}
\newcommand{\ZKP}{\ac{ZKP}\xspace}
\newcommand{\ZKPs}{\ac{ZKPs}\xspace}
\newcommand{\SNARK}{\ac{SNARK}\xspace}
\newcommand{\SNARKs}{\acp{SNARK}\xspace}
\newcommand{\ZKLP}{\ac{ZKLP}\xspace}
\newcommand{\LPPM}{\ac{LPPM}\xspace}
\newcommand{\MPC}{\ac{MPC}\xspace}
\newcommand{\FPSingle}{\ac{FP32}\xspace}
\newcommand{\FPDouble}{\ac{FP64}\xspace}

\newcommand*{\Computation}{\ensuremath{\var{H}}}
\newcommand*{\Predicate}{\ensuremath{\var{P}}}
\newcommand*{\Circuit}{\ensuremath{\mathcal{C}}}

\newcommand*{\ProtocolBaselineZKLP}{\ensuremath{\Pi_{\text{Base}}}\xspace}

\DeclareMathOperator{\abs}{abs}
\newcommand*{\shl}{\ensuremath{\mathbin{<\!\!<}}}
\newcommand*{\shr}{\ensuremath{\mathbin{>\!\!>}}}
\newcommand*{\concat}{\ensuremath{\mathbin\Vert}}

\newcommand*{\var}[1]{\ensuremath{{#1}}\xspace}
\newcommand*{\varsign}{\var{s}}
\newcommand*{\varexponent}{\var{e}}
\newcommand*{\varmantissa}{\var{m}}
\newcommand*{\varabnormal}{\var{a}}

\newcommand*{\CircuitIsEq}{\ensuremath{\Circuit_\mathsf{IsEq}}\xspace}
\newcommand*{\AssertBitLength}{\ensuremath{\Circuit_{\mathsf{RC}}}\xspace}
\newcommand*{\CircuitAbs}{\ensuremath{\Circuit_{\mathsf{Abs}}}\xspace}
\newcommand*{\CircuitGreaterThanZero}{\ensuremath{\Circuit_{\mathsf{GEZ}}}\xspace}

\newcommand*{\CircuitMin}{\ensuremath{\Circuit_{\mathsf{Min}}}\xspace}
\newcommand*{\CircuitMax}{\ensuremath{\Circuit_{\mathsf{Max}}}\xspace}
\newcommand*{\CircuitInitFloat}{\ensuremath{\Circuit_{\mathsf{FpInit}}}\xspace}
\newcommand*{\CircuitRoundFloat}{\ensuremath{\Circuit_{\mathsf{FpRound}}}\xspace}
\newcommand*{\signBitShort}{\varsign}
\newcommand*{\absoluteValueShort}{\ensuremath{\var{abs}}}

\newcommand*{\CircuitZKLP}{\ensuremath{\Circuit_{\mathsf{ZKLP}}}\xspace}
\newcommand*{\CircuitRadial}{\ensuremath{\Circuit_{\mathsf{rad}}}\xspace}
\newcommand*{\CircuitPolar}{\ensuremath{\Circuit_{\mathsf{Hex2D}}}\xspace}
\newcommand*{\CircuitIJK}{\ensuremath{\Circuit_{\mathsf{IJK}}}\xspace}
\newcommand*{\CircuitNormalizeIJK}{\ensuremath{\Circuit_{\mathsf{Normalize}}}\xspace}
\newcommand{\latitudeShort}{\theta\xspace} 
\newcommand{\longitudeShort}{\phi\xspace}

\newcommand{\azimuthShort}{\zeta}
\newcommand{\resolutionShort}{\text{res}\xspace}
\newcommand{\scalingFactorGnomonic}{c_{G}}
\newcommand{\scalingFactorPower}{c_{\rho}}
\newcommand*{\Angle}{\ensuremath{\sigma}}
\newcommand*{\userPoint}{\ensuremath{P_u}}

\newcommand{\empirical}[1]{{\color{black}#1}}

\newcommand{\factorSingleBetterThanBaselineConstraints}{\empirical{15.9}\times\xspace} 
\newcommand{\factorDoubleBetterThanBaselineConstraints}{\empirical{12.2}\times\xspace} 
\newcommand{\numberProofsPerSecond}{\empirical{470}\xspace} 

\newcommand{\numberConstraintsFPSixBNGSixteenForCZKLP}{309.6} 
\newcommand{\numberConstraintsSinglePrecisionBNGSixteenForCZKLP}{19.5} 
\newcommand{\numberConstraintsDoublePrecisionBNGSixteenForCZKLP}{25.5} 
\newcommand{\numberConstraintsFPSixBNPlonkForCZKLP}{570.9} 
\newcommand{\numberConstraintsSinglePrecisionBNPlonkForCZKLP}{55.5} 
\newcommand{\numberConstraintsDoublePrecisionBNPlonkForCZKLP}{81.3} 

\newcommand{\ProverTimeFPSixBNGSixteenForCZKLP}{\SI{1.75}{\second}} 
\newcommand{\ProverTimeSinglePrecisionBNGSixteenForCZKLP}{\SI{0.26}{\second}} 
\newcommand{\ProverTimeDoublePrecisionBNGSixteenForCZKLP}{\SI{0.32}{\second}} 
\newcommand{\ProverTimeFPSixBNPlonkForCZKLP}{\SI{38.42}{\second}} 
\newcommand{\ProverTimeSinglePrecisionBNPlonkForCZKLP}{\SI{2.19}{\second}} 
\newcommand{\ProverTimeDoublePrecisionBNPlonkForCZKLP}{\SI{4.41}{\second}} 

\newcommand{\MemoryFPSixBNGSixteenForCZKLP}{517.3}  
\newcommand{\MemorySinglePrecisionBNGSixteenForCZKLP}{248.7} 
\newcommand{\MemoryDoublePrecisionBNGSixteenForCZKLP}{246.7} 
\newcommand{\MemoryFPSixBNPlonkForCZKLP}{1654.1}  
\newcommand{\MemorySinglePrecisionBNPlonkForCZKLP}{249.1} 
\newcommand{\MemoryDoublePrecisionBNPlonkForCZKLP}{247.1} 

\newcommand{\SRSFPSixBNGSixteenForCZKLP}{52.4}
\newcommand{\SRSFPSixBNPlonkForCZKLP}{33.6}
\newcommand{\SRSSinglePrecisionBNGSixteenForCZKLP}{4.6} 
\newcommand{\SRSDoublePrecisionBNGSixteenForCZKLP}{6.2} 
\newcommand{\SRSSinglePrecisionBNPlonkForCZKLP}{2.1} 
\newcommand{\SRSDoublePrecisionBNPlonkForCZKLP}{4.2} 

\newcommand{\TimeProximityTesting}{\SI{581.87}{\micro\second}} 

%% file: Figures/grid.tex
{
\setlength{\belowcaptionskip}{-1em} 
\begin{figure}[!t]
    \centering
    \begin{minipage}{0.23\textwidth}
        \centering
        \includegraphics[trim={7cm 3cm 7cm 3cm}, clip, width=1.2\linewidth]{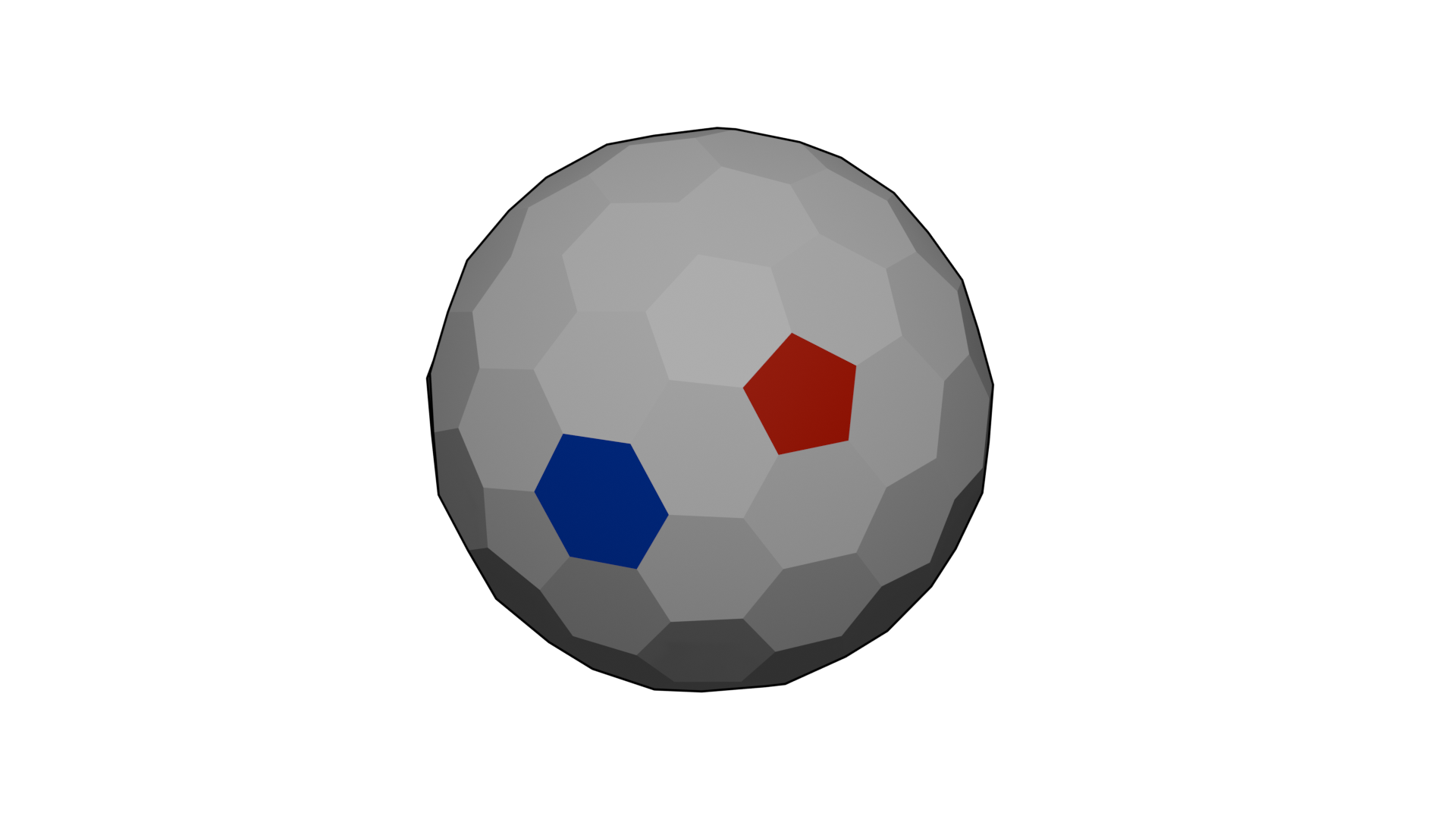}
    \end{minipage}\hfill
    \begin{minipage}{0.23\textwidth}
        \centering
        \begin{tikzpicture}[rotate=-30, scale=0.7]
            \draw[line width=1.1pt] (0:1.5cm) \foreach \x in {60,120,...,360} { -- (\x:1.5cm) } -- cycle;
            
            \foreach \i in {0,60,...,300} {
                \draw (\i:1.5cm) node[circle, fill, inner sep=1.5pt] (v\i) {};
            }
            \node[below right] at (v0) {$v_1$};
            \node[above right] at (v60) {$v_2$};
            \node[above left] at (v120) {$v_3$};
            \node[above left] at (v180) {$v_4$};
            \node[below left] at (v240) {$v_5$};
            \node[below right] at (v300) {$v_6$};
            
            \coordinate (center) at (0,0);
            \path (0:1.5cm) -- (60:1.5cm) coordinate[midway] (midpoint);

            \draw[dotted] (center) -- (midpoint);

            \foreach \n in {1, ..., 9}{
                \pgfmathsetmacro{\logpos}{1.3 * (ln(\n) / ln(20))}
                \coordinate (tickpos) at ($(center)!\logpos!(midpoint)$);
                \draw [red, thick, rotate around={30:(tickpos)}] (tickpos) -- ++(90:3pt) -- ++(270:6pt);
            }
            
            \coordinate (Pu) at (center);
            \coordinate (Pv) at ($(midpoint)!-1.5cm!(-30:0cm)$); 
            \draw (Pu) node[circle, fill, inner sep=1.5pt, red] {};
            \draw (Pv) node[circle, fill, inner sep=1.5pt, black] {};
            \node[left] at (Pu) {$P_v$};
            \node[right] at (Pv) {$P_u$};

            \draw[<->, >=stealth, black, thick] (Pv) -- (midpoint) node[midway, above, black] {$\Delta_{\text{min}}$};           
            
        \end{tikzpicture}
    \end{minipage}
    \caption{\emph{(i)} Icosahedral Polyhedron. A hexagon is highlighted in blue, and a pentagon is highlighted in red. \emph{(ii) Hexagon $\mathbf{h} = [v_1, ..., v_{6}]$}.
    $P_u$ and $P_v$ denote the points of verifier $u$ and prover $v$ respectively.
    $\Delta_{min}$ is the privacy-preserving  distance of $u$ to $v$. The red ticks depict our methodology to evaluate \ZKLP in Section~\ref{section:evaluation}.
    }
    \label{figure:sphere}
\end{figure}
}

%% file: Protocols/gadget_InitFloat.tex
\algrenewcommand\alglinenumber[1]{\footnotesize\textbf{#1:}}

\begin{figure}[!t]
  \centering
  \footnotesize
  \fbox{
  \begin{minipage}{\columnwidth}
    \underline{$\CircuitInitFloat(\hat{\varsign}, \hat{\varexponent}, \hat{\varmantissa})$}
  \begin{algorithmic}[1] 
    \State $\hat{\varsign}(1-\hat{\varsign})=0; \AssertBitLength(\hat{\varexponent}, E); \AssertBitLength(\hat{\varmantissa}, M)$
    \State $\varsign \coloneqq \hat{\varsign}; \varexponent \coloneqq \hat{\varexponent}; \varmantissa \coloneqq \hat{\varmantissa}$
    \State $\var{m\_is\_0} \coloneqq \CircuitIsEq(\varmantissa, 0)$
    \State $\var{e\_is\_min} \coloneqq \CircuitIsEq(\varexponent, 0); \var{e\_is\_max} \coloneqq \CircuitIsEq(\varexponent, 2^E - 1)$
    \State Receive hint $\var{d} = \Computation_{\mathsf{Norm}}(\var{m})$
    \State $\AssertBitLength((\varmantissa \shl \var{d}) - (\var{m\_is\_0} \mathbin{?} 0 : 2^M), M)$
    \State $\varexponent \coloneqq -2^{E - 1} + 1 + (\var{e\_is\_min} \mathbin{?} (\var{m\_is\_0} \mathbin{?} -M : 1 - \var{d}) : \varexponent$)
    \State $\varmantissa \coloneqq (\var{e\_is\_max} \mathbin{\land}  \neg \var{m\_is\_0}) \mathbin{?} 0 : (\var{e\_is\_min} \mathbin{?} \varmantissa \shl \var{d} : \varmantissa + 2^M)$
    \State $\varabnormal \coloneqq \var{e\_is\_max}$
  \end{algorithmic}
  \end{minipage}
  }
  \caption{Circuit for initializing floating-point numbers}
  \label{figure:init_float}
\end{figure}

%% file: Protocols/gadget_RoundFloat.tex
\algrenewcommand\alglinenumber[1]{\footnotesize\textbf{#1:}}

\begin{figure}[!t]
  \centering
  \footnotesize
  \fbox{
  \begin{minipage}{\columnwidth}
    \underline{$\CircuitRoundFloat(\varexponent, \varmantissa, \Delta \varexponent, \var{aux} = 1)$}
  \begin{algorithmic}[1] 
    \Require $\varmantissa \in [0, 2^N - 1], \Delta \varexponent \in [0, K], 2^{N + K} < p$
    \State Receive hints $\var{u}', \var{b}_1, \var{b}_2, \var{v}' = \Computation_{\mathsf{Split}}(\varmantissa \shl (K - \Delta \varexponent))$
    \State $\AssertBitLength(\var{u}', M); \var{b}_1(1 - \var{b}_1) = 0; \var{b}_2(1 - \var{b}_2) = 0; \AssertBitLength(\var{v}', N - M - 2 + K)$
    \State $\varmantissa \shl K = (\var{u}' \concat \var{b}_1 \concat \var{b}_2 \concat \var{v}') \shl \Delta \varexponent$
    \State $\var{u} \coloneqq \var{u}' \concat \var{b}_1; \var{v} \coloneqq \var{b}_2 \concat \var{v}'$
    \State $\var{half} \coloneqq \CircuitIsEq(\var{v}, 2^{N - M - 2 + K}) \land \var{aux}$
    \State $\varmantissa' \coloneqq (\var{u} + (\var{half} \mathbin{?} \var{b}_1 : \var{b}_2)) \shl \Delta\varexponent$
    \State $\var{overflow} \coloneqq \CircuitIsEq(\varmantissa', 2^{M + 1})$
    \State $\varexponent' \coloneqq \varexponent + \var{overflow}$
    \State $\varmantissa' \coloneqq \var{overflow} \mathbin{?} 2^M : \varmantissa'$
    \State \Return $\varexponent', \varmantissa'$
  \end{algorithmic}
  \end{minipage}
  }
  \caption{Circuit for rounding floating-point numbers}
  \label{figure:round_float}
\end{figure}

%% file: Protocols/circuit_AddFloat.tex
\algrenewcommand\alglinenumber[1]{\footnotesize\textbf{#1:}}

\begin{figure}[!t]
  \centering
  \footnotesize
  \fbox{
  \begin{minipage}{0.8\columnwidth}
    \underline{$\var{\alpha} + \var{\beta}$}
  \begin{algorithmic}[1] 
    \State $\var{c}, \var{abs} \coloneqq \CircuitAbs(\varexponent_\var{\beta} - \varexponent_\var{\alpha}, E + 1)$
    \State $\varexponent \coloneqq \var{c} \mathbin{?} \varexponent_\var{\beta} : \varexponent_\var{\alpha}$
    \State $\var{abs} \coloneqq \CircuitMin(\var{abs}, M + 3)$
    \State $\var{x} \coloneqq (\var{c} \mathbin{?} \varsign_\var{\beta} \varmantissa_\var{\beta} : \varsign_\var{\alpha} \varmantissa_\var{\alpha}) \shl L$
    \State $\var{y} \coloneqq (\var{c} \mathbin{?} \varsign_\var{\alpha} \varmantissa_\var{\alpha} : \varsign_\var{\beta} \varmantissa_\var{\beta}) \shl (L - \var{abs})$
    \State $\var{z} \coloneqq \var{x} + \var{y}$
    \State $\neg \varsign, \varmantissa \coloneqq \CircuitAbs(\var{z}, 2M + 5)$
    \State $\varexponent \coloneqq \varexponent + 1$
    \State $\varabnormal \coloneqq \varabnormal_\var{\alpha} \lor \varabnormal_\var{\beta}$
    \State $\var{m\_is\_0} \coloneqq \CircuitIsEq(\varmantissa, 0)$
    \State Receive hint $\Computation_{\mathsf{Norm}}(\var{v}) = \var{d}$
    \State $\varmantissa \coloneqq \varmantissa \shl \var{d}; \varexponent \coloneqq \varexponent - \var{d}$
    \State $\AssertBitLength(\varmantissa - (\var{m\_is\_0} \mathbin{?} 0 : 2^{2M + 4}), 2M + 4)$
    \State $\varexponent', \varmantissa' \coloneqq \CircuitRoundFloat(\varexponent, \varmantissa, 0)$
    \State $\varabnormal' \coloneqq \varabnormal \lor \CircuitGreaterThanZero(\varexponent' - 2^{E - 1}, E + 1)$
    \State $\varsign' \coloneqq \CircuitIsEq(\varsign_{\var{\alpha}}, \varsign_{\var{\beta}}) \mathbin{?} \varsign_{\var{\alpha}} : \varsign$
    \State $\varexponent' \coloneqq \varabnormal' \mathbin{?} 2^{E - 1} : (\var{m\_is\_0} \mathbin{?} -2^{E - 1} + 1 - M : \varexponent')$
    \State $\varmantissa'_1 \coloneqq (\neg\varabnormal_{\var{\beta}} \lor \CircuitIsEq(\varsign_\var{\alpha}\varmantissa_\var{\alpha}, \varsign_\var{\beta}\varmantissa_\var{\beta})) \mathbin{?} \varmantissa_\var{\alpha} : 0$
    \State $\varmantissa'_2 \coloneqq \varabnormal_\var{\beta} \mathbin{?} \varmantissa_\var{\beta} : (\varabnormal' \mathbin{?} 2^M : \varmantissa')$
    \State $\varmantissa' \coloneqq \varabnormal_{\var{\alpha}} \mathbin{?} \varmantissa'_1 : \varmantissa'_2$
    \State \Return $\varsign', \varexponent', \varmantissa', \varabnormal'$
  \end{algorithmic}
    \end{minipage}
  }
  \caption{Circuit for floating-point addition}
  \label{figure:add_float}
\end{figure}

%% file: Protocols/circuit_MulFloat.tex
\algrenewcommand\alglinenumber[1]{\footnotesize\textbf{#1:}}

\begin{figure}[!t]
  \centering
  \footnotesize
  \fbox{
  \begin{minipage}{0.9\columnwidth}
    \underline{$\var{\alpha} \cdot \var{\beta}$}
  \begin{algorithmic}[1] 
    \State $\varsign \coloneqq \varsign_\var{\alpha} \oplus \varsign_\var{\beta}; \varexponent \coloneqq \varexponent_\var{\alpha} + \varexponent_\var{\beta}; \varmantissa \coloneqq \varmantissa_\var{\alpha} \cdot \varmantissa_\var{\beta}; \varabnormal \coloneqq \varabnormal_\var{\alpha} \lor \varabnormal_\var{\beta}$
    \State Receive hint $\var{b} = \Computation_{\mathsf{MSB}}(\varmantissa)$
    \State $\var{b}(1 - \var{b}) = 1; \AssertBitLength(\varmantissa - (\var{b} \shl (2M + 1)), 2M + 1)$
    \State $\varmantissa \coloneqq \var{b} \mathbin{?} \varmantissa : \varmantissa \shl 1; \varexponent \coloneqq \varexponent + \var{b}$
    \State $\Delta \varexponent \coloneqq \CircuitMax(\CircuitMin(-2^{E - 1} + 2 - \varexponent, M + 2, E + 1), E + 1)$
    \State $\varexponent', \varmantissa' \coloneqq \CircuitRoundFloat(\varexponent, \varmantissa, \Delta \varexponent)$
    \State $\varabnormal' \coloneqq \varabnormal \lor \CircuitGreaterThanZero(\varexponent' - 2^{E - 1}, E + 1)$
    \State $\var{m'\_is\_0} \coloneqq \CircuitIsEq(\varmantissa', 0)$
    \State $\varexponent' \coloneqq \varabnormal' \mathbin{?} 2^{E - 1} : (\var{m'\_is\_0} \mathbin{?} -2^{E - 1} + 1 - M : \varexponent')$
    \State $\varmantissa' \coloneqq (\varabnormal' \land \neg \var{m'\_is\_0}) \mathbin{?} 2^M : \varmantissa'$
    \State \Return $\varsign, \varexponent', \varmantissa', \varabnormal'$
  \end{algorithmic}
  \end{minipage}
  }
  \caption{Circuit for floating-point multiplication}
  \label{figure:mul_float}
\end{figure}

%% file: Protocols/circuit_SqrtFloat.tex
\algrenewcommand\alglinenumber[1]{\footnotesize\textbf{#1:}}

\begin{figure}[!t]
  \centering
  \footnotesize
  \fbox{
  \begin{minipage}{0.8\columnwidth}
    \underline{$\sqrt{\var{\alpha}}$}
  \begin{algorithmic}[1] 
    \State $\varsign \coloneqq \varsign_{\var{\alpha}}$
    \State Receive hint $\var{b} = \Computation_{\mathsf{LSB}}(\varexponent_\var{\alpha})$
    \State $\varexponent \coloneqq (\varexponent_\var{\alpha} - \var{b}) / 2$
    \State $\var{b}(1 - \var{b}) = 1; \CircuitAbs(\varexponent, E - 1)$
    \State Receive hint $\var{n} = \Computation_{\mathsf{Sqrt}}(\varmantissa_\var{\alpha} \shl (M + 4 + \var{b}))$
    \State $\var{r} \coloneqq (\varmantissa_\var{\alpha} \shl (M + 4 + \var{b})) - \var{n}^2$
    \State $\AssertBitLength(\var{r}, M + 4), \AssertBitLength(2\var{n} - \var{r}, M + 4)$
    \State $\varmantissa \coloneqq \var{n}$
    \State $\var{m\_is\_0} \coloneqq \CircuitIsEq(\varmantissa_\var{\alpha}, 0)$
    \State $\varabnormal \coloneqq \varabnormal_\var{\alpha} \lor (\varsign_\var{\alpha} \land \neg \var{m\_is\_0})$
    \State $\varexponent', \varmantissa' \coloneqq \CircuitRoundFloat(\varexponent, \varmantissa, 0, \CircuitIsEq(r, 0))$
    \State $\varexponent' \coloneqq \varabnormal \mathbin{?} 2^{E - 1} : (\var{m\_is\_0} \mathbin{?} -2^{E - 1} + 1 - M : \varexponent')$
    \State $\varmantissa' \coloneqq \varsign \mathbin{?} 0 : \varmantissa'$
    \State \Return $\varsign, \varexponent', \varmantissa', \varabnormal$
  \end{algorithmic}
  \end{minipage}
  }
  \caption{Circuit for floating-point square root}
  \label{figure:sqrt_float}
\end{figure}

%% file: Protocols/circuit_LessFloat.tex
\algrenewcommand\alglinenumber[1]{\footnotesize\textbf{#1:}}

\begin{figure}[!t]
  \centering
  \footnotesize
  \fbox{
  \begin{minipage}{\columnwidth}
    \underline{$\var{\alpha} < \var{\beta}$}
  \begin{algorithmic}[1] 
    \State $\var{e\_ge} \coloneqq \CircuitGreaterThanZero(\varexponent_\var{\alpha} - \varexponent_\var{\beta}, E + 1); \var{m\_ge} \coloneqq \CircuitGreaterThanZero(\varmantissa_\var{\alpha} - \varmantissa_\var{\beta}, M + 1)$
    \State $\var{s\_lt} \coloneqq (\CircuitIsEq(\varmantissa_\var{\alpha}, 0) \land \CircuitIsEq(\varmantissa_\var{\beta}, 0)) \mathbin{?} 0 : \varexponent_\var{\alpha}$
    \State $\var{e\_lt} \coloneqq \varsign_\var{\alpha} \mathbin{?} \var{e\_ge} : \neg\var{e\_ge}$
    \State $\var{m\_lt} \coloneqq \CircuitIsEq(\varmantissa_\var{\alpha}, \varmantissa_\var{\beta}) \mathbin{?} 0 : (\varsign_\var{\alpha} \mathbin{?} \var{m\_ge} : \neg\var{m\_ge})$
    \State $\var{b} \coloneqq \CircuitIsEq(\varsign_\var{\alpha}, \varsign_\var{\beta}) \mathbin{?} (\CircuitIsEq(\varexponent_\var{\alpha}, \varexponent_\var{\beta}) \mathbin{?}  \var{m\_lt} : \var{e\_lt}) : \var{s\_lt}$
    \State $\var{b}' \coloneqq ((\varabnormal_\var{\alpha} \land \CircuitIsEq(\varmantissa_\var{\alpha}, 0)) \lor (\varabnormal_\var{\beta} \land \CircuitIsEq(\varmantissa_\var{\beta}, 0))) \mathbin{?} 0 : \var{b}$
    \State \Return $\var{b}'$
  \end{algorithmic}
  \end{minipage}
  }
  \caption{Circuit for floating-point comparison}
  \label{figure:less_float}
\end{figure}

%% file: Protocols/zkLocation_Baseline.tex
\algrenewcommand\alglinenumber[1]{\footnotesize\textbf{#1:}}

\begin{figure}[t!]
  \centering
  \footnotesize
  \fbox{
  \begin{minipage}{\columnwidth}
  \underline{$\ProtocolBaselineZKLP(\latitudeShort, \longitudeShort, \resolutionShort)$}
    \begin{algorithmic}[1] 
        \State  $(\latitudeShort_{\text{rad}}, \longitudeShort_{\text{rad}}) \xleftarrow{} \text{ToRadians}(\latitudeShort, \longitudeShort)$ \Comment{Transform to Radians}
        \State $(x_u,y_u,z_u) \xleftarrow{} \text{ToCartesian3D}(\latitudeShort_{\text{rad}}, \longitudeShort_{\text{rad}})$ \Comment{Transform to Cartesian}
        \State $d^2 = (x_F - x_u)^2 + (y_F - y_u)^2 + (z_F - z_u)^2$ \Comment{Closest Face}
        \State $r \xleftarrow{} \text{CalculateRadialDist}(d^2, \resolutionShort)$ \Comment{Calculate the radial distance}
        \State $\Angle \xleftarrow{} \text{CalculateAngle}(\latitudeShort_F, \longitudeShort_F, \latitudeShort, \longitudeShort)$ \Comment{Calculate angle to closest Face}
        \State $(x,y) \xleftarrow{} \text{ToCartesian2D}(r, \Angle)$ \Comment{Calculate 2D Cartestian}
        \State $(\var{I}, \var{J}, \var{K}) = \text{TransformToIJK}(x, y)$ \Comment{Transform to ``I,J,K''}
        \State \Return $(\var{I}, \var{J}, \var{K})$
    \end{algorithmic}
  \end{minipage}
  }
  \caption{
  Baseline Protocol for deriving $(i, j, k)$ from $(\latitudeShort, \longitudeShort)$.
  }
  \label{figure:baseline}
\end{figure}

%% file: Protocols/circuit_ZKLP.tex
\algrenewcommand\alglinenumber[1]{\footnotesize\textbf{#1:}}

\begin{figure}[t!]
  \centering
  \footnotesize
  \fbox{
  \begin{minipage}{0.9\columnwidth}
  \underline{$\CircuitZKLP(\latitudeShort_{\text{rad}}, \longitudeShort_{\text{rad}} \mathbin{;} \resolutionShort, \var{I}, \var{J}, \var{K}$)}
  \begin{algorithmic}[1] 
    \State Receive $\Computation_{\mathsf{ZKLP}}(i) = [\alpha_i, \beta_i, \gamma_i , \delta_i]$,  $i \in \{\latitudeShort_{\text{rad}}, \longitudeShort_{\text{rad}}\}$
    \State $\gamma_i^2 + d_i^2 = 1 \mathbin{;} \delta_i \cdot a_i = \gamma_i \mathbin{;} 2\gamma_i \cdot d_i = \beta_i$,  $i \in \{\latitudeShort_{\text{rad}}, \longitudeShort_{\text{rad}}\}$
    \State $r = \sqrt{1 - b_{\latitudeShort_{\text{rad}}}^2}\hspace{2em} ; \hspace{1em} z = b_{\latitudeShort_{\text{rad}}} $
    \State $x = \sqrt{1 - b_{\longitudeShort_{\text{rad}}}^2} \cdot r \hspace{0.5em} ; \hspace{1em} y = b_{\longitudeShort_{\text{rad}}} \cdot r $
    \For{$i \in \{0, \ldots, 19\}$}
        \State $d_i^2 = (x_{F_i} - x)^2 + (y_{F_i} - y)^2 + (z_{F_i} - z)^2$
        \State $d^2 = (d_i^2 \stackrel{?}{<} d^2) \mathbin{?} d_i^2 : d^2$
    \EndFor
    \State $r \xleftarrow{} \CircuitRadial(d^2, \resolutionShort)$
    \State $\sin_i = \beta_i \mathbin{;} \cos_i = \delta_i^2 - \gamma_i^2$ for $i \in \{\latitudeShort_{\text{rad}}, \longitudeShort_{\text{rad}}\}$
    \State $(x,y) \xleftarrow{} \CircuitPolar(r, \resolutionShort, \sin_{\latitudeShort_{\text{rad}}}, \cos_{\latitudeShort_{\text{rad}}}, \sin_{\longitudeShort_{\text{rad}}}, \cos_{\longitudeShort_{\text{rad}}}, F_i)$ 
    \State $(i,j,k) \xleftarrow{} \CircuitIJK(x,y)$ 
    \State $\CircuitIsEq(i, \var{I}) \mathbin{;} \CircuitIsEq(j, \var{J}) \mathbin{;} \CircuitIsEq(k, \var{K})$
  \end{algorithmic}
  \end{minipage}
  }
  \caption{Optimized Circuit for computing ZKLP. 
  }
  \label{figure:ZKLP}
\end{figure}

%% file: Protocols/circuit_R.tex
\algrenewcommand\alglinenumber[1]{\footnotesize\textbf{#1:}}

\begin{figure}[t!]
  \centering
  \footnotesize
  \fbox{
  \begin{minipage}{0.6\columnwidth}
  \underline{$\CircuitRadial(d^2, \resolutionShort)$}
  \begin{algorithmic}[1] 
    \State $r = \sqrt{\frac{-(d^2 - 4) \cdot d^2}{(2 - d^2)}} \mathbin{;} r = \frac{r}{\scalingFactorGnomonic}$
    \State $\gamma \coloneqq 1.0 \mathbin{;} \scalingFactorPower \coloneqq \sqrt{7}$
    \For{$i \in \{1, \ldots, R\}$}
        \State $\gamma \coloneqq (\var{\resolutionShort[i]}) \mathbin{?} \gamma \cdot \scalingFactorPower : \gamma$
        \State $\scalingFactorPower \coloneqq \scalingFactorPower^2$
    \EndFor
    \State \Return $(r \cdot \gamma)$
  \end{algorithmic}
  \end{minipage}
  }
  \caption{Optimized Sub-Circuit for computing the radial distance $r$. $R$ represents the number of bits of $\resolutionShort$. 
  }
  \label{figure:CircuitRadial}
\end{figure}

%% file: Protocols/circuit_Theta.tex
\algrenewcommand\alglinenumber[1]{\footnotesize\textbf{#1:}}

\begin{figure}[t!]
  \centering
  \footnotesize
  \fbox{
  \begin{minipage}{0.9\columnwidth}
  \underline{$\CircuitPolar(r, \resolutionShort, \sin_{\latitudeShort_{\text{rad}}}, \cos_{\latitudeShort_{\text{rad}}}, \sin_{\longitudeShort_{\text{rad}}}, \cos_{\longitudeShort_{\text{rad}}}, F_i)$}
  \begin{algorithmic}[1] 
    \State $a \coloneqq \sin_{\longitudeShort_{\text{rad}}} \cdot \cos_{\longitudeShort_{F_i, \text{rad}}} \mathbin{;} \hspace{0.5em} b \coloneqq \cos_{\longitudeShort_{\text{rad}}} \cdot \sin_{\longitudeShort_{F_i, \text{rad}}}$
    \State $c \coloneqq  \cos_{\latitudeShort_{F_i, \text{rad}}} \cdot \sin_{\latitudeShort_{\text{rad}}}  \mathbin{;} \hspace{0.5em} d \coloneqq \sin_{\latitudeShort_{F_i, \text{rad}}} \cdot \cos_{\latitudeShort_{\text{rad}}}$
    \State $e \coloneqq   \cos_{\longitudeShort_{\text{rad}}} \cdot \cos_{\longitudeShort_{F_i, \text{rad}}}  \mathbin{;} \hspace{0.5em} f \coloneqq \sin_{\longitudeShort_{\text{rad}}} \cdot \sin_{\longitudeShort_{F_i, \text{rad}}}$
    \State $x \coloneqq c - d \cdot (e+g) \mathbin{;} y \coloneqq \cos_{\latitudeShort_{\text{rad}}} \cdot (a-b) \mathbin{;} z \coloneqq \sqrt{x^2 + y^2}$ 
    \State $\sin_{\azimuthShort_{F_i}} \coloneqq (\var{\resolutionShort[0]}) \mathbin{?} (\sin_{\azimuthShort_{F_i}}-\arcsin(\sqrt{\frac{3}{28}})) : \sin_{\azimuthShort_{F_i}}$
    \State $\cos_{\azimuthShort_{F_i}} \coloneqq (\var{\resolutionShort[0]}) \mathbin{?} (\cos_{\azimuthShort_{F_i}}-\arcsin(\sqrt{\frac{3}{28}})) : \cos_{\azimuthShort_{F_i}}$
    \State $\sin_{\Angle} \coloneqq (\sin_{\azimuthShort_{F_i}} \cdot \frac{x}{z}) - (\cos_{\azimuthShort_{F_i}} \cdot \frac{y}{z})$
    \State $\cos_{\Angle} \coloneqq (\cos_{\azimuthShort_{F_i}} \cdot \frac{x}{z}) - (\sin_{\azimuthShort_{F_i}} \cdot \frac{y}{z})$
    \State \Return $(r \cdot \sin_{\Angle}, r \cdot \cos{\Angle})$
  \end{algorithmic}
  \end{minipage}
  }
  \caption{Optimized Sub-Circuit for computing two-dimensional cartesian coordinates. Variables related to the face center $F_i$ are constant floating-point numbers. 
  }
  \label{figure:Hex2D}
\end{figure}

%% file: Tables/table_evaluation_float.tex
\begin{table}[!t]
\centering
\caption{Number of R1CS constraints for in-circuit floating-point primitive operations, with $|\mathcal{T}_\mathsf{RC}| = 2^8, |\mathcal{T}_\mathsf{Pow2}| = 1 + E + M$}
\label{table:fp_constraints}
\resizebox{\columnwidth}{!}{
\begin{tabular}{cccccccc}
\toprule
\multicolumn{2}{c}{\textbf{FP32 Operation}} & \textbf{Init} & \textbf{Add/Sub} & \textbf{Mul} & \textbf{Div} & \textbf{Sqrt} & \textbf{Cmp} \\
\cmidrule{1-8}
\multicolumn{2}{c}{\# Native Constraints}    & 13   & 42 & 31 & 38 & 23 & 26 \\
\cmidrule{1-8}
\multirow{2}{*}[-0.3em]{\shortstack{\# Lookup\\ Constraints}} & (i)   & 17 & 43 & 33 & 38 & 22 & 7 \\
\cmidrule{2-8}
                        & (ii, iii) & \multicolumn{6}{c}{\cellcolor{gray!15}291} \\
\midrule
\multicolumn{2}{c}{\textbf{FP64 Operation}} & \textbf{Init} & \textbf{Add/Sub} & \textbf{Mul} & \textbf{Div} & \textbf{Sqrt} & \textbf{Cmp} \\
\cmidrule{1-8}
\multicolumn{2}{c}{\# Native Constraints}    & 13   & 42 & 31 & 38 & 23 & 26 \\
\cmidrule{1-8}
\multirow{2}{*}[-0.3em]{\shortstack{\# Lookup\\ Constraints}} & (i)   & 32 & 71 & 57 & 60 & 38 & 11 \\
\cmidrule{2-8}
                        & (ii, iii) & \multicolumn{6}{c}{\cellcolor{gray!15}323} \\
\bottomrule
\end{tabular}
}
\vspace{-2mm}
\end{table}

%% file: Figures/graph_float.tex
\definecolor{darkblue}{rgb}{0.0, 0.28, 0.67}
\definecolor{dartmouthgreen}{rgb}{0.05, 0.5, 0.06}
\definecolor{harvardcrimson}{rgb}{0.81, 0.06, 0.13}
\definecolor{darkred}{RGB}{128,0,0}
\definecolor{darkgreen}{RGB}{0,102,0}
\definecolor{darkpurple}{RGB}{51,0,51}
\definecolor{darkorange}{RGB}{204,85,0}
\definecolor{darkbrown}{RGB}{102,51,0}

{
\setlength{\belowcaptionskip}{-1em}
\begin{figure}[!t]
    \centering
    \begin{tikzpicture}
        \begin{axis}[
            width=0.9\columnwidth,
            height=0.55\columnwidth,
            name=boundary,
            title={},
            xlabel={\textbf{\# FP Multiplications}},
            ylabel style={align=center},
            ylabel={\begin{tabular}{@{}c@{}} \textbf{\# Constraints} \\ per Multiplication \end{tabular}},
            xmode=log,
            log basis x={2},
            ymode=log,
            log basis y={10},
            xmin=2, xmax=32768,
            ymin=50, ymax=500,
            xtick={2,4,8,16,32,64,128,256,512,1024,2048,4096,8192,16384,32768},
            xticklabels={\(2^1\),\(2^2\),\(2^3\),\(2^4\),\(2^5\),\(2^6\),\(2^7\),\(2^8\),\(2^9\),\(2^{10}\),\(2^{11}\),\(2^{12}\),\(2^{13}\),\(2^{14}\),\(2^{15}\)},
            ytick={50,60,70,80,90, 100, 200, 500},
            yticklabels={50,,,,,100, 200, 500},
            log ticks with fixed point,
            grid style=dashed,
            grid=major,
            legend pos=outer north east,
            legend style={draw=none, fill=none},
            label style={font=\normalfont}, 
            title style={font=\bfseries},
            tick label style={font=\small},
            every axis plot/.append style={thick},
            enlargelimits=0.05
        ]
        
        \addplot[
            color=darkblue,
            mark=*,
            mark size=0.7pt,
            ]
            coordinates {
            (2, 209)(4, 136)(8, 100)(16, 82)(32, 73)(64, 68)(128, 66)(256, 65)(512, 64)(1024, 64)(2048, 64)(4096, 64)(8192, 64)(16384, 64)(32768, 64)
            };\label{pgfplots:f32_t8}
        \addplot[
            color=harvardcrimson,
            mark=*,
            mark size=0.7pt,
            ]
            coordinates {
            (2, 2117)(4, 1084)(8, 568)(16, 310)(32, 181)(64, 116)(128, 84)(256, 68)(512, 60)(1024, 56)(2048, 54)(4096, 53)(8192, 52)(16384, 52)(32768, 52)
            };\label{pgfplots:f32_t12}
        \addplot[
            color=dartmouthgreen,
            mark=*,
            mark size=0.7pt,
            ]
            coordinates {
            (2, 32834)(4, 16441)(8, 8245)(16, 4147)(32, 2098)(64, 1073)(128, 561)(256, 305)(512, 177)(1024, 113)(2048, 81)(4096, 65)(8192, 57)(16384, 53)(32768, 51)
            };\label{pgfplots:f32_t16}
        
        \addplot[
            color=darkblue,
            mark=square*,
            mark size=0.7pt,
            ]
            coordinates {
            (2, 249)(4, 168)(8, 128)(16, 108)(32, 98)(64, 93)(128, 90)(256, 89)(512, 88)(1024, 88)(2048, 88)(4096, 88)(8192, 88)(16384, 88)(32768, 88)
            };\label{pgfplots:f64_t8}
        \addplot[
            color=harvardcrimson,
            mark=square*,
            mark size=0.7pt,
            ]
            coordinates {
            (2, 2149)(4, 1108)(8, 588)(16, 328)(32, 198)(64, 133)(128, 100)(256, 84)(512, 76)(1024, 72)(2048, 70)(4096, 69)(8192, 68)(16384, 68)(32768, 68)
            };\label{pgfplots:f64_t12}
        \addplot[
            color=dartmouthgreen,
            mark=square*,
            mark size=0.7pt,
            ]
            coordinates {
            (2, 32862)(4, 16461)(8, 8261)(16, 4161)(32, 2111)(64, 1086)(128, 573)(256, 317)(512, 189)(1024, 125)(2048, 93)(4096, 77)(8192, 69)(16384, 65)(32768, 63)
            };\label{pgfplots:f64_t16}
        \end{axis}


        \node[draw, fill=white, inner sep=1pt, below left=0.5em] at (boundary.north east) {
            \scriptsize 
            \setlength{\tabcolsep}{1pt} 
            \renewcommand{\arraystretch}{0.8} 
            \begin{tabular}{ccc}
                FP32 & FP64 & $|\mathcal{T}_\mathsf{RC}|$ \\
                \ref{pgfplots:f32_t8} & \ref{pgfplots:f64_t8} & $2^8$\\
                \ref{pgfplots:f32_t12} & \ref{pgfplots:f64_t12} & $2^{12}$\\
                \ref{pgfplots:f32_t16} & \ref{pgfplots:f64_t16} & $2^{16}$
            \end{tabular}
        };
    \end{tikzpicture}
    \caption{Number of Constraints per Multiplication vs. Number of Multiplications, for Groth16. $|\mathcal{T}_\mathsf{RC}|$ is the size of the lookup table for the range check (cf. Section~\ref{section:fp-primitive-ops}). }
    \label{figure:amortization}
\end{figure}
}

%% file: Tables/table_evaluation_zklp.tex
\begin{table}[!t]
\centering
\caption{
Evaluation of $\CircuitZKLP$ over BN254 for the floating-point implementation opposed to the baseline protocol $\ProtocolBaselineZKLP$ (cf. Figure~\ref{figure:baseline}), implemented with fixed-point arithmetic P20.
For Groth16, the SRS size is the size of the prover key.
}
\label{table:ZKLP}
\resizebox{\columnwidth}{!}{%
\begin{tabular}{
  l
  l 
  c 
  S[table-format=3.2]
  S[table-format=3.2]
  S[table-format=3.2]
  S[table-format=3.2]
}
\toprule
\textbf{Circuit} & \textbf{Type} & \makecell{\textbf{Proof} \\ \textbf{System}} & {\makecell{\textbf{\# Constraints} \\ \textbf{(x$10^3$)}}} & {\makecell{\textbf{Prover} \\ \textbf{Time}}} & {\makecell{\textbf{Memory} \\ \textbf{(MB)}}} &  {\makecell{\textbf{SRS} \\ \textbf{(MB)}}} \\
\midrule
Baseline & P20 & Groth16 & \numberConstraintsFPSixBNGSixteenForCZKLP & \ProverTimeFPSixBNGSixteenForCZKLP & \MemoryFPSixBNGSixteenForCZKLP & \SRSFPSixBNGSixteenForCZKLP \\
\cmidrule{2-7}
\multirow{2}{*}{\textbf{$\CircuitZKLP$}} & FP32 & Groth16 & \numberConstraintsSinglePrecisionBNGSixteenForCZKLP & \ProverTimeSinglePrecisionBNGSixteenForCZKLP & \MemorySinglePrecisionBNGSixteenForCZKLP & \SRSSinglePrecisionBNGSixteenForCZKLP\\
& FP64 & Groth16 & \numberConstraintsDoublePrecisionBNGSixteenForCZKLP & \ProverTimeDoublePrecisionBNGSixteenForCZKLP & \MemoryDoublePrecisionBNGSixteenForCZKLP & \SRSDoublePrecisionBNGSixteenForCZKLP \\
\midrule
Baseline & P20 & Plonk & \numberConstraintsFPSixBNPlonkForCZKLP & \ProverTimeFPSixBNPlonkForCZKLP & \MemoryFPSixBNPlonkForCZKLP & \SRSFPSixBNPlonkForCZKLP \\
\cmidrule{2-7}
\multirow{2}{*}{\textbf{$\CircuitZKLP$}} & FP32 & Plonk & \numberConstraintsSinglePrecisionBNPlonkForCZKLP & \ProverTimeSinglePrecisionBNPlonkForCZKLP & \MemorySinglePrecisionBNPlonkForCZKLP & \SRSSinglePrecisionBNPlonkForCZKLP \\
& FP64 & Plonk & \numberConstraintsDoublePrecisionBNPlonkForCZKLP & \ProverTimeDoublePrecisionBNPlonkForCZKLP & \MemoryDoublePrecisionBNPlonkForCZKLP & \SRSDoublePrecisionBNPlonkForCZKLP \\
\bottomrule 
\end{tabular}
}
\end{table}

%% file: Tables/table_test_new.tex
\newcommand{\customarrowLeft}[1]{%
  \makebox[#1]{%
    \hspace*{-4pt}
    $\xleftarrow{\hspace*{#1}}$%
    \hspace*{-4pt}
  }
}
\newcommand{\customarrowRight}[1]{%
  \makebox[#1]{%
    \hspace*{-4pt}
    $\xrightarrow{\hspace*{#1}}$%
    \hspace*{-4pt}
  }
}

\begin{figure*}[!ht]
    \raggedright
    \makebox[\textwidth][c]{
\resizebox{\textwidth}{!}{\begin{tabular}{l*{64}{p{0.1mm}|}}
\cmidrule{1-65}
& \multicolumn{64}{c}{\textbf{Resolution}} \\
\cmidrule{2-65}
& \multicolumn{16}{c}{\customarrowLeft{8.5em} \textbf{0} \customarrowRight{8.5em}} & \multicolumn{16}{c}{\customarrowLeft{8.5em} \textbf{1} \customarrowRight{8.5em}} & \multicolumn{16}{c}{\customarrowLeft{8.5em} \textbf{2} \customarrowRight{8.5em}} & \multicolumn{16}{c}{\customarrowLeft{8.5em} \textbf{3} \customarrowRight{8.5em}} \\
\cmidrule{2-65}
\textbf{P20}  & \cellcolor{red!16!yellow!25} & \cellcolor{green!25} & \cellcolor{green!25} & \cellcolor{green!25} & \cellcolor{green!25} & \cellcolor{green!25} & \cellcolor{red!1!yellow!25} & \cellcolor{red!1!yellow!25} & \cellcolor{red!1!yellow!25} & \cellcolor{green!25} & \cellcolor{green!25} & \cellcolor{green!25} & \cellcolor{green!25} & \cellcolor{green!25} & \cellcolor{green!25} & \cellcolor{green!25} & \cellcolor{red!4!yellow!25} & \cellcolor{red!1!yellow!25} & \cellcolor{red!7!yellow!25} & \cellcolor{red!2!yellow!25} & \cellcolor{green!25} & \cellcolor{red!1!yellow!25} & \cellcolor{red!1!yellow!25} & \cellcolor{red!3!yellow!25} & \cellcolor{red!2!yellow!25} & \cellcolor{green!25} & \cellcolor{green!25} & \cellcolor{red!1!yellow!25} & \cellcolor{red!4!yellow!25} & \cellcolor{red!4!yellow!25} & \cellcolor{red!25!yellow!25} & \cellcolor{red!27!yellow!25} & \cellcolor{red!8!yellow!25} & \cellcolor{green!25} & \cellcolor{red!2!yellow!25} & \cellcolor{red!1!yellow!25} & \cellcolor{green!25} & \cellcolor{green!25} & \cellcolor{green!25} & \cellcolor{red!1!yellow!25} & \cellcolor{green!25} & \cellcolor{green!25} & \cellcolor{green!25} & \cellcolor{red!3!yellow!25} & \cellcolor{red!6!yellow!25} & \cellcolor{red!28!yellow!25} & \cellcolor{red!36!yellow!25} & \cellcolor{red!43!yellow!25} & \cellcolor{red!5!yellow!25} & \cellcolor{green!25} & \cellcolor{green!25} & \cellcolor{green!25} & \cellcolor{green!25} & \cellcolor{green!25} & \cellcolor{green!25} & \cellcolor{green!25} & \cellcolor{green!25} & \cellcolor{green!25} & \cellcolor{red!1!yellow!25} & \cellcolor{red!8!yellow!25} & \cellcolor{red!25!yellow!25} & \cellcolor{red!41!yellow!25} & \cellcolor{red!49!yellow!25} & \cellcolor{red!59!yellow!25} \\
\cmidrule{2-65}
\textbf{P40}  & \cellcolor{red!15!yellow!25} & \cellcolor{red!2!yellow!25} & \cellcolor{red!1!yellow!25} & \cellcolor{red!1!yellow!25} & \cellcolor{red!1!yellow!25} & \cellcolor{green!25} & \cellcolor{green!25} & \cellcolor{green!25} & \cellcolor{green!25} & \cellcolor{green!25} & \cellcolor{green!25} & \cellcolor{green!25} & \cellcolor{green!25} & \cellcolor{green!25} & \cellcolor{red!1!yellow!25} & \cellcolor{red!1!yellow!25} & \cellcolor{red!8!yellow!25} & \cellcolor{red!4!yellow!25} & \cellcolor{red!5!yellow!25} & \cellcolor{red!2!yellow!25} & \cellcolor{red!3!yellow!25} & \cellcolor{red!1!yellow!25} & \cellcolor{red!3!yellow!25} & \cellcolor{red!1!yellow!25} & \cellcolor{red!3!yellow!25} & \cellcolor{green!25} & \cellcolor{red!1!yellow!25} & \cellcolor{red!1!yellow!25} & \cellcolor{red!3!yellow!25} & \cellcolor{red!3!yellow!25} & \cellcolor{red!1!yellow!25} & \cellcolor{red!3!yellow!25} & \cellcolor{red!12!yellow!25} & \cellcolor{red!3!yellow!25} & \cellcolor{red!4!yellow!25} & \cellcolor{green!25} & \cellcolor{green!25} & \cellcolor{green!25} & \cellcolor{red!1!yellow!25} & \cellcolor{green!25} & \cellcolor{red!2!yellow!25} & \cellcolor{green!25} & \cellcolor{green!25} & \cellcolor{red!1!yellow!25} & \cellcolor{red!1!yellow!25} & \cellcolor{green!25} & \cellcolor{green!25} & \cellcolor{green!25} & \cellcolor{green!25} & \cellcolor{green!25} & \cellcolor{red!1!yellow!25} & \cellcolor{green!25} & \cellcolor{green!25} & \cellcolor{green!25} & \cellcolor{green!25} & \cellcolor{green!25} & \cellcolor{green!25} & \cellcolor{green!25} & \cellcolor{green!25} & \cellcolor{red!1!yellow!25} & \cellcolor{green!25} & \cellcolor{green!25} & \cellcolor{green!25} & \cellcolor{green!25} \\
\cmidrule{2-65}
\textbf{F32}  & \cellcolor{red!5!yellow!25} & \cellcolor{green!25} & \cellcolor{green!25} & \cellcolor{green!25} & \cellcolor{green!25} & \cellcolor{green!25} & \cellcolor{green!25} & \cellcolor{green!25} & \cellcolor{green!25} & \cellcolor{green!25} & \cellcolor{green!25} & \cellcolor{green!25} & \cellcolor{green!25} & \cellcolor{green!25} & \cellcolor{green!25} & \cellcolor{green!25} & \cellcolor{red!1!yellow!25} & \cellcolor{green!25} & \cellcolor{green!25} & \cellcolor{green!25} & \cellcolor{green!25} & \cellcolor{green!25} & \cellcolor{green!25} & \cellcolor{green!25} & \cellcolor{green!25} & \cellcolor{green!25} & \cellcolor{green!25} & \cellcolor{green!25} & \cellcolor{green!25} & \cellcolor{green!25} & \cellcolor{green!25} & \cellcolor{green!25} & \cellcolor{red!8!yellow!25} & \cellcolor{green!25} & \cellcolor{green!25} & \cellcolor{green!25} & \cellcolor{green!25} & \cellcolor{green!25} & \cellcolor{green!25} & \cellcolor{green!25} & \cellcolor{green!25} & \cellcolor{green!25} & \cellcolor{green!25} & \cellcolor{green!25} & \cellcolor{green!25} & \cellcolor{green!25} & \cellcolor{red!1!yellow!25} & \cellcolor{green!25} & \cellcolor{red!3!yellow!25} & \cellcolor{green!25} & \cellcolor{green!25} & \cellcolor{green!25} & \cellcolor{green!25} & \cellcolor{green!25} & \cellcolor{green!25} & \cellcolor{green!25} & \cellcolor{green!25} & \cellcolor{green!25} & \cellcolor{green!25} & \cellcolor{green!25} & \cellcolor{green!25} & \cellcolor{green!25} & \cellcolor{green!25} & \cellcolor{green!25} \\
\cmidrule{2-65}
\textbf{F64}  & \cellcolor{green!25} & \cellcolor{green!25} & \cellcolor{green!25} & \cellcolor{green!25} & \cellcolor{green!25} & \cellcolor{green!25} & \cellcolor{green!25} & \cellcolor{green!25} & \cellcolor{green!25} & \cellcolor{green!25} & \cellcolor{green!25} & \cellcolor{green!25} & \cellcolor{green!25} & \cellcolor{green!25} & \cellcolor{green!25} & \cellcolor{green!25} & \cellcolor{green!25} & \cellcolor{green!25} & \cellcolor{green!25} & \cellcolor{green!25} & \cellcolor{green!25} & \cellcolor{green!25} & \cellcolor{green!25} & \cellcolor{green!25} & \cellcolor{green!25} & \cellcolor{green!25} & \cellcolor{green!25} & \cellcolor{green!25} & \cellcolor{green!25} & \cellcolor{green!25} & \cellcolor{green!25} & \cellcolor{green!25} & \cellcolor{green!25} & \cellcolor{green!25} & \cellcolor{green!25} & \cellcolor{green!25} & \cellcolor{green!25} & \cellcolor{green!25} & \cellcolor{green!25} & \cellcolor{green!25} & \cellcolor{green!25} & \cellcolor{green!25} & \cellcolor{green!25} & \cellcolor{green!25} & \cellcolor{green!25} & \cellcolor{green!25} & \cellcolor{green!25} & \cellcolor{green!25} & \cellcolor{green!25} & \cellcolor{green!25} & \cellcolor{green!25} & \cellcolor{green!25} & \cellcolor{green!25} & \cellcolor{green!25} & \cellcolor{green!25} & \cellcolor{green!25} & \cellcolor{green!25} & \cellcolor{green!25} & \cellcolor{green!25} & \cellcolor{green!25} & \cellcolor{green!25} & \cellcolor{green!25} & \cellcolor{green!25} & \cellcolor{green!25} \\
\cmidrule{2-65}
& \multicolumn{16}{c}{\customarrowLeft{8.5em} \textbf{4} \customarrowRight{8.5em}} & \multicolumn{16}{c}{\customarrowLeft{8.5em} \textbf{5} \customarrowRight{8.5em}} & \multicolumn{16}{c}{\customarrowLeft{8.5em} \textbf{6} \customarrowRight{8.5em}} & \multicolumn{16}{c}{\customarrowLeft{8.5em} \textbf{7} \customarrowRight{8.5em}} \\
\cmidrule{2-65}
\textbf{P20}  & \cellcolor{red!2!yellow!25} & \cellcolor{green!25} & \cellcolor{green!25} & \cellcolor{green!25} & \cellcolor{green!25} & \cellcolor{green!25} & \cellcolor{green!25} & \cellcolor{red!1!yellow!25} & \cellcolor{green!25} & \cellcolor{red!4!yellow!25} & \cellcolor{red!18!yellow!25} & \cellcolor{red!38!yellow!25} & \cellcolor{red!41!yellow!25} & \cellcolor{red!56!yellow!25} & \cellcolor{red!48!yellow!25} & \cellcolor{red!49!yellow!25} & \cellcolor{red!1!yellow!25} & \cellcolor{green!25} & \cellcolor{green!25} & \cellcolor{green!25} & \cellcolor{green!25} & \cellcolor{green!25} & \cellcolor{green!25} & \cellcolor{red!1!yellow!25} & \cellcolor{red!10!yellow!25} & \cellcolor{red!32!yellow!25} & \cellcolor{red!49!yellow!25} & \cellcolor{red!47!yellow!25} & \cellcolor{red!53!yellow!25} & \cellcolor{red!57!yellow!25} & \cellcolor{red!56!yellow!25} & \cellcolor{red!59!yellow!25} & \cellcolor{green!25} & \cellcolor{green!25} & \cellcolor{green!25} & \cellcolor{green!25} & \cellcolor{green!25} & \cellcolor{red!1!yellow!25} & \cellcolor{red!3!yellow!25} & \cellcolor{red!12!yellow!25} & \cellcolor{red!39!yellow!25} & \cellcolor{red!41!yellow!25} & \cellcolor{red!59!yellow!25} & \cellcolor{red!51!yellow!25} & \cellcolor{red!52!yellow!25} & \cellcolor{red!52!yellow!25} & \cellcolor{red!54!yellow!25} & \cellcolor{red!54!yellow!25} & \cellcolor{green!25} & \cellcolor{green!25} & \cellcolor{green!25} & \cellcolor{green!25} & \cellcolor{red!1!yellow!25} & \cellcolor{red!6!yellow!25} & \cellcolor{red!20!yellow!25} & \cellcolor{red!38!yellow!25} & \cellcolor{red!49!yellow!25} & \cellcolor{red!57!yellow!25} & \cellcolor{red!67!yellow!25} & \cellcolor{red!62!yellow!25} & \cellcolor{red!58!yellow!25} & \cellcolor{red!63!yellow!25} & \cellcolor{red!62!yellow!25} & \cellcolor{red!68!yellow!25} \\
\cmidrule{2-65}
\textbf{P40}  & \cellcolor{green!25} & \cellcolor{red!2!yellow!25} & \cellcolor{green!25} & \cellcolor{green!25} & \cellcolor{green!25} & \cellcolor{green!25} & \cellcolor{green!25} & \cellcolor{green!25} & \cellcolor{green!25} & \cellcolor{green!25} & \cellcolor{green!25} & \cellcolor{green!25} & \cellcolor{green!25} & \cellcolor{green!25} & \cellcolor{green!25} & \cellcolor{green!25} & \cellcolor{green!25} & \cellcolor{red!1!yellow!25} & \cellcolor{green!25} & \cellcolor{green!25} & \cellcolor{green!25} & \cellcolor{green!25} & \cellcolor{green!25} & \cellcolor{green!25} & \cellcolor{green!25} & \cellcolor{green!25} & \cellcolor{green!25} & \cellcolor{green!25} & \cellcolor{green!25} & \cellcolor{green!25} & \cellcolor{green!25} & \cellcolor{green!25} & \cellcolor{green!25} & \cellcolor{green!25} & \cellcolor{green!25} & \cellcolor{green!25} & \cellcolor{green!25} & \cellcolor{green!25} & \cellcolor{green!25} & \cellcolor{green!25} & \cellcolor{green!25} & \cellcolor{green!25} & \cellcolor{green!25} & \cellcolor{green!25} & \cellcolor{green!25} & \cellcolor{green!25} & \cellcolor{green!25} & \cellcolor{green!25} & \cellcolor{green!25} & \cellcolor{green!25} & \cellcolor{green!25} & \cellcolor{green!25} & \cellcolor{green!25} & \cellcolor{green!25} & \cellcolor{green!25} & \cellcolor{green!25} & \cellcolor{green!25} & \cellcolor{green!25} & \cellcolor{green!25} & \cellcolor{green!25} & \cellcolor{green!25} & \cellcolor{green!25} & \cellcolor{green!25} & \cellcolor{green!25} \\
\cmidrule{2-65}
\textbf{F32}  & \cellcolor{red!1!yellow!25} & \cellcolor{green!25} & \cellcolor{green!25} & \cellcolor{green!25} & \cellcolor{green!25} & \cellcolor{green!25} & \cellcolor{green!25} & \cellcolor{red!1!yellow!25} & \cellcolor{green!25} & \cellcolor{green!25} & \cellcolor{green!25} & \cellcolor{green!25} & \cellcolor{red!1!yellow!25} & \cellcolor{green!25} & \cellcolor{green!25} & \cellcolor{red!11!yellow!25} & \cellcolor{green!25} & \cellcolor{green!25} & \cellcolor{green!25} & \cellcolor{green!25} & \cellcolor{green!25} & \cellcolor{green!25} & \cellcolor{green!25} & \cellcolor{green!25} & \cellcolor{green!25} & \cellcolor{green!25} & \cellcolor{green!25} & \cellcolor{green!25} & \cellcolor{green!25} & \cellcolor{red!4!yellow!25} & \cellcolor{red!19!yellow!25} & \cellcolor{red!39!yellow!25} & \cellcolor{green!25} & \cellcolor{green!25} & \cellcolor{green!25} & \cellcolor{green!25} & \cellcolor{green!25} & \cellcolor{green!25} & \cellcolor{green!25} & \cellcolor{green!25} & \cellcolor{green!25} & \cellcolor{green!25} & \cellcolor{green!25} & \cellcolor{green!25} & \cellcolor{red!9!yellow!25} & \cellcolor{red!31!yellow!25} & \cellcolor{red!41!yellow!25} & \cellcolor{red!59!yellow!25} & \cellcolor{green!25} & \cellcolor{green!25} & \cellcolor{green!25} & \cellcolor{green!25} & \cellcolor{green!25} & \cellcolor{green!25} & \cellcolor{green!25} & \cellcolor{green!25} & \cellcolor{green!25} & \cellcolor{green!25} & \cellcolor{red!1!yellow!25} & \cellcolor{red!12!yellow!25} & \cellcolor{red!36!yellow!25} & \cellcolor{red!62!yellow!25} & \cellcolor{red!57!yellow!25} & \cellcolor{red!66!yellow!25} \\
\cmidrule{2-65}
\textbf{F64}  & \cellcolor{green!25} & \cellcolor{green!25} & \cellcolor{green!25} & \cellcolor{green!25} & \cellcolor{green!25} & \cellcolor{green!25} & \cellcolor{green!25} & \cellcolor{green!25} & \cellcolor{green!25} & \cellcolor{green!25} & \cellcolor{green!25} & \cellcolor{green!25} & \cellcolor{green!25} & \cellcolor{green!25} & \cellcolor{green!25} & \cellcolor{green!25} & \cellcolor{green!25} & \cellcolor{green!25} & \cellcolor{green!25} & \cellcolor{green!25} & \cellcolor{green!25} & \cellcolor{green!25} & \cellcolor{green!25} & \cellcolor{green!25} & \cellcolor{green!25} & \cellcolor{green!25} & \cellcolor{green!25} & \cellcolor{green!25} & \cellcolor{green!25} & \cellcolor{green!25} & \cellcolor{green!25} & \cellcolor{green!25} & \cellcolor{green!25} & \cellcolor{green!25} & \cellcolor{green!25} & \cellcolor{green!25} & \cellcolor{green!25} & \cellcolor{green!25} & \cellcolor{green!25} & \cellcolor{green!25} & \cellcolor{green!25} & \cellcolor{green!25} & \cellcolor{green!25} & \cellcolor{green!25} & \cellcolor{green!25} & \cellcolor{green!25} & \cellcolor{green!25} & \cellcolor{green!25} & \cellcolor{green!25} & \cellcolor{green!25} & \cellcolor{green!25} & \cellcolor{green!25} & \cellcolor{green!25} & \cellcolor{green!25} & \cellcolor{green!25} & \cellcolor{green!25} & \cellcolor{green!25} & \cellcolor{green!25} & \cellcolor{green!25} & \cellcolor{green!25} & \cellcolor{green!25} & \cellcolor{green!25} & \cellcolor{green!25} & \cellcolor{green!25} \\
\cmidrule{2-65}
& \multicolumn{16}{c}{\customarrowLeft{8.5em} \textbf{8} \customarrowRight{8.5em}} & \multicolumn{16}{c}{\customarrowLeft{8.5em} \textbf{9} \customarrowRight{8.5em}} & \multicolumn{16}{c}{\customarrowLeft{8.5em} \textbf{10} \customarrowRight{8.5em}} & \multicolumn{16}{c}{\customarrowLeft{8.5em} \textbf{11} \customarrowRight{8.5em}} \\
\cmidrule{2-65}
\textbf{P20}  & \cellcolor{green!25} & \cellcolor{green!25} & \cellcolor{green!25} & \cellcolor{green!25} & \cellcolor{red!10!yellow!25} & \cellcolor{red!44!yellow!25} & \cellcolor{red!38!yellow!25} & \cellcolor{red!46!yellow!25} & \cellcolor{red!57!yellow!25} & \cellcolor{red!52!yellow!25} & \cellcolor{red!63!yellow!25} & \cellcolor{red!54!yellow!25} & \cellcolor{red!52!yellow!25} & \cellcolor{red!53!yellow!25} & \cellcolor{red!49!yellow!25} & \cellcolor{red!54!yellow!25} & \cellcolor{green!25} & \cellcolor{red!2!yellow!25} & \cellcolor{red!3!yellow!25} & \cellcolor{red!18!yellow!25} & \cellcolor{red!37!yellow!25} & \cellcolor{red!56!yellow!25} & \cellcolor{red!47!yellow!25} & \cellcolor{red!60!yellow!25} & \cellcolor{red!60!yellow!25} & \cellcolor{red!61!yellow!25} & \cellcolor{red!73!yellow!25} & \cellcolor{red!65!yellow!25} & \cellcolor{red!58!yellow!25} & \cellcolor{red!67!yellow!25} & \cellcolor{red!61!yellow!25} & \cellcolor{red!68!yellow!25} & \cellcolor{red!1!yellow!25} & \cellcolor{red!8!yellow!25} & \cellcolor{red!36!yellow!25} & \cellcolor{red!47!yellow!25} & \cellcolor{red!48!yellow!25} & \cellcolor{red!60!yellow!25} & \cellcolor{red!53!yellow!25} & \cellcolor{red!51!yellow!25} & \cellcolor{red!58!yellow!25} & \cellcolor{red!50!yellow!25} & \cellcolor{red!63!yellow!25} & \cellcolor{red!56!yellow!25} & \cellcolor{red!56!yellow!25} & \cellcolor{red!54!yellow!25} & \cellcolor{red!50!yellow!25} & \cellcolor{red!51!yellow!25} & \cellcolor{red!41!yellow!25} & \cellcolor{red!47!yellow!25} & \cellcolor{red!61!yellow!25} & \cellcolor{red!63!yellow!25} & \cellcolor{red!68!yellow!25} & \cellcolor{red!76!yellow!25} & \cellcolor{red!55!yellow!25} & \cellcolor{red!67!yellow!25} & \cellcolor{red!66!yellow!25} & \cellcolor{red!64!yellow!25} & \cellcolor{red!71!yellow!25} & \cellcolor{red!71!yellow!25} & \cellcolor{red!67!yellow!25} & \cellcolor{red!65!yellow!25} & \cellcolor{red!67!yellow!25} & \cellcolor{red!68!yellow!25} \\
\cmidrule{2-65}
\textbf{P40}  & \cellcolor{green!25} & \cellcolor{green!25} & \cellcolor{green!25} & \cellcolor{green!25} & \cellcolor{green!25} & \cellcolor{green!25} & \cellcolor{green!25} & \cellcolor{green!25} & \cellcolor{green!25} & \cellcolor{green!25} & \cellcolor{green!25} & \cellcolor{green!25} & \cellcolor{green!25} & \cellcolor{green!25} & \cellcolor{green!25} & \cellcolor{green!25} & \cellcolor{green!25} & \cellcolor{green!25} & \cellcolor{green!25} & \cellcolor{green!25} & \cellcolor{green!25} & \cellcolor{green!25} & \cellcolor{green!25} & \cellcolor{green!25} & \cellcolor{green!25} & \cellcolor{green!25} & \cellcolor{green!25} & \cellcolor{green!25} & \cellcolor{green!25} & \cellcolor{green!25} & \cellcolor{green!25} & \cellcolor{green!25} & \cellcolor{green!25} & \cellcolor{green!25} & \cellcolor{green!25} & \cellcolor{green!25} & \cellcolor{green!25} & \cellcolor{green!25} & \cellcolor{green!25} & \cellcolor{green!25} & \cellcolor{green!25} & \cellcolor{green!25} & \cellcolor{green!25} & \cellcolor{green!25} & \cellcolor{green!25} & \cellcolor{green!25} & \cellcolor{green!25} & \cellcolor{green!25} & \cellcolor{green!25} & \cellcolor{green!25} & \cellcolor{green!25} & \cellcolor{green!25} & \cellcolor{green!25} & \cellcolor{green!25} & \cellcolor{green!25} & \cellcolor{green!25} & \cellcolor{green!25} & \cellcolor{green!25} & \cellcolor{green!25} & \cellcolor{green!25} & \cellcolor{green!25} & \cellcolor{green!25} & \cellcolor{green!25} & \cellcolor{green!25} \\
\cmidrule{2-65}
\textbf{F32}  & \cellcolor{green!25} & \cellcolor{green!25} & \cellcolor{green!25} & \cellcolor{green!25} & \cellcolor{green!25} & \cellcolor{green!25} & \cellcolor{green!25} & \cellcolor{green!25} & \cellcolor{green!25} & \cellcolor{red!6!yellow!25} & \cellcolor{red!34!yellow!25} & \cellcolor{red!46!yellow!25} & \cellcolor{red!58!yellow!25} & \cellcolor{red!58!yellow!25} & \cellcolor{red!66!yellow!25} & \cellcolor{red!62!yellow!25} & \cellcolor{green!25} & \cellcolor{green!25} & \cellcolor{green!25} & \cellcolor{green!25} & \cellcolor{green!25} & \cellcolor{green!25} & \cellcolor{green!25} & \cellcolor{red!1!yellow!25} & \cellcolor{red!14!yellow!25} & \cellcolor{red!41!yellow!25} & \cellcolor{red!54!yellow!25} & \cellcolor{red!59!yellow!25} & \cellcolor{red!62!yellow!25} & \cellcolor{red!68!yellow!25} & \cellcolor{red!64!yellow!25} & \cellcolor{red!65!yellow!25} & \cellcolor{green!25} & \cellcolor{green!25} & \cellcolor{green!25} & \cellcolor{green!25} & \cellcolor{green!25} & \cellcolor{green!25} & \cellcolor{red!7!yellow!25} & \cellcolor{red!33!yellow!25} & \cellcolor{red!41!yellow!25} & \cellcolor{red!51!yellow!25} & \cellcolor{red!62!yellow!25} & \cellcolor{red!64!yellow!25} & \cellcolor{red!62!yellow!25} & \cellcolor{red!69!yellow!25} & \cellcolor{red!59!yellow!25} & \cellcolor{red!66!yellow!25} & \cellcolor{green!25} & \cellcolor{green!25} & \cellcolor{green!25} & \cellcolor{green!25} & \cellcolor{green!25} & \cellcolor{red!17!yellow!25} & \cellcolor{red!35!yellow!25} & \cellcolor{red!55!yellow!25} & \cellcolor{red!60!yellow!25} & \cellcolor{red!63!yellow!25} & \cellcolor{red!67!yellow!25} & \cellcolor{red!65!yellow!25} & \cellcolor{red!68!yellow!25} & \cellcolor{red!66!yellow!25} & \cellcolor{red!66!yellow!25} & \cellcolor{red!67!yellow!25} \\
\cmidrule{2-65}
\textbf{F64}  & \cellcolor{green!25} & \cellcolor{green!25} & \cellcolor{green!25} & \cellcolor{green!25} & \cellcolor{green!25} & \cellcolor{green!25} & \cellcolor{green!25} & \cellcolor{green!25} & \cellcolor{green!25} & \cellcolor{green!25} & \cellcolor{green!25} & \cellcolor{green!25} & \cellcolor{green!25} & \cellcolor{green!25} & \cellcolor{green!25} & \cellcolor{green!25} & \cellcolor{green!25} & \cellcolor{green!25} & \cellcolor{green!25} & \cellcolor{green!25} & \cellcolor{green!25} & \cellcolor{green!25} & \cellcolor{green!25} & \cellcolor{green!25} & \cellcolor{green!25} & \cellcolor{green!25} & \cellcolor{green!25} & \cellcolor{green!25} & \cellcolor{green!25} & \cellcolor{green!25} & \cellcolor{green!25} & \cellcolor{green!25} & \cellcolor{green!25} & \cellcolor{green!25} & \cellcolor{green!25} & \cellcolor{green!25} & \cellcolor{green!25} & \cellcolor{green!25} & \cellcolor{green!25} & \cellcolor{green!25} & \cellcolor{green!25} & \cellcolor{green!25} & \cellcolor{green!25} & \cellcolor{green!25} & \cellcolor{green!25} & \cellcolor{green!25} & \cellcolor{green!25} & \cellcolor{green!25} & \cellcolor{green!25} & \cellcolor{green!25} & \cellcolor{green!25} & \cellcolor{green!25} & \cellcolor{green!25} & \cellcolor{green!25} & \cellcolor{green!25} & \cellcolor{green!25} & \cellcolor{green!25} & \cellcolor{green!25} & \cellcolor{green!25} & \cellcolor{green!25} & \cellcolor{green!25} & \cellcolor{green!25} & \cellcolor{green!25} & \cellcolor{green!25} \\
\cmidrule{2-65}
& \multicolumn{16}{c}{\customarrowLeft{8.5em} \textbf{12} \customarrowRight{8.5em}} & \multicolumn{16}{c}{\customarrowLeft{8.5em} \textbf{13} \customarrowRight{8.5em}} & \multicolumn{16}{c}{\customarrowLeft{8.5em} \textbf{14} \customarrowRight{8.5em}} & \multicolumn{16}{c}{\customarrowLeft{8.5em} \textbf{15} \customarrowRight{8.5em}} \\
\cmidrule{2-65}
\textbf{P20}  & \cellcolor{red!91!yellow!25} & \cellcolor{red!91!yellow!25} & \cellcolor{red!89!yellow!25} & \cellcolor{red!90!yellow!25} & \cellcolor{red!91!yellow!25} & \cellcolor{red!91!yellow!25} & \cellcolor{red!87!yellow!25} & \cellcolor{red!84!yellow!25} & \cellcolor{red!83!yellow!25} & \cellcolor{red!84!yellow!25} & \cellcolor{red!88!yellow!25} & \cellcolor{red!86!yellow!25} & \cellcolor{red!87!yellow!25} & \cellcolor{red!90!yellow!25} & \cellcolor{red!87!yellow!25} & \cellcolor{red!88!yellow!25} & \cellcolor{red!100!yellow!25} & \cellcolor{red!98!yellow!25} & \cellcolor{red!98!yellow!25} & \cellcolor{red!98!yellow!25} & \cellcolor{red!100!yellow!25} & \cellcolor{red!98!yellow!25} & \cellcolor{red!98!yellow!25} & \cellcolor{red!100!yellow!25} & \cellcolor{red!100!yellow!25} & \cellcolor{red!96!yellow!25} & \cellcolor{red!98!yellow!25} & \cellcolor{red!99!yellow!25} & \cellcolor{red!99!yellow!25} & \cellcolor{red!98!yellow!25} & \cellcolor{red!99!yellow!25} & \cellcolor{red!99!yellow!25} & \cellcolor{red!99!yellow!25} & \cellcolor{red!100!yellow!25} & \cellcolor{red!100!yellow!25} & \cellcolor{red!100!yellow!25} & \cellcolor{red!100!yellow!25} & \cellcolor{red!100!yellow!25} & \cellcolor{red!100!yellow!25} & \cellcolor{red!99!yellow!25} & \cellcolor{red!100!yellow!25} & \cellcolor{red!100!yellow!25} & \cellcolor{red!100!yellow!25} & \cellcolor{red!98!yellow!25} & \cellcolor{red!100!yellow!25} & \cellcolor{red!100!yellow!25} & \cellcolor{red!99!yellow!25} & \cellcolor{red!100!yellow!25} & \cellcolor{red!100!yellow!25} & \cellcolor{red!100!yellow!25} & \cellcolor{red!100!yellow!25} & \cellcolor{red!100!yellow!25} & \cellcolor{red!100!yellow!25} & \cellcolor{red!100!yellow!25} & \cellcolor{red!100!yellow!25} & \cellcolor{red!100!yellow!25} & \cellcolor{red!100!yellow!25} & \cellcolor{red!100!yellow!25} & \cellcolor{red!100!yellow!25} & \cellcolor{red!100!yellow!25} & \cellcolor{red!100!yellow!25} & \cellcolor{red!100!yellow!25} & \cellcolor{red!100!yellow!25} & \cellcolor{red!100!yellow!25} \\
\cmidrule{2-65}
\textbf{P40}  & \cellcolor{green!25} & \cellcolor{green!25} & \cellcolor{green!25} & \cellcolor{green!25} & \cellcolor{green!25} & \cellcolor{green!25} & \cellcolor{green!25} & \cellcolor{green!25} & \cellcolor{green!25} & \cellcolor{green!25} & \cellcolor{green!25} & \cellcolor{green!25} & \cellcolor{green!25} & \cellcolor{green!25} & \cellcolor{green!25} & \cellcolor{green!25} & \cellcolor{green!25} & \cellcolor{green!25} & \cellcolor{green!25} & \cellcolor{green!25} & \cellcolor{green!25} & \cellcolor{green!25} & \cellcolor{green!25} & \cellcolor{green!25} & \cellcolor{green!25} & \cellcolor{green!25} & \cellcolor{green!25} & \cellcolor{green!25} & \cellcolor{green!25} & \cellcolor{green!25} & \cellcolor{green!25} & \cellcolor{green!25} & \cellcolor{green!25} & \cellcolor{green!25} & \cellcolor{green!25} & \cellcolor{green!25} & \cellcolor{green!25} & \cellcolor{green!25} & \cellcolor{green!25} & \cellcolor{green!25} & \cellcolor{green!25} & \cellcolor{green!25} & \cellcolor{green!25} & \cellcolor{green!25} & \cellcolor{green!25} & \cellcolor{green!25} & \cellcolor{red!1!yellow!25} & \cellcolor{red!16!yellow!25} & \cellcolor{green!25} & \cellcolor{green!25} & \cellcolor{green!25} & \cellcolor{green!25} & \cellcolor{green!25} & \cellcolor{green!25} & \cellcolor{green!25} & \cellcolor{green!25} & \cellcolor{green!25} & \cellcolor{green!25} & \cellcolor{green!25} & \cellcolor{green!25} & \cellcolor{green!25} & \cellcolor{red!3!yellow!25} & \cellcolor{red!27!yellow!25} & \cellcolor{red!26!yellow!25} \\
\cmidrule{2-65}
\textbf{F32}  & \cellcolor{green!25} & \cellcolor{green!25} & \cellcolor{red!1!yellow!25} & \cellcolor{red!4!yellow!25} & \cellcolor{red!30!yellow!25} & \cellcolor{red!52!yellow!25} & \cellcolor{red!50!yellow!25} & \cellcolor{red!54!yellow!25} & \cellcolor{red!62!yellow!25} & \cellcolor{red!60!yellow!25} & \cellcolor{red!68!yellow!25} & \cellcolor{red!68!yellow!25} & \cellcolor{red!56!yellow!25} & \cellcolor{red!60!yellow!25} & \cellcolor{red!61!yellow!25} & \cellcolor{red!61!yellow!25} & \cellcolor{green!25} & \cellcolor{red!2!yellow!25} & \cellcolor{red!17!yellow!25} & \cellcolor{red!43!yellow!25} & \cellcolor{red!51!yellow!25} & \cellcolor{red!70!yellow!25} & \cellcolor{red!57!yellow!25} & \cellcolor{red!61!yellow!25} & \cellcolor{red!64!yellow!25} & \cellcolor{red!67!yellow!25} & \cellcolor{red!71!yellow!25} & \cellcolor{red!61!yellow!25} & \cellcolor{red!61!yellow!25} & \cellcolor{red!69!yellow!25} & \cellcolor{red!60!yellow!25} & \cellcolor{red!68!yellow!25} & \cellcolor{red!36!yellow!25} & \cellcolor{red!48!yellow!25} & \cellcolor{red!54!yellow!25} & \cellcolor{red!63!yellow!25} & \cellcolor{red!63!yellow!25} & \cellcolor{red!77!yellow!25} & \cellcolor{red!62!yellow!25} & \cellcolor{red!58!yellow!25} & \cellcolor{red!70!yellow!25} & \cellcolor{red!59!yellow!25} & \cellcolor{red!71!yellow!25} & \cellcolor{red!70!yellow!25} & \cellcolor{red!67!yellow!25} & \cellcolor{red!62!yellow!25} & \cellcolor{red!62!yellow!25} & \cellcolor{red!62!yellow!25} & \cellcolor{red!86!yellow!25} & \cellcolor{red!90!yellow!25} & \cellcolor{red!87!yellow!25} & \cellcolor{red!90!yellow!25} & \cellcolor{red!89!yellow!25} & \cellcolor{red!96!yellow!25} & \cellcolor{red!88!yellow!25} & \cellcolor{red!90!yellow!25} & \cellcolor{red!95!yellow!25} & \cellcolor{red!90!yellow!25} & \cellcolor{red!92!yellow!25} & \cellcolor{red!91!yellow!25} & \cellcolor{red!88!yellow!25} & \cellcolor{red!86!yellow!25} & \cellcolor{red!91!yellow!25} & \cellcolor{red!89!yellow!25} \\
\cmidrule{2-65}
\textbf{F64}  & \cellcolor{green!25} & \cellcolor{green!25} & \cellcolor{green!25} & \cellcolor{green!25} & \cellcolor{green!25} & \cellcolor{green!25} & \cellcolor{green!25} & \cellcolor{green!25} & \cellcolor{green!25} & \cellcolor{green!25} & \cellcolor{green!25} & \cellcolor{green!25} & \cellcolor{green!25} & \cellcolor{green!25} & \cellcolor{green!25} & \cellcolor{green!25} & \cellcolor{green!25} & \cellcolor{green!25} & \cellcolor{green!25} & \cellcolor{green!25} & \cellcolor{green!25} & \cellcolor{green!25} & \cellcolor{green!25} & \cellcolor{green!25} & \cellcolor{green!25} & \cellcolor{green!25} & \cellcolor{green!25} & \cellcolor{green!25} & \cellcolor{green!25} & \cellcolor{green!25} & \cellcolor{green!25} & \cellcolor{green!25} & \cellcolor{green!25} & \cellcolor{green!25} & \cellcolor{green!25} & \cellcolor{green!25} & \cellcolor{green!25} & \cellcolor{green!25} & \cellcolor{green!25} & \cellcolor{green!25} & \cellcolor{green!25} & \cellcolor{green!25} & \cellcolor{green!25} & \cellcolor{green!25} & \cellcolor{green!25} & \cellcolor{green!25} & \cellcolor{green!25} & \cellcolor{green!25} & \cellcolor{green!25} & \cellcolor{green!25} & \cellcolor{green!25} & \cellcolor{green!25} & \cellcolor{green!25} & \cellcolor{green!25} & \cellcolor{green!25} & \cellcolor{green!25} & \cellcolor{green!25} & \cellcolor{green!25} & \cellcolor{green!25} & \cellcolor{green!25} & \cellcolor{green!25} & \cellcolor{green!25} & \cellcolor{green!25} & \cellcolor{green!25} \\
\cmidrule{2-65}
\end{tabular}}}
\caption{Testing proof generation for ZKLP with resolutions $0$ to $15$ for fixed-point (P20, P40), single precision (FP32) and double precision (FP64) floating-point values. For a given resolution, we test 16 different distances from the test case to the boundary, with 100 randomly sampled test cases at each distance.
All tests are for Groth16 over BN254. Green $=$ pass all tests, Yellow and Red $=$ fail some tests. A cell with deeper red indicates more failures.}
\label{figure:resolutionComparison}
\end{figure*}

%% file: Protocols/gadget_AssertBitLength.tex
\algrenewcommand\alglinenumber[1]{\footnotesize\textbf{#1:}}


\begin{figure}[t!]
  \centering
  \footnotesize
  \fbox{
  \begin{minipage}{0.7\columnwidth}
  \underline{$\AssertBitLength(\var{v}, L)$, based on bit decomposition}
  \begin{algorithmic}[1] 
    \Require $2^{L} < p$
    \State Receive hint $\{\var{v}_0, \dots, \var{v}_{L - 1}\} = \Computation_{\mathsf{Dec}}(\var{v})$ 
    \For{$i \in [0, L - 1]$}
        \State $\var{v}_i(1 - \var{v}_i) = 0$
    \EndFor
    \State $\var{v} = \sum_{i = 0}^{L - 1}2^i \var{v}_i$
  \end{algorithmic}
  \underline{$\AssertBitLength(\var{v}, L)$, based on lookup argument}
  \begin{algorithmic}[1] 
    \Require $2^{L} < p$
    \State Receive hint $\{\var{v}'_0, \dots, \var{v}'_{L / T - 1}\} = \Computation_{\mathsf{Dec}'}(\var{v})$ 
    \For{$i \in [0, L / T - 1]$}
        \State $\mathbf{t}_\mathsf{RC} \coloneqq \mathbf{t}_\mathsf{RC} \cup \var{v}'_i$
    \EndFor
    \State $\var{v} = \sum_{i = 0}^{L/T - 1}2^{iT} \var{v}'_i$
  \end{algorithmic}
  \end{minipage}
  }
  \caption{
  Circuit for checking $\var{v} \in [0, 2^L - 1]$.
  }
  \label{figure:assert_bit_length}
\end{figure}

%% file: Protocols/gadget_Abs_Max_Min.tex
\algrenewcommand\alglinenumber[1]{\footnotesize\textbf{#1:}}

\begin{figure}[t!]
  \centering
  \footnotesize
  \fbox{
  \begin{minipage}{0.45\columnwidth}
    \underline{$\CircuitAbs(\var{v}, L)$}
      \begin{algorithmic}[1] 
        \Require $2^L < (p - 1) / 2$
        \State Receive hint $\signBitShort = \Computation_{\mathsf{GEZ}}(\var{v})$
        \State $\signBitShort(1-\signBitShort)=0$
        \State $\absoluteValueShort \coloneqq \signBitShort \mathbin{?} \var{v} : -\var{v}$
        \State $\AssertBitLength(\absoluteValueShort, L)$
        \State \Return $\signBitShort, \absoluteValueShort$
      \end{algorithmic}
  \end{minipage}
  }
  \hfill
  \fbox{
  \begin{minipage}{0.4\columnwidth}
    \underline{$\CircuitMax(\var{x}, \var{y}, L)$}
  \begin{algorithmic}[1] 
    \Require $2^L < (p - 1) / 2$
    \State $\signBitShort, \_ \coloneqq \CircuitAbs(\var{x} - \var{y}, L)$
    \State \Return $(\signBitShort) \, ? \, \var{x} \, : \, \var{y}$
  \end{algorithmic}
  \underline{$\CircuitMin(\var{x}, \var{y}, L)$}
  \begin{algorithmic}[1] 
    \Require $2^L < (p - 1) / 2$
    \State $\signBitShort, \_ \coloneqq \CircuitAbs(\var{x} - \var{y}, L)$
    \State \Return $(\signBitShort) \, ? \, \var{y} \, : \, \var{x}$
  \end{algorithmic}
  \end{minipage}
  }
  \caption{Circuits for computing sign \& absolute value $(\CircuitAbs)$ and maximum $(\CircuitMax)$ \& minimum values $(\CircuitMin)$.}
  \label{figure:combined_circuits}
\end{figure}

%% file: Protocols/gadget_Shift.tex
\algrenewcommand\alglinenumber[1]{\footnotesize\textbf{#1:}}

\begin{figure}[t!]
  \footnotesize
  \makebox[\columnwidth][r]{\fbox{
  \begin{minipage}{0.45\columnwidth}
    \underline{$2^\var{d} = 1 \shl \var{d}$}    
  \begin{algorithmic}[1] 
    \Require $\var{d} \in [0, K], 2^{K} < p$
    \State Receive hint $\var{r} = \Computation_\mathsf{Pow2}(\var{d})$
    \State $\mathbf{t}_\mathsf{Pow2} \coloneqq \mathbf{t}_\mathsf{Pow2} \cup (\var{d}, \var{r})$
    \State \Return $\var{r}$
  \end{algorithmic}
    \underline{$\var{v} \shl \var{d}$}    
  \begin{algorithmic}[1] 
    \Require $\var{v} \in [0, 2^L - 1]$, $\var{d} \in [0, K]$, $2^{L + K} < p$
    \State \Return $\var{v} \cdot 2^\var{d}$
  \end{algorithmic}
  \end{minipage}
  }
  \fbox{
  \begin{minipage}{0.55\columnwidth}
    \underline{$\var{v} \shr \var{d}$}
  \begin{algorithmic}[1] 
    \Require $\var{v} \in [0, 2^L - 1]$, $\var{d} \in [0, K]$, $2^{L + K} < p$
    \State $\var{v'} \coloneqq \var{v} \cdot 2^{K - \var{d}}$
    \State Receive hints $\var{q}, \var{r} = \Computation_{\mathsf{Div}}(\var{v}', 2^{K})$ 
    \State $\AssertBitLength(\var{q}, L)$
    \State $\AssertBitLength(\var{r}, K)$
    \State $\var{v}' = \var{q} \cdot 2^K + \var{r}$
    \State \Return $\var{q}$
  \end{algorithmic}
  \end{minipage}}}
  \caption{Circuits for shift operations} 
  \label{figure:shl_shr}
\end{figure}

%% file: Protocols/circuit_DivFloat.tex
\algrenewcommand\alglinenumber[1]{\footnotesize\textbf{#1:}}

\begin{figure}[!t]
  \centering
  \footnotesize
  \fbox{
  \begin{minipage}{0.9\columnwidth}
    \underline{$\var{\alpha} \div \var{\beta}$}
  \begin{algorithmic}[1] 
    \State $\varsign \coloneqq \varsign_\var{\alpha} \oplus \varsign_\var{\beta}$
    \State $\varexponent \coloneqq \varexponent_\var{\alpha} - \varexponent_\var{\beta}$
    \State Receive hints $\var{q}, \var{r} = \Computation_{\mathsf{Div}}(\varmantissa_\var{\alpha} \shl (M + 2), \varmantissa_\var{\beta})$
    \State $\AssertBitLength(\var{r}, M + 1)$
    \State $\AssertBitLength(\varmantissa_\var{\beta} - \var{r} - 1, M + 1)$
    \State $\varmantissa_\var{\alpha} \shl (M + 2) = \var{q} \cdot \varmantissa_\var{\beta} + \var{r}$
    \State $\varmantissa \coloneqq \var{q}$
    \State $\varabnormal \coloneqq \varabnormal_\var{\alpha} \lor \CircuitIsEq(\varmantissa_\var{\beta}, 0)$
    \State Receive hint $\var{b} = \Computation_{\mathsf{MSB}}(\varmantissa)$
    \State $\var{b}(1 - \var{b}) = 1$
    \State $\AssertBitLength(\varmantissa - (\var{b} \shl (M + 2)), M + 2)$
    \State $\varmantissa \coloneqq \var{b} \mathbin{?} \varmantissa : \varmantissa \shl 1$
    \State $\varexponent \coloneqq \varexponent + \var{b} - 1$
    \State $\Delta \varexponent \coloneqq \CircuitMax(\CircuitMin(-2^{E - 1} + 2 - \varexponent, M + 2, E + 1), E + 1)$
    \State $\varexponent', \varmantissa' \coloneqq \CircuitRoundFloat(\varexponent, \varmantissa, \Delta \varexponent, \CircuitIsEq(r, 0))$
    \State $\varabnormal' \coloneqq \varabnormal \lor \CircuitGreaterThanZero(\varexponent' - 2^{E - 1}, E + 1)$
    \State $\var{m'\_is\_0} \coloneqq \CircuitIsEq(\varmantissa', 0)$
    \State $\varexponent' \coloneqq \varabnormal' \mathbin{?} 2^{E - 1} : ((\var{m'\_is\_0} \lor \varabnormal_\var{\beta}) \mathbin{?} -2^{E - 1} + 1 - M : \varexponent')$
    \State $\varmantissa' \coloneqq \varabnormal_\var{\beta} \mathbin{?} 0 : (\varabnormal' \mathbin{?} 2^M : \varmantissa')$
    \State \Return $\varsign, \varexponent', \varmantissa', \varabnormal'$
  \end{algorithmic}
  \end{minipage}
  }
  \caption{Circuit for floating-point division}
  \label{figure:div_float}
\end{figure}

%% file: Protocols/circuit_IJK.tex
\algrenewcommand\alglinenumber[1]{\footnotesize\textbf{#1:}}

\begin{figure}[t!]
  \centering
  \footnotesize
  \fbox{
  \begin{minipage}{1.0\columnwidth}
  \underline{$\CircuitIJK(x, y)$}
  \begin{algorithmic}[1]
    \State $a_1 \coloneqq \lvert x \rvert \mathbin{;} \hspace{0.5em} a_2 \coloneqq \lvert y \rvert$
    \State $x_2 \coloneqq \frac{a_2}{\sin(\frac{\pi}{3})} \mathbin{;} \hspace{0.5em} x_1 \coloneqq a_1 + \frac{x_2}{2}$
    \State $m_1 \coloneqq \lfloor x_1 \rfloor \mathbin{;} \hspace{0.5em} m_2 \coloneqq \lfloor x_2 \rfloor$
    \State $r_1 \coloneqq x_1 - m_1 \mathbin{;} \hspace{0.5em} r_2 \coloneqq x_2 - m_2$
    \State $  r_{1, \text{A}} = (r_1 < \frac{1}{2}) \mathbin{?} \text{1} : \text{0} \mathbin{;} \hspace{0.5em}  r_{1, \text{A1}} = (r_1 < \frac{1}{3}) \mathbin{?} \text{1} : \text{0}$
    \State $r_{1, \text{B1}} = (r_1 < \frac{2}{3}) \mathbin{?} \text{1} : \text{0}$
    \State $  i_{\text{A2}, 1} = (1 - r_1 \leq r_2) \mathbin{?} \text{1} : \text{0} \mathbin{;} \hspace{0.5em}  i_{\text{A2}, 2} = (2 \cdot r_1 > r_2) \mathbin{?} \text{1} : \text{0}$
    \State $  i_{\text{B1}, 1} = (r_2 > 2 \cdot r_1 - 1) \mathbin{?} \text{1} : \text{0} \mathbin{;} \hspace{0.5em}  i_{\text{B1}, 2} = (1 - r_1 > r_2) \mathbin{?} \text{1} : \text{0}$
    \State $ i_{\text{A}} = (i_{\text{A2}, 2} \mathbin{?} m_1 + 1 : m_1) $ 
    \State $i_{\text{B}} = (i_{\text{B1}, 1} \mathbin{?} (i_{\text{B1}, 2} \mathbin{?} m_1 : m_1 + 1) : m_1 + 1)$
    \State $ i = r_{1, \text{A}} \mathbin{?} (r_{1, \text{A1}} \mathbin{?} m_1 : i_{\text{A}}) : (r_{1, \text{B1}} \mathbin{?} i_{\text{B}} : m_1 + 1)$
    \State $ j_A = (\frac{r_1 + 1}{2} > r_2) \mathbin{?} \text{1} : \text{0}\mathbin{;} \hspace{0.5em} j_B = (1 - r_1 > r_2) \mathbin{?} \text{1} : \text{0}$
    \State $ j_C = (\frac{r_1}{2} > r_2) \mathbin{?} \text{1} : \text{0}$
    \State $ j_{\text{A}} = r_{1, \text{A1}} \mathbin{?} ((j_A) \mathbin{?} m_2 : m_2 + 1) : ((j_B) \mathbin{?} m_2 : m_2 + 1)$
    \State $ j_{\text{B}} = r_{1, \text{B1}} \mathbin{?} ((j_B) \mathbin{?} m_2 : m_2 + 1) : ((j_C) \mathbin{?} m_2 : m_2 + 1)$
    \State $ j = r_{1, \text{A}} \mathbin{?} j_{\text{A}} : j_{\text{B}}$
    \State $  i_{>} = (j < i) \mathbin{?} \text{1} : \text{0} $
    \State $  i_{-} = (x < 0) \mathbin{?} ((y < 0) \mathbin{?} 1 : i_{>}) : ((y < 0) \mathbin{?} (1 - i_{>}) : 0)$
    \State $ i_A = (y < 0) \mathbin{?} (i_{>} \mathbin{?} (i - j) : (j - i)) : -i\mathbin{;} \hspace{0.5em} i_B = (i_{>} \mathbin{?} (i - j) : (j - i))$
    \State $ i = (x < 0) \mathbin{?} i_A : ((y < 0) \mathbin{?} i_B : i)$
    \State \Return $\CircuitNormalizeIJK(i_{-}, i, y.S, j, 0, 0)$
  \end{algorithmic}
  \end{minipage}
  }
  \caption{Sub-Circuit for computing the conversion of two dimensional hexagon coordinates $x,y$ to three dimensional coordinates $i,j,k$. Primitive Operations are floating-point.}
  \label{figure:IJK}
\end{figure}

%% file: Protocols/circuit_NormalizeIJK.tex
\algrenewcommand\alglinenumber[1]{\footnotesize\textbf{#1:}}

\begin{figure}[t!]
  \centering
  \footnotesize
  \fbox{
  \begin{minipage}{1.0\columnwidth}
  \underline{$\CircuitNormalizeIJK(i_{-}, i, j_{-}, j, k_{-}, k)$}
  \begin{algorithmic}[1]
    \State $  i_{>j} = (j < i) \mathbin{?} \text{1} : \text{0}\mathbin{;} \hspace{0.5em}  i_{>k} = (k < i) \mathbin{?} \text{1} : \text{0}$
    \State $  j_{A} = (j_{-}) \mathbin{?} ((i_{>j}) \mathbin{?} (i - j) : (j - i)) : (i + j)\mathbin{;} $
    \State $j_{A-} = (j_{-}) \mathbin{?} (1 - i_{>j}) : 0$
    \State $  k_{A} = (k_{-}) \mathbin{?} ((i_{>k}) \mathbin{?} (i - k) : (k - i)) : (i + k)\mathbin{;} $
    \State $ k_{A-} = (k_{-}) \mathbin{?} (1 - i_{>k}) : 0$
    \State $  i = (i_{-}) \mathbin{?} 0 : i$
    \State $  j = (i_{-}) \mathbin{?} j_{A} : j\mathbin{;} \hspace{0.5em}  j_{-} = (i_{-}) \mathbin{?} j_{A-} : j_{-}$
    \State $k = (i_{-}) \mathbin{?} k_{A} : k\mathbin{;} \hspace{0.5em}  k_{-} = (i_{-}) \mathbin{?} k_{A-} : k_{-}$
    \State $  i = (j_{-}) \mathbin{?} (i + j) : i\mathbin{;} \hspace{0.5em}  j = (j_{-}) \mathbin{?} 0 : j\mathbin{;} \hspace{0.5em}  k = (j_{-}) \mathbin{?} k_{A} : k $
    \State $k_{-} = (j_{-}) \mathbin{?} k_{A-} : k_{-}$
    \State $  i = (k_{-}) \mathbin{?} (i + k) : i\mathbin{;} \hspace{0.5em}  j = (k_{-}) \mathbin{?} (j + k) : j\mathbin{;} \hspace{0.5em}  k = (k_{-}) \mathbin{?} 0 : k$
    \State $  \text{min} = (i_{>j}) \mathbin{?} j : i$
    \State $  \text{min} = (k < \text{min}) \mathbin{?} k : \text{min}$
    \State $  i = i - \text{min}\mathbin{;} \hspace{0.5em}  j = j - \text{min}\mathbin{;} \hspace{0.5em}  k = k - \text{min}$
    \State \Return $[i, j, k]$
  \end{algorithmic}
  \end{minipage}
  }
  \caption{Normalization Sub-Circuit for adjusting $i, j, k$ coordinates. Primitive Operations are floating-point.
  }
  \label{figure:CircuitNormalizeIJK}
\end{figure}